\documentclass[preprints,article,accept,moreauthors,pdftex]{Definitions/mdpi}
\pdfoutput=1
\usepackage{todonotes}
\reversemarginpar
\usepackage{comment}
\usepackage{bm}
\usepackage[normalem]{ulem}  % for striking out
\usepackage{xcolor}
\usepackage{tikz}
\usepackage{[caption}
\usetikzlibrary{arrows}
\usetikzlibrary{decorations.markings}
\usetikzlibrary{decorations.pathmorphing}

\firstpage{1}
\pubvolume{1}
\issuenum{1}
\articlenumber{0}
\pubyear{2021}
\copyrightyear{2021}
\datereceived{}
\dateaccepted{}
\datepublished{}
\hreflink{https://doi.org/} % If needed use \linebreak

\graphicspath{{figures/}}
\definecolor{ForestGreen}{RGB}{30,139,30}
\newcommand {\Ekin}       {\mbox{$E_{\rm kin}$}}
\newcommand {\GeVc}       {\mbox{$\rm GeV\!/\!c$}}

\newcommand {\keVc}       {\mbox{$\rm keV\!/\!c$}}
\newcommand {\kmax}       {\mbox{$k_{\rm max}$}}
\newcommand {\MeVc}       {\mbox{$\rm MeV\!/\!c$}}

\newcommand {\MuToEm}     {\mbox{$\mu^- \ra e^-$}}
\newcommand {\MuToEmConv} {\mbox{$\mu^- A \ra e^- A$}}

\newcommand {\ra}         {\rightarrow}
\newcommand {\Rmue}       {\mbox{$R_{\mu e}$}}

\Title{Mu2e Run I Sensitivity Projections for the Neutrinoless \MuToEm\ Conversion Search in Aluminum.\\
}

\TitleCitation{Mu2e Run I Sensitivity Projections for the Neutrinoless Muon to Electron Conversion Search in Aluminum}

\Author{
F. Abdi $^{26}$ ,
R. Abrams $^{28}$ ,
J. Adentunji $^{11}$ ,
W. Ahmed $^{15}$ ,
R. Alber $^{11}$ ,
D. Alexander $^{15}$ ,
D. Allen $^{11}$ ,
D. Allspach $^{11}$ ,
C. Alvarez-Garcia $^{23}$ ,
D. Ambrose $^{26}$
% ...

\secondnote{Deceased.}

  \hspace{3.in} (Mu2e Collaboration)
}

\AuthorNames{Mu2e Collaboration}

\AuthorCitation{Mu2e Collaboration}

\address{$^{1}$ \quad Argonne National Laboratory;            \\
% ...
$^{11}$ \quad Fermi National Accelerator Laboratory;        \\
% ...
$^{15}$ \quad University of Houston;                         \\
% ...
$^{23}$ \quad University of Manchester;        \\
% ...
$^{26}$ \quad University of Minnesota;        \\
% ...
$^{28}$ \quad Muons Inc.;        \\
% ...
}

\corres{Correspondence: {murat@fnal.gov}}

\abstract{
  The Mu2e experiment at Fermilab will search for the
  neutrinoless \MuToEm\ conversion in the field of an aluminum nucleus.
  The Mu2e data-taking plan assumes two running periods,
  Run I and Run II, separated by an approximately two-year-long shutdown.
  This paper presents an estimate of the expected Mu2e Run I search sensitivity and
  includes a detailed discussion of the background sources, uncertainties of their prediction,
  analysis procedures, and the optimization of the experimental sensitivity.
  The expected Run I $5 \sigma$ discovery sensitivity is $\Rmue = 1.2 \times 10^{-15}$, 
  with a total expected background of $0.11 \pm 0.03$ events.
  In the absence of a signal, the expected upper limit is $\Rmue < 6.2 \times 10^{-16}$
  at 90\% CL. This represents a three order of magnitude improvement over the current experimental
  limit of $\Rmue < 7 \times 10^{-13}$ at 90\% CL set by the SINDRUM II experiment.
}


\begin{document}
\section {Introduction}

Experimental observation of quark mixing and neutrino oscillations proves that interactions of the Standard Model (SM)
fermions are non-diagonal in flavor.
%% \del{
%% Large mixing angles in the quark and neutrino sectors,  $sin^2 \theta_C {\sim} 0.25 ,
%% \theta_{13} {\sim} 0.1$ {\red (insert references, put more accurate numbers)}, make interactions of charged leptons,
%% which do not show any evidence of flavor non-conservation and cross-generation mixing, to stand out.
%% }
Cross-generational mixing in the quark and neutrino sectors is large,
$|V_{us}| {\sim} 0.2$ \cite{TOOLS_2020_RPP_CKM} and \allowbreak $ \sin^2 \theta_{23} {\sim} 0.6$~\cite{TOOLS_2020_NUFIT}.
In striking contrast, no indication of flavor mixing has been observed in the charged lepton sector.
In the SM with massive neutrinos,  charged lepton flavor is  only approximately conserved. Virtual loops with mixing neutrinos result in charged lepton flavor violating (CLFV) transitions, regardless of whether
neutrinos are Dirac or Majorana particles \cite{CLFV_2014_HAMBYE,1974_CLFV_Eliezer_Ross}.
The branching fractions of the corresponding processes are suppressed by factors proportional to ${(\Delta m_{\nu}^2})^2/M_W^4$
to a level below $10^{-50}$ \cite{2008_CLFV_Marciano}, significantly lower than the sensitivity
of any current or planned experiment. Experimental observation of any CLFV process would therefore imply
the presence of
%% \del{non-SM physics beyond that leading to neutrino masses.}
physics beyond the SM.
Many extensions of the SM predict much higher rates of CLFV processes \cite{2013_CLFV_DEGOUVEA_VOGEL},
falling within the reach of the new generation of CLFV experiments coming online within the next few years
\cite{2010_CLFV_DEEME_PROPOSAL, 2013_CLFV_MU3E_PROPOSAL, 2015_Mu2e_TDR, 2018_CLFV_MEG_II_DETECTOR, 2020_CLFV_COMET_TDR}.
The process of coherent neutrinoless muon to electron conversion  in a nuclear field, \MuToEmConv,
probes a wide spectrum of new physics models (see Ref.~\cite{2001_CLFV_BSM_Kuno} for general calculations).
The present experimental limit on the rate of this process

$$
R_{\mu e} = \frac{\Gamma(\mu^{-} + N(A,Z) \rightarrow e^{-} + N(A,Z))}{\Gamma(\mu^{-} + N(A,Z) \rightarrow \nu_{\mu} + N(A,Z-1))} ~<~ 7 \times  10^{-13} ~ (\rm 90\% ~CL)
$$
\noindent
has been set by the SINDRUM II experiment on a gold target \cite{2006_CLFV_SINDRUM_II_GOLD}.

The Mu2e experiment at Fermilab \cite{2015_Mu2e_TDR} will search for \MuToEmConv\ on an aluminum target with an improved sensitivity of
about four orders of magnitude below the SINDRUM II limit. The current Mu2e run plan assumes two data-taking periods,
Run I and Run II, separated by an approximately two-year-long shutdown. Run I is anticipated to start in 2025 and collect
about 10\% of the total expected muon flux, improving the search sensitivity by three orders of magnitude.
Run II will further enhance the search sensitivity by another order of magnitude.

This article details estimates of the expected backgrounds and the sensitivity projections for Mu2e Run I.
The material is organized as follows. Section~\ref{mu2e_detector} describes the Mu2e experiment and the run plan.
%
% \todo[color=green!20,linecolor=green,tickmarkheight=10pt] {Sophie - "running plan" or "run plan"?}
%
Section~\ref{sec:simulation} presents an overview of the event simulation framework.
Sections \ref{sec:reconstruction}, \ref{trigger}, and \ref{sec:track-selection_cuts_summary}
contain discussion of the event reconstruction, trigger simulation, and event selection, respectively.
Section~\ref{sec:background_estimates} describes the background processes, details of their simulation,
and gives the estimated contributions from each background source.
Section~\ref{sec:sensitivity_optimization} presents the sensitivity optimization procedure and discussion of the results.

%%% Local Variables:
%%% mode: latex
%%% TeX-master: t
%%% End:

%%%%%%%%%%%%%%%%%%%%%%%%%%%%%%%%%%%%%%%%%%%%%%%%%%%%%%%%%%%%%%%%%%%%%%%%%%%%%% 
\section {Mu2e Experiment}
\label{mu2e_detector}

%%%%%%%%%%%%%%%%%%%%%%%%%%%%%%%%%%%%%%%%%%%%%%%%%%%%%%%%%%%%%%%%%%%%%%%%%%%%%% 
\subsection{Muon Beamline}

\begin{figure}[ht]
  \hspace{-0.15in}
  \begin{tikzpicture}
    \node[anchor=south west,inner sep=0] at (0,0.) {
      % \node[shift={(0 cm,0.cm)},inner sep=0,rotate={90}] at (0,0) {}
      % \makebox[\textwidth][c] {
      \includegraphics[width=1.06\linewidth]{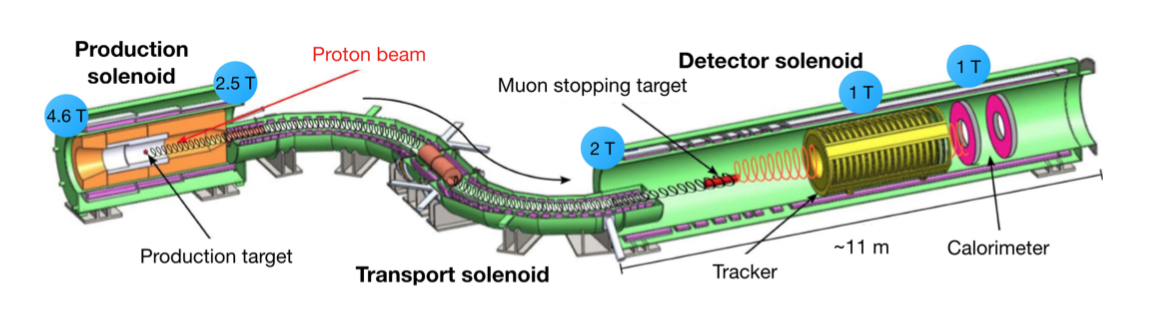}
      % }
    };
    \draw [-stealth, line width=0.5pt](5.35,1.90) -- (5.35,2.55);
    \draw [-stealth, line width=0.5pt](5.35,1.90) -- (5.95,2.00);
    \draw [-stealth, line width=0.5pt](5.35,1.90) -- (5.05,2.20);                  % (5.36,1.48);

    \path
    (5.75, 2.55)  node[left ] {$y$}
    (5.90, 1.93)  node[above] {$z$}
    (5.15, 2.10)  node[left ] {$x$};

    \path
    (2.85,2.10) node[below]{\bf\small COL1}
    (5.25,1.70) node[left ]{\bf\small COL3}
    (7.35,1.20) node[below]{\bf\small COL5};

  \end{tikzpicture}
  \caption{
    \label{fig:mu2e_layout}
    Schematic view of the Mu2e apparatus.
    The center of the Mu2e reference frame is located in the COL3 collimator center,
    its $y$-axis points upwards, the $z$-axis is parallel to the DS axis and points downstream,
    and the $x$-axis completes the right-handed reference frame. The particle detectors,
    the tracker and the calorimeter, are located in the downstream part of the DS,
    in a uniform magnetic field of 1 T.
  }
\end{figure}
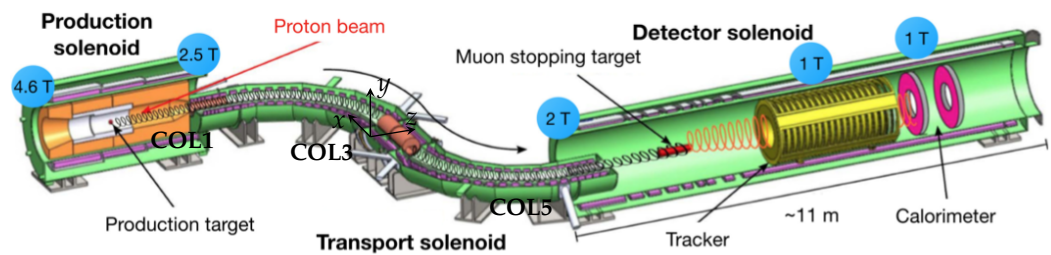

The Mu2e experiment is based upon a concept proposed in Ref.~\cite{1989_CLFV_Dzhilkibaev_Lobashev}.
A schematic view of the experiment is shown in Figure~\ref{fig:mu2e_layout}.
Formation of the Mu2e muon beam proceeds as follows.
A primary proton beam with \Ekin\ = 8 GeV is extracted from the Fermilab Delivery Ring
using the slow resonant extraction technique \cite{TOOLS_2019_PhysRevAccelBeams.22.043501}. 
The beam has a pulsed timing structure, with 250 ns-wide proton pulses separated by 1695 ns. 
% Jim Miller v1.4
During each 1.4 s main injector cycle, the proton pulses are delivered continuously 
for about 0.4 seconds, then the beam is off for the remainder of the cycle.
On a millisecond time scale, slow resonant extraction results in significant 
proton pulse intensity variations \cite{2012_TOOLS_arxiv.1207.6621}.
The spill duty factor ${\rm SDF} = 1/(1+\sigma_{I}^2/I_0^2)$,
where $\sigma_I^2$ is the variance of the pulse intensity distribution
and $I_0$ is the mean pulse intensity, is expected to be above 60\%.

The beam interacts with the ${\sim} 1.6$ interaction lengths-long tungsten
production target positioned in the center of the superconducting production
solenoid (PS).
The PS graded magnetic field reaches its maximal strength of 4.6 T downstream of the production target.
Most of the particles produced in $pW$ interactions are pions.
Particles produced backwards as well as reflected in the PS magnetic mirror
travel through the S-shaped superconducting transport solenoid (TS) towards
the superconducting detector solenoid (DS).
Muons are mainly produced in $\pi^- \ra \mu^- \nu$ decays, which occur in both the PS and TS.
The TS magnetic field is also graded, from ${\sim} 2.5$ T at the entrance to about 2.1 T
in the region where particles exit the TS and enter the DS.
Collimators at the entrance, center, and exit of the TS (COL1, COL3, and COL5) define
the TS momentum acceptance, greatly reducing the transport efficiency for particles
with momenta above ${\sim} 100~\MeVc$.
The curved magnetic field of the TS causes the charged particles of opposite signs to
drift vertically in opposite directions -- see, for example, Ref. \cite{BOOKS_1998_Jackson_Electrodynamics}.
The vertical separation reaches its maximum in the center of the TS.
A vertically offset opening of the rotatable COL3 collimator selects 
% \todo[color=gray!20,linecolor=gray,tickmarkheight=10pt] {
%   (Michael) Perhaps ``allows for the selection of'' or ``allows one to select?''
% }
the beam sign, passing through either negative or positive particles.
The DS magnetic field has two regions -- an upstream region with a graded magnetic
field and a downstream region with a uniform field of 1 T.

The inner volumes of all three solenoids are kept at near vacuum.
Exposed to the intense proton beam, the radiatively cooled production target
will operate at temperatures above 1000$^o~$C.
Maintaining a low tungsten oxidation rate requires the pressure in the PS region to be kept at ${\sim}10^{-5}$ torr.
To optimize the transport efficiency, suppress backgrounds from secondary interactions,
and improve the momentum reconstruction accuracy, the pumping system for the DS region 
is designed to achieve $10^{-4}$ torr.
A thin window in the TS center separates the two vacuum regions.

The stopping target is positioned in the graded B-field region of the DS.
The average momentum of the muons entering the DS is ${\sim} 50 ~\MeVc$,
and about 1/3 of them stop in the stopping target made of 37 Al annular foils
spaced 2.2 cm apart.
Each foil is 105 $\mu$m thick and has an inner and an outer radii of 2.2 cm and 7.5 cm
respectively. The foils are arranged co-axially along the DS axis.

Muons reaching the stopping target and stopping there come from decays of pions with
an average momentum $p {\sim}100$ \MeVc.
The average number of stopped muons per primary proton, that is the {\em stopped muon rate},
determined from the muon beam simulations is $N^{\mu^-}_{\rm POT} = 1.6\times 10^{-3}$.
This number highly depends on the pion production cross section for the protons interacting
on the tungsten target. Published measurements of the low-momentum pion
production \cite{PION_PRODUCTION_2008_PhysRevC.77.055207,PION_PRODUCTION_1991_Armutliiski}
are not consistent with each other, so the simulation-based estimate of $N^{\mu^-}_{POT}$
has a large uncertainty. The impact of this uncertainty on the expected sensitivity is discussed
in Section \ref{sec:sensitivity_estimate}.

In addition to charged pions, interactions of the proton beam with the production target 
also produce a large number of $\pi^0$'s.
Photons from $\pi^0 \to \gamma \gamma$ decays converting in the target 
result in a flash of low momentum electrons and positrons traveling through the TS
and reaching the detector within 150-200 ns from production, as seen in Figure~\ref{fig:beam_struc}.
Upon arrival to the DS, the beam flash overwhelms the detector, producing spikes
in the detector occupancy.
Another consequence of the beam flash is long-term radiation damage to the detectors.
Both effects are primarily due to electron bremsstrahlung in the stopping target 
foils.
A significant fraction of the beam flash particles pass through the holes in the foils,
reducing the radiation dose absorbed by the detectors by about 30\%.
% \textcolor{blue}{
%   The annular structure of the tracker and the calorimeter helps to
%   further reduce the dose absorbed by these two detectors.
% } 

%%%%%%%%%%%%%%%%%%%%%%%%%%%%%%%%%%%%%%%%%%%%%%%%%%%%%%%%%%%%%%%%%%%%%%%%%%%%%%
\subsection{Signal and Main Backgrounds}

\begin{figure}[H]
  \centering
  \includegraphics[width=1.0\linewidth]{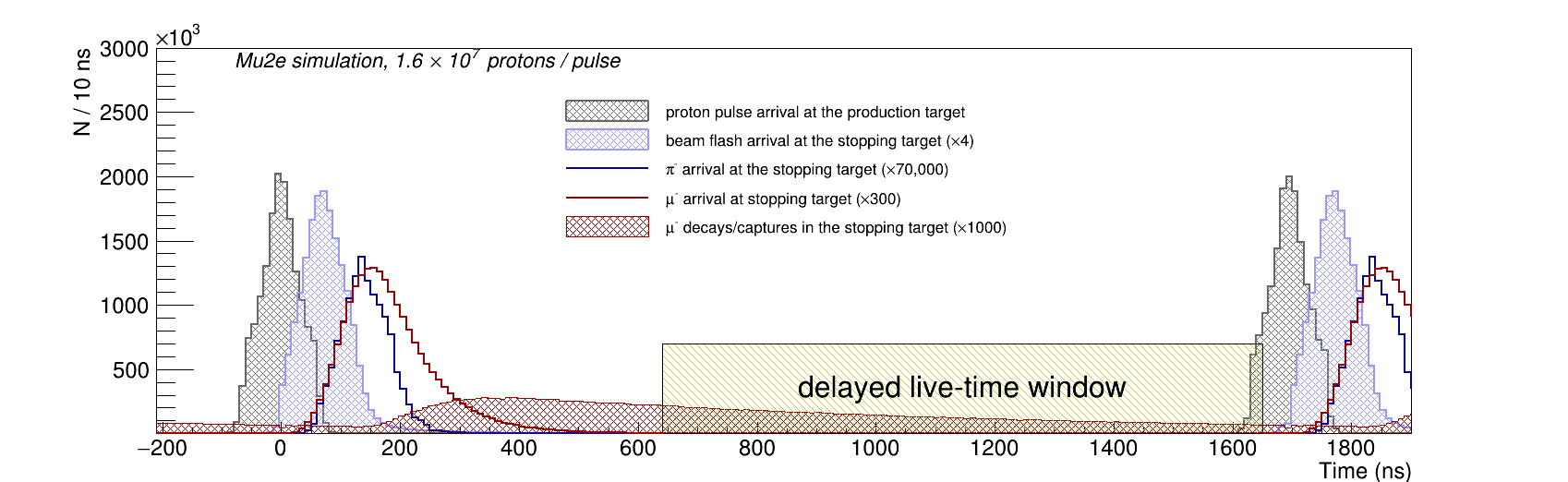}
  \caption{Proton pulses arrive at the production solenoid 1695 ns apart.
    A delayed live-time window suppresses the beam-related background. }
  \label{fig:beam_struc}
\end{figure}

Muons stopped in the target foils rapidly cascade to a 1s orbit in the Al atoms
and could undergo the process of \MuToEm\ conversion.
Because in the process of coherent conversion the outgoing nucleus remains
in the ground state, the experimental signature of the process is a monochromatic
conversion electron (CE) with energy

\begin{equation}
E_{CE} = m_{\mu} - E_{recoil} - E_{bind},
\end{equation}
where $m_{\mu}$ is the muon mass, $E_{recoil}$ is the recoil energy of the target
nucleus, and $E_{bind}$ is the binding energy of the 1s state of the muonic atom.
For the Mu2e stopping target material, $^{27}$Al, $E_{CE} = 104.97$ MeV \cite{2011_DIO_Czarnecki}.
Radiative corrections to the conversion electron spectrum have been calculated and are discussed in Ref.~\cite{szafron}.
105 MeV electrons could also come from a number of background processes.
\begin{itemize}
\item
  Cosmic particles interacting and decaying in the detector volume are a source of electrons
  whose momentum spectrum covers the region around 100 \MeVc. Most cosmic particles entering
  the detector are muons; suppression of the cosmic background requires identifying
  muons and vetoing them.
\item
  Decays in orbit (DIO) of muons stopped in the stopping target
  and captured by the Al atoms produce electrons with a momentum 
  spectrum extending up to $E_{CE}$ and rapidly falling towards the spectrum endpoint.
  Observing a peak from \MuToEm\ conversion in the presence of the DIO background
  requires searching for the signal in a 1-2 \MeVc\ wide momentum window
  and a detector with an excellent momentum resolution $\Delta{p}$,
  full width at half maximum (FWHM), $\lesssim$ 1 \MeVc.
\item
  Antiprotons produced by the proton beam and annihilating either in the stopping target
  or the TS also generate ${\sim} 100~\MeVc$ electrons. The antiproton background is suppressed
  by several absorption elements installed in the TS.
\item
  Radiative capture of pions (RPC) contaminating the muon beam and stopping in the Al target
  generates a significant background which rapidly falls in time.
  Suppressing the RPC background requires the live-time window to be delayed with respect
  to the proton pulse arrival at the production target by several hundred nanoseconds,
  as schematically shown in Figure~\ref{fig:beam_struc}.
  The delayed live-time window
  technique is not efficient against secondary particles produced by
  protons arriving at the production target between the proton pulses.
  Suppressing the contribution of those protons requires the proton beam
  extinction $\zeta < 1 \times 10^{-10}$, where $\zeta$ is 
  the relative fraction of the beam protons between the pulses.
\item
  Electrons with momenta ${\sim} 100~\MeVc$ entering the DS and scattering in the Al stopping target.
  Similar to RPC, suppressing this background requires the delayed live-time window and
  an excellent proton beam extinction.
\item
  Decays in flight of negative muons and pions entering the DS and producing electrons with $p > 100$ \MeVc.
\item
  Radiative muon capture (RMC), a process analogous to RPC, but with a lower maximal energy. In aluminum this energy is $\sim$ 102 MeV.
\end{itemize}

The physics processes listed above have very different timing dependencies.
The rates of RPC, beam electrons, and decays in flight are strongly correlated
with the time of the proton pulse arrival at the production target.
The time dependence of the \MuToEm\ conversion signal, DIO, and RMC
are all determined by the lifetime of a muonic Al atom, $864 \pm 1$ ns \cite{RMC_1987_MUON_CAPTURE_RATES_SUZUKI}.
Cosmic background events are distributed uniformly in time.

%%%%%%%%%%%%%%%%%%%%%%%%%%%%%%%%%%%%%%%%%%%%%%%%%%%%%%%%%%%%%%%%%%%%%%%%%%%%%%
\subsection{Detector}

Momenta of the secondary charged particles produced by decays of nuclear interactions
of muons stopped in the stopping target are measured by the
straw tracker, located about 3 m downstream of the stopping target in the uniform 1 T region
of the DS magnetic field. The tracker is approximately 3 m long and consists of 18 tracking stations,
covering radii between 38~cm and 68~cm.
It is constructed out of 5~mm diameter straw tubes of different lengths,  20,736 straws in total,
filled with a 80\%:20\% Ar:CO$_2$ mixture at a pressure of 1 atm.
Each straw is read out from both ends, providing two timing measurements for each hit.
The difference between the two measured times is used to reconstruct the hit coordinate along the straw.
For 100 \MeVc\ electrons, the intrinsic momentum resolution of the tracker
is expected to be $\Delta p_{\rm trk} < 300 ~\keVc$ FWHM. For muons of the same momentum,
the resolution is slightly worse due to higher energy losses.

Protons from muon captures in the stopping target generate a significant charge
load on the tracker. The charge load is reduced by a cylindrical-shaped polyethylene
proton absorber placed approximately half-way between the stopping target
and the tracker. The proton absorber is 0.5 mm thick, with a radius of 30 cm
and a length of 100 cm. Fluctuations of energy losses in the stopping target
and the proton absorber dominate the expected momentum resolution
in the production vertex $\Delta p {\sim} 950 ~\keVc$ FWHM at 100 \MeVc.

The electromagnetic calorimeter, constructed out of two annular disks
covering radii from 37 cm to 66 cm and separated by 70 cm,
is positioned immediately downstream of the tracker.
Each disk is assembled from 674 undoped CsI crystals, 3.3$\times$3.4$\times$20 cm$^3$ in size
and read out by two silicon photomultipliers (SiPMs).
Tests of the calorimeter prototype using an electron beam have demonstrated, at 100 MeV,
energy resolution $\Delta{E}/E = 16.4\%$ FWHM, dominated by energy leakage,
and timing resolution $\sigma_T = 110$ ps \cite{2017_TOOLS_MU2E_CALORIMETER}.
The inner radius of the instrumented detector region is limited by the rapidly
increasing occupancy due to DIO and the radiation damage induced by the beam flash.

Combined together, measurements in the tracker and in the calorimeter provide efficient particle
identification and are expected to reduce the background from muons misidentified as electrons
down to a negligible level.

\begin{figure}[H]
  \centering
  \includegraphics[width=0.8\linewidth]{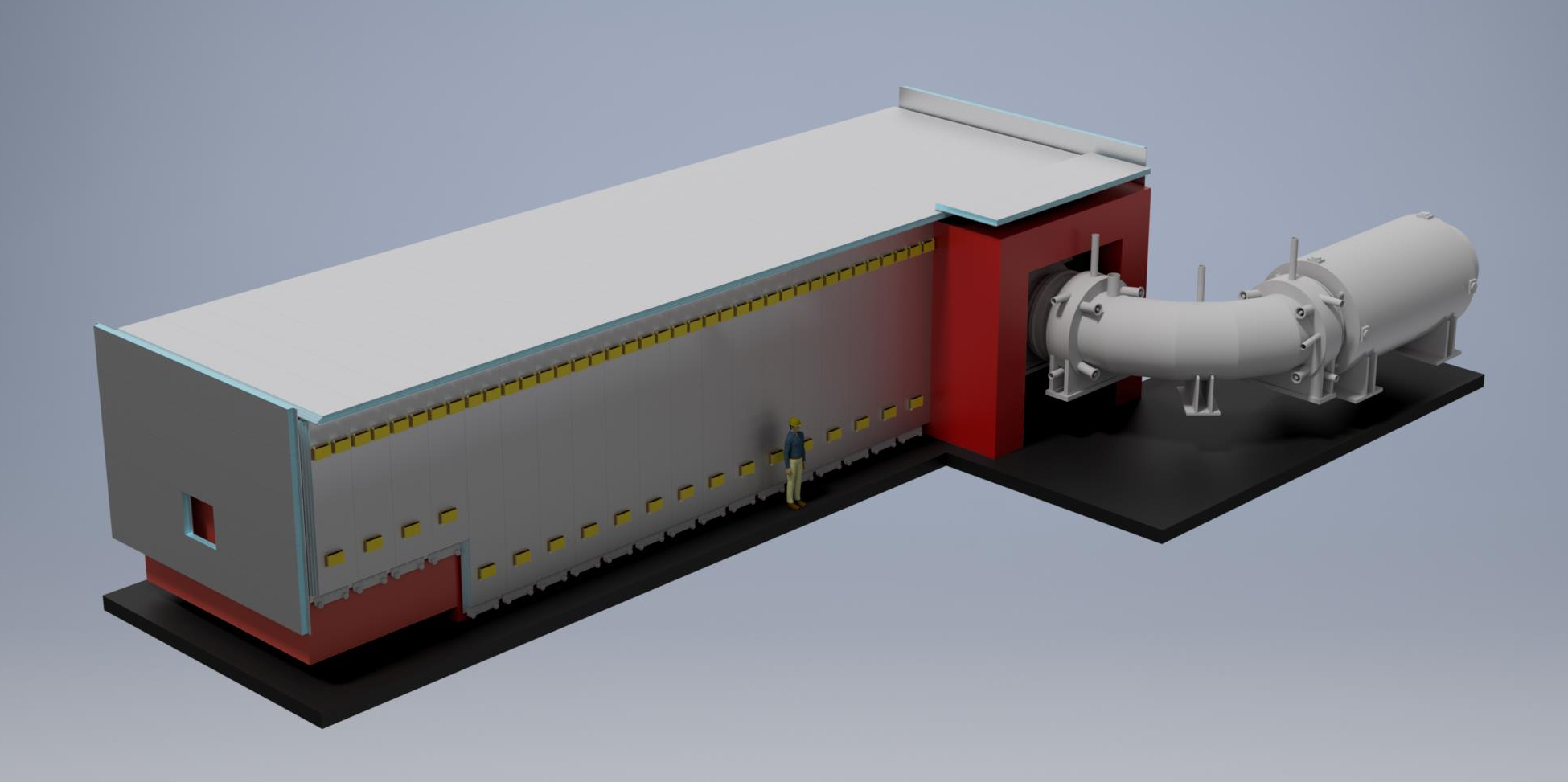}
  \caption{View of the CRV enclosing the Mu2e detector region. The
    Transport Solenoid region is also shown. Note the gap in the
    CRV coverage to permit the entrance of the TS cryostat.}
  \label{fig:crv}
\end{figure}

For the experiment to reach its design sensitivity, the Cosmic Ray Veto
system (CRV), shown in Figure~\ref{fig:crv},  must suppress the cosmic ray background by four orders of magnitude.
The CRV consists of four layers of extruded plastic scintillation counters outfitted
with wavelength-shifting fibers \cite{2018_TOOLS_CRV_PE_YIELD} and read out by SiPMs.
%
% The scintillators are glued into CRV modules.

The proton beam extinction is monitored using a magnetic spectrometer with silicon pixel detectors
positioned downstream and off-axis of the primary proton beam. The extinction monitor is
described in more detail in Ref. \cite{2015_Mu2e_TDR}.
The stopped muon flux is measured by a high purity Ge detector and a LaBr$_{3}$ detector,
located about 30 m downstream of the stopping target, which detect photons emitted
in the process of $\mu^-$ capture in Al.

The data read out from the Mu2e subdetectors are digitized and zero-suppressed
by the front-end electronics and transmitted from the detector via optical fibers
to the data acquisition system (DAQ).
The Mu2e event builder combines the data read out between the two consecutive proton
pulses into one event and sends assembled events to a one-level software trigger.
To reduce the DAQ rates, the detector readout starts about 500 ns after the proton pulse
arrival at the production target when the flux of beam flash particles have already subsided.

A detailed description of the apparatus can be found in the Mu2e Technical Design
Report (TDR) \cite{2015_Mu2e_TDR}.

%%%%%%%%%%%%%%%%%%%%%%%%%%%%%%%%%%%%%%%%%%%%%%%%%%%%%%%%%%%%%%%%%%%%%%%%%%%%%%
\subsection{Mu2e Run I Data-Taking Plan}
\label{experiment_config}
The Mu2e data-taking plan assumes two running periods, Run I and Run II,
separated by an approximately two-year-long shutdown.
According to the Run I plan, the experiment will start taking data using a
low intensity proton beam with a mean intensity of $1.6 \times 10^7$ protons/pulse.
Starting at a lower beam intensity facilitates the commissioning of the experiment.
During the second part of Run I, the delivered beam will have a higher intensity,
with a mean of $3.9\times 10^7$ protons/pulse.
About 75\% of the total number of protons on target will be delivered in the 
low intensity running mode, and about 25\% in the high intensity running mode. 
Table \ref{tab:running_time} summarizes the expected Run~I conditions for the two running modes.
\begin{specialtable}[H]
  \centering
  \small
  \caption{
    \label{tab:running_time}
    Expected running time, proton counts, and stopped muon counts for Mu2e Run I. The running time is the
    time, in seconds, during which the experiment is running and taking data.
    The numbers in the last two columns do not include the trigger, reconstruction, and selection efficiency.
  }
  \begin{tabular}{lcccc}
    \toprule
    Running mode        & Mean proton pulse    &  Running time (s)  & N(POT)                & N(stopped muons)    \\
                        &  intensity           &                    &                       &                     \\
    \midrule
    Low intensity       & $1.6 \times 10^7$    & $ 9.5 \times 10^6 $ & $2.9 \times 10^{19}$   & $4.6 \times 10^{16}$ \\
    \midrule
    High intensity      & $3.9 \times 10^7$    & $ 1.6 \times 10^6 $ & $9.0 \times 10^{18}$   & $1.4 \times 10^{16}$ \\
    \midrule
    Total               &                      & $ 11.1 \times 10^6 $ & $3.8 \times 10^{19}$   & $6.0 \times 10^{16}$ \\
    \bottomrule
  \end{tabular}
%    1              & $1.6 \times 10^7$     & $ 2.237\times10^6 $ & $2.86 \times 10^{19}$   & $4.55 \times 10^{16}$ \\
%    2              & $1.6 \times 10^7$     & $ 6.650\times10^6 $ &                       &                     \\
\end{specialtable}

%%% Local Variables:
%%% mode: latex
%%% TeX-master: t
%%% End:

%%%%%%%%%%%%%%%%%%%%%%%%%%%%%%%%%%%%%%%%%%%%%%%%%%%%%%%%%%%%%%%%%%%%%%%%%%%%%%
\section{Simulation Framework}
\label{sec:simulation}
The Mu2e simulation framework is based on Geant4 \cite{TOOLS_2003_GEANT4,TOOLS_2006_GEANT4,TOOLS_2016_GEANT4}.
The framework takes into consideration cross sections and time dependencies
of the physics processes, timing response of the subdetectors,
and effects of hit readout and digitization. 
Geant4 v10.5 with the "ShieldingM" physics list has been used 
as an underlying simulation engine.
All simulations and reconstruction assume perfectly aligned and calibrated detector 
with no dead channels.

\subsection{Pileup Simulation}
\begin{figure}[H]
  \centering
  \includegraphics[width=0.9\linewidth]{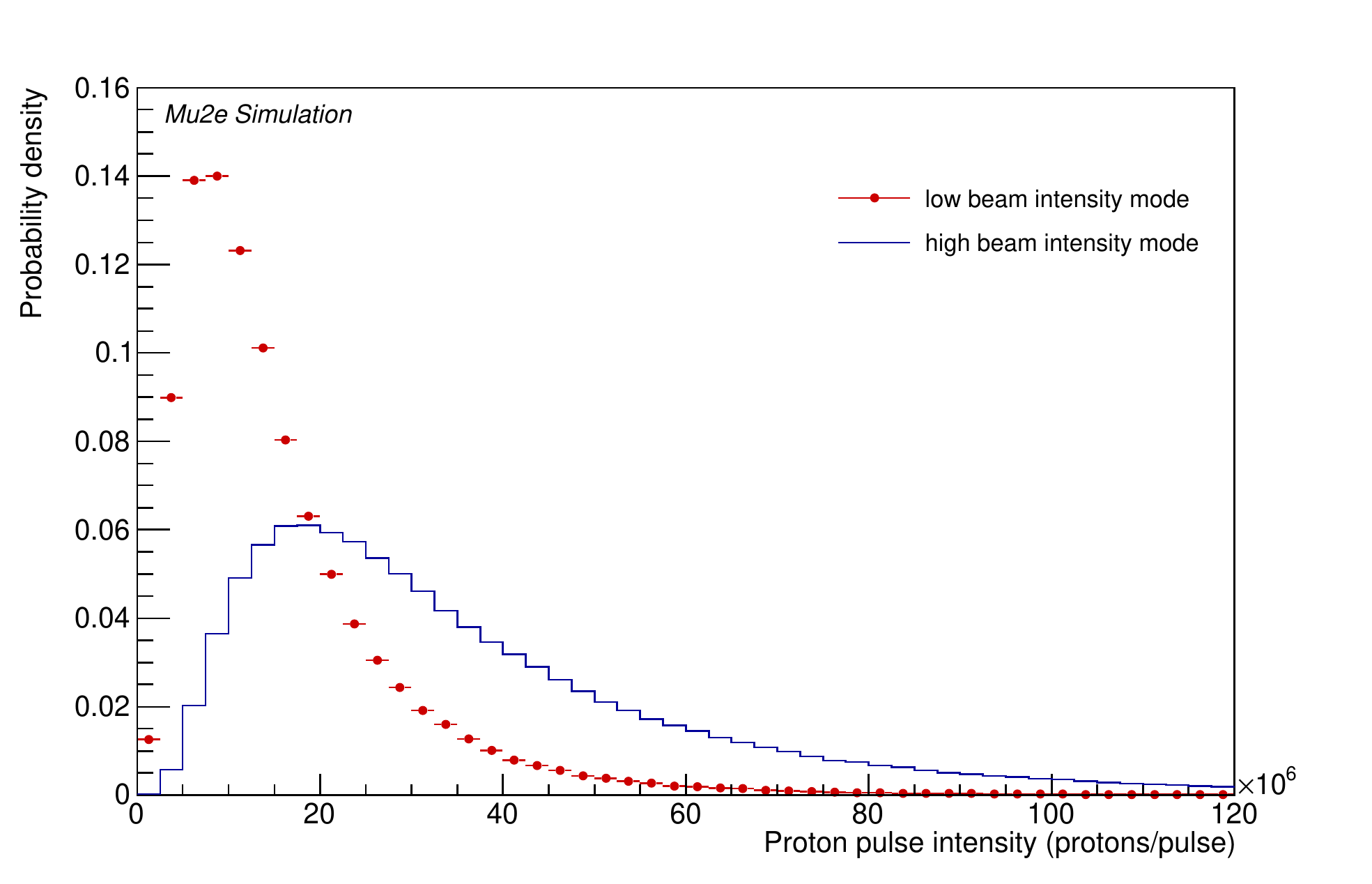}
  \caption{
    \label{fig:pulse_intensity}
    Simulated proton pulse intensity distributions for low and high beam intensity modes.
    The distributions have SDF = 60\%, an upper cut-off at $1.2\times 10^{8}$ protons per pulse,
    and are normalized to a unit area.
  }
\end{figure}

Electron events with $p_e {\sim} 100~\MeVc$ are extremely rare. In addition to hits produced
by signal-like particle, an event accepted by the Mu2e trigger is expected to have multiple
background hits produced by lower momentum particles.
Moreover, the Mu2e readout event window is about 1200 ns long, and a realistic detector simulation
has to handle particles producing hits in the detector at different times. For the low intensity
running mode with the mean intensity of $1.6 \times 10^7$ protons/pulse,
about 25,000 muons per proton pulse stop in the Al stopping target.
About 39\% of muons decay in orbit, and about 61\% are captured by the Al nuclei,
so an average "zero bias" Mu2e event includes ${\sim} 10,000$ muon DIO
and ${\sim} 15,000$ nuclear muon captures.
For the high intensity mode, the corresponding numbers are about 2.5 times higher.
The impact of the proton pulse intensity variations is taken into account 
by approximating them with the log-normal distribution with SDF = 60\%.
The simulated proton pulse intensity distributions for the low and high intensity
running modes are shown in Figure \ref{fig:pulse_intensity}.
The highest simulated pulse intensity is $1.2 \times 10^8$ protons per pulse.
The upper cutoff is taken into account in the evaluation of the systematic uncertainties.

The DIO simulation relies on the DIO electron spectrum on Al calculated in the leading logarithmic
accuracy in Ref. \cite{2016_Szafron_DIO_LL_PhysRev}.
Production of different particle species in ordinary nuclear muon captures is simulated
using custom event generators tuned to the data to reproduce the inclusive yields.
Simulation of protons and deuterons produced in nuclear muon captures relies on their inclusive
yields in Al reported in Refs. \cite{2020_DIO_TWIST_PhysRevC.101.035502, TOOLS_2022_ALCAP_PhysRevC.105.035501_EDMONDS}.
As there are no published neutron spectra on Al, the simulation of neutrons relies on
the neutron spectrum on Ca \cite{RMC_1978_OMC_NEUTRONS_KOZLOWSKI} and assumes 1.2 neutrons
emitted per muon capture, in agreement with Ref. \cite{RMC_1965_OMC_NEUTRONS_MACDONALD}.
Low energy photons produced in ordinary muon capture are assumed to have a uniform
energy distribution from 0 to 7 MeV, with two photons per capture produced on average.
The pileup simulation also includes simulation of the beam flash.
% \del{-- a flux of mostly low energy
% electrons and positrons coming from the production target. Most of them result from $\pi^0$ production
% and cross the DS within the first 200 ns after the proton pulse arrival at the production target.
% However, the beam flash timing distribution has a long tail due to effects like nuclear activation
% of the production target and muon captures in the upstream part of the detector.
% The tail extends into the readout window and is a source of additional detector occupancy.}
%
% \todo[color=green!20,linecolor=green,tickmarkheight=10pt] {
% Sophie - you already have beam flash discussion - is this repeatition, if not perhaps combine?
% }
%%% Local Variables:
%%% mode: latex
%%% TeX-master: "Mu2e_Run1_Sensitivity"
%%% End:

%%%%%%%%%%%%%%%%%%%%%%%%%%%%%%%%%%%%%%%%%%%%%%%%%%%%%%%%%%%%%%%%%%%%%%%%%%%%%%
\section{Event Reconstruction}
\label{sec:reconstruction}

% This section presents an overview of the offline reconstruction algorithms in the tracker, the calorimeter,
% and the CRV. The calorimeter and the CRV reconstruction rely on standalone algorithms.
% The track reconstruction uses the reconstructed calorimeter clusters to seed the pattern recognition
% and to improve the robustness of the Kalman fit.

In contrast to most collider and fixed target experiments, where particles coming from the primary vertex
are produced at a known time, the Mu2e event reconstruction has to deal with particles with unknown production
times. Timing of all reconstructed primitives -- tracks, calorimeter clusters, CRV stubs introduced later
in this section -- is therefore a parameter determined by the reconstruction, which could vary within hundreds
of nanoseconds with respect to the proton pulse arrival at the production target.

% The coordinate system used by the event reconstruction has its origin in the middle of the tracker.
% The Z-axis is parallel to the DS axis and points away from the stopping target, the Y-axis points upwards,
% and the X-axis completes the right-handed coordinate system.

%%%%%%%%%%%%%%%%%%%%%%%%%%%%%%%%%%%%%%%%%%%%%%%%%%%%%%%%%%%%%%%%%%%%%%%%%%%%%%
\subsection{Calorimeter Reconstruction}

The Mu2e calorimeter reconstruction processes the digitized waveforms from the calorimeter SiPMs
and reconstructs times and energy deposits of the corresponding hits. A single hit waveform
is ${\sim} 250$ ns long, so resolving hits with overlapping waveforms is an important part
of the data processing. Hits with $E > 10$ MeV are used to seed a two-pass clustering procedure.
For 105 MeV simulated electrons produced at the stopping target, ${\sim} 95\%$ of electrons
with a reconstructed track also have a reconstructed calorimeter cluster with $E > 10$ MeV.
The remaining ${\sim} 5\%$ of electrons go through the central hole or close to the edge of both
calorimeter disks and do not deposit enough energy in the calorimeter for a cluster to be reconstructed.
The calorimeter reconstruction runs before the track reconstruction.
That allows the found clusters to be used to seed the pattern recognition.

%%%%%%%%%%%%%%%%%%%%%%%%%%%%%%%%%%%%%%%%%%%%%%%%%%%%%%%%%%%%%%%%%%%%%%%%%%%%%%
\subsection{Track Reconstruction}

% In Mu2e, track reconstruction searches for particle tracks in the straw tracker
% by grouping straw hits which are coincident in space and time as coming from a single particle.
%
In the momentum region of primary interest, $p {\sim} 100~\MeVc$, different charged particle
species producing hits in the Mu2e tracker -- electrons, muons, and protons -- behave very differently.
Electrons are ultra-relativistic and have their velocity very close to the speed of the light,
$\beta_e=v_e/c {\sim} 1$. Muons are significantly slower, $\beta_\mu {\sim} 0.7$,
and the difference between the electron and muon propagation
times through the tracker is large on a scale of a single straw timing resolution.
For both electrons and muons, however, the average energy losses in the tracker are on the order
of 1-2 MeV, significantly smaller than the particle energy. This is not true for 100 \MeVc\ protons
which are highly non-relativistic and in most cases lose all their energy in the tracker because
of the ionization energy losses.
These differences require introducing particle mass-specific corrections at a very early reconstruction stage.

Particles produced at the stopping target pass through the tracker with $p_Z > 0$, and their reconstructed tracks
are referred to as downstream tracks. Cosmic ray-induced events often have particles traversing
the tracker with $p_Z < 0$. Efficient rejection of the cosmic background therefore requires
reconstructing tracks of such particles and tagging them as upstream tracks.

To handle all these different cases, the offline track reconstruction performs several passes.
Each reconstruction pass assumes a specific hypothesis about the particle mass
and the propagation direction and proceeds in three steps:
pattern recognition, fast Kalman fit, and full Kalman fit.
Two pattern recognition algorithms, a standalone pattern recognition and
a calorimeter-seeded one, are run in parallel.
The standalone pattern recognition associates hits with helical trajectories
and searches for the track candidates relying only on the straw hit information.
The calorimeter-seeded pattern recognition uses reconstructed energetic calorimeter
clusters to initiate the track candidate search.
It also exploits an assumption that a track
corresponds to a particle coming from the stopping target and, by doing that,
improves the track finding efficiency for the $\mu^- \to e^-$ conversion signal.

The fast Kalman fit does not take into account effects of multiple scattering, energy losses,
and the drift times reconstructed in individual straws. It converges within ${\sim} 1$ ms/event
providing a momentum resolution of ${\sim} 3\%$~FWHM.
If an event has a reconstructed calorimeter cluster with a position and time consistent
with the track, the cluster is included into the Kalman fit, which determines the Z-coordinate
of the cluster and its timing and coordinate residuals.
A general overview of the first two track reconstruction steps
is given in Ref.~\cite{TRIGGER_2020_GIANI_CTD2020}.
The final track reconstruction step, a full Kalman fit, provides the electron track momentum
resolution of $\Delta p_{\rm trk}/p {\sim} 0.3\%$~FWHM at $p = 100$ \MeVc.
About 33\% of the simulated \MuToEm\ conversion electron events have reconstructed tracks.

\subsection{CRV Reconstruction}

Similar to the calorimeter crystals, the CRV counters are read out by SiPMs,
and the times and energies of hits in the CRV counters are reconstructed from the
digitized waveforms of the SiPM signals. For counters read out from both ends, the time difference of signals read
out from the two ends is used to determine the hit coordinate along the counter.
The signature of a cosmic muon entering the Mu2e detector is a CRV stub --
hits in at least 3 out of 4 CRV layers with a pattern consistent
with the pattern of hits produced by a single relativistic particle.

%%% Local Variables:
%%% mode: latex
%%% TeX-master: "Mu2e_Run1_Sensitivity"
%%% End:

\section{Trigger Simulation}
\label{trigger}

The Mu2e trigger system is a one-level online software trigger system.
Multiple triggers are implemented as multiple independent
reconstruction paths, each path running one or several reconstruction algorithms
followed by a software filter to make the trigger decision.
The trigger uses the offline reconstruction algorithms with settings optimizing
the timing performance.
The online track reconstruction path includes two algorithmic steps -
a pattern recognition followed by the fast Kalman track fit.
The fast Kalman fit provides sufficient, for the trigger, momentum resolution,
making it unnecessary to use the full Kalman fit, which is significantly slower.
That improves the trigger timing and reduces dependence of the trigger performance
on the tracker calibrations.

To improve the trigger efficiency, the two track reconstruction paths
exploiting two pattern recognition algorithms introduced in Section \ref{sec:reconstruction},
are run in parallel.
The conversion electron trigger selects events with at least one
reconstructed downstream electron track with $p > 80$ \MeVc.
The trigger accepts tracks in a wide enough momentum range to enable an
analysis of both low-momentum and high-momentum sidebands of the
\MuToEm\ conversion signal.

\begin{figure}[H]
%  \begin{tikzpicture}
%    \node[anchor=south west,inner sep=0] at (0.,0.) {
%    \node[shift={(0 cm,0.cm)},inner sep=0,rotate={90}] at (0,0) {}
  \makebox[\linewidth][c] {
    \includegraphics[width=0.5\textwidth]{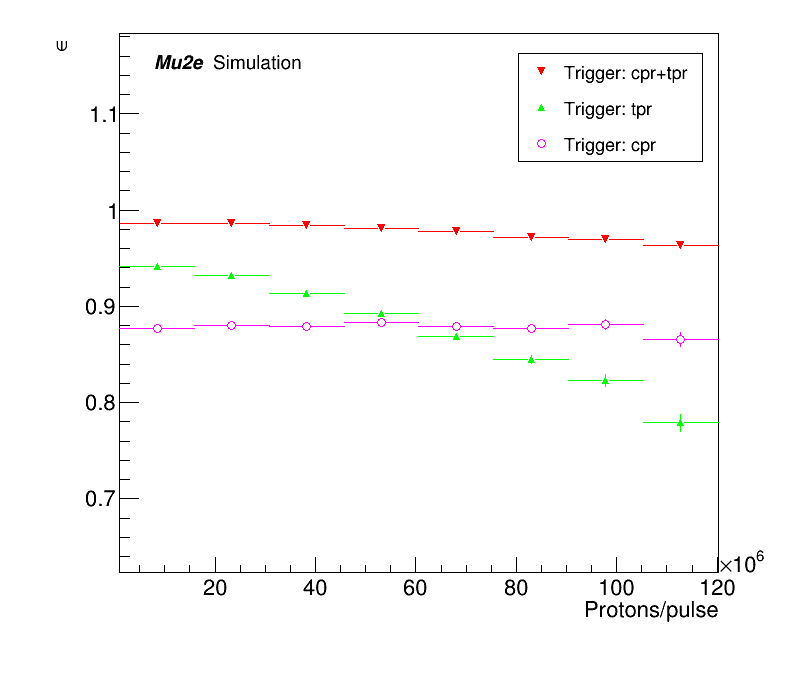}
  }
  % };
  % \end{tikzpicture}

  \caption{
    \label{fig:global_trig_eff_rel}
    Trigger efficiency for \MuToEm\ conversion on Al (red markers)
    relative to the offline reconstruction efficiency as a function of the proton pulse 
    intensity.  Also shown are the efficiencies of the online triggers
    running the individual pattern recognition algorithms : the standalone ({\bf tpr})
    and the calorimeter-seeded ({\bf cpr}).
  }
\end{figure}

Figure~\ref{fig:global_trig_eff_rel} shows the trigger efficiency for
the simulated conversion electron events which have a reconstructed
track passing the offline selections.
% v1_01_Andy
Plotted as a function of the proton pulse intensity, the trigger efficiency
varies from 99\% at zero beam intensity to 97\% at $1.2\times 10^{8}$ protons/pulse,
the highest simulated pulse intensity.
Also shown in Figure~\ref{fig:global_trig_eff_rel} are the trigger
efficiency curves corresponding to the use of the individual pattern
recognition algorithms.
For the calorimeter-seeded track finding, the trigger efficiency is limited
by the calorimeter acceptance and the trigger requirement on the seed
cluster energy, $E > 50$ MeV.
However, the efficiency is almost independent of
the beam intensity.  In comparison, the efficiency of the trigger
based on the standalone tracker pattern recognition at $1.2 \times 10^{8}$
protons/pulse drops by ${\sim} 15\%$. Stable performance of the trigger
based on the OR of the two pattern recognition algorithms illustrates
the importance of using both for the online track finding.
The expected instantaneous trigger rate
% , which does not take into account the duty factor of the proton delivery system,
is about 60~Hz for the low beam intensity mode.

% {\bf \red trigger rate for high intensity mode ? - ~ 150 Hz? - can mention no fakes..}

%

% The difference between the algorithms' efficiency at low intensity
% relies on the intrinsic limit of the calorimeter geometrical
% acceptance and the inefficiency of the calorimeter online-clustering
% algorithm that includes also an energy threshold at 50 MeV.
%
% The expected trigger rate is about 60~Hz.
%
A more complete description of the Mu2e trigger system
can be found in Refs.~\cite{TRIGGER_2015_GIANI_IEEE,TRIGGER_2020_GIANI_CTD2020}.
% \todo[color=magenta!10,inline] { Stefano \\
%   ``The resulting ... resulted ...''}
% \todo[color=green!10,inline]{Giani\\
%   fixed.
% }
%
% \todo[color=magenta!10,inline] { Stefano \\
%   What's the trigger rate with 2 batches? Do we need to quote the trigger rate at all?}
% \todo[color=green!10,inline]{Giani\\
%   I removed the reference to the batch mode of the sample used. I think that the Trigger section should state the expected trigger rate. Many times chatting with COMET people they were curios about our trigger performance in terms of expected trigger rate.
% }

\section{Event Selection}
\label{sec:track-selection_cuts_summary}

The selection of \MuToEm\ conversion electron event candidates proceeds in several steps.
First , selected event candidates are required to have a track passing the following pre-selection cuts:
%which have an overall efficiency of 84\% for conversion electrons:

\begin{itemize}
  \label{lst:track_selection_cuts}
\item
  {\bf N(hits) $\geq$ 20}: the track has a sufficient number of hits in the tracker.
\item
  {\bf $\bm{|D_0|}$ < 100 mm}: the reconstructed track impact parameter, $D_0$, is consistent
  with the particle coming from the stopping target.
\item
  {\bf R(max) < 680 mm}: the maximal distance from the reconstructed trajectory to the DS axis
  is less than the radius of the tracker, so the reconstructed trajectory is fully contained
  within the tracker fiducial volume.
\item
  $\pmb {0.5 < \cot\theta < 1.0}$: the angle  $\pmb{\theta}$ between the track momentum
  vector and the DS axis,
  at the tracker entrance, is consistent with a track of a particle produced at
  the stopping target. As the DS magnetic field is graded and is higher at the DS entrance,
  typical values of $\cot\theta$ for particles entering the DS from the TS are greater than 1.0.
\item
  $\bm{\sigma_{T_0} < 0.9}$ {\bf ns}: the uncertainty on the reconstructed track time, $T_0$,
  returned by the fit is consistent with a downstream electron hypothesis.
  This requirement implies that the Kalman fit with the calorimeter cluster
  included has successfully converged (see Section \ref{sec:reconstruction}).
\end{itemize}

\begin{figure}[H]
  \centering
  \includegraphics[width=1.0\linewidth]{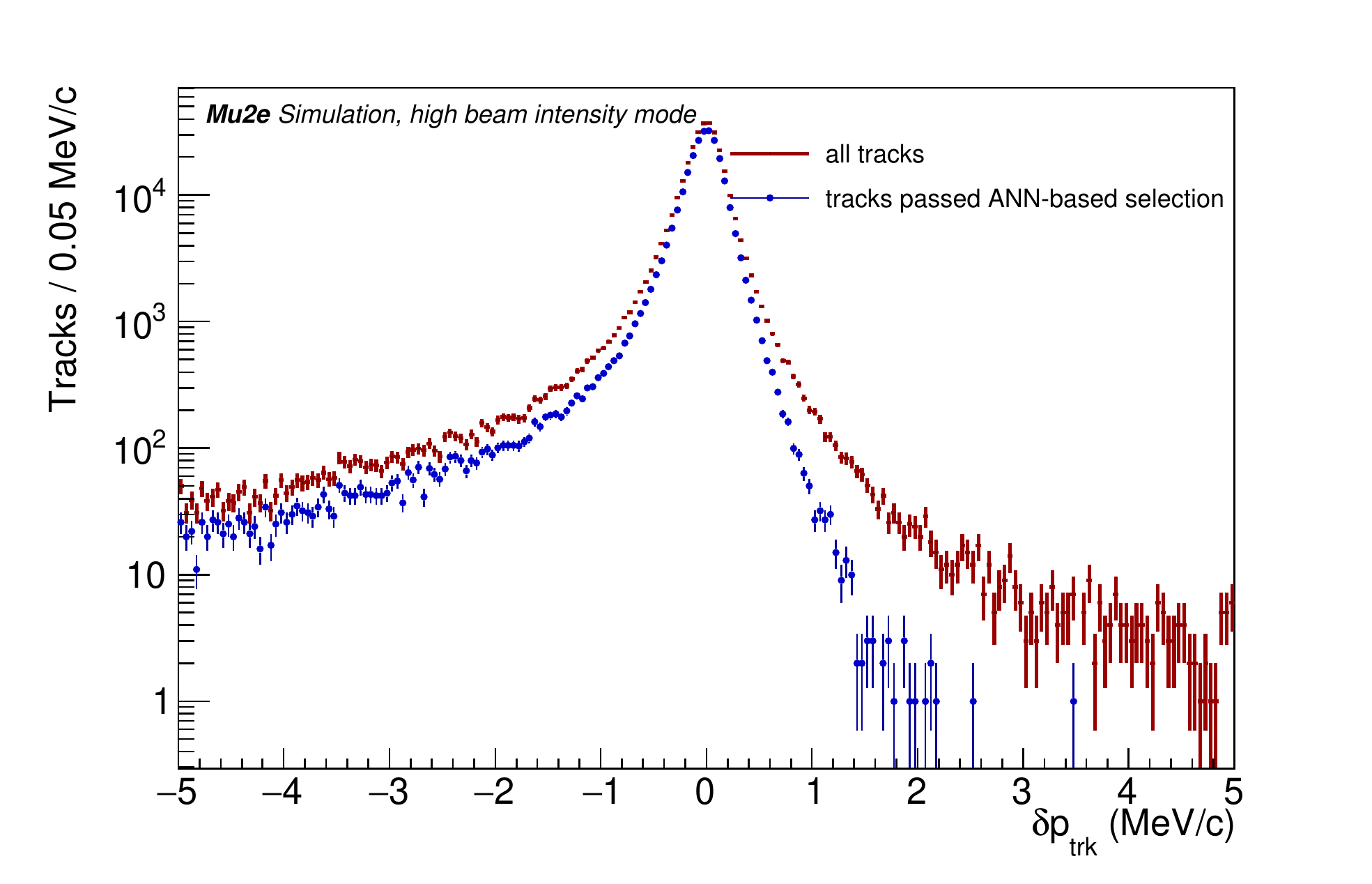}
  \caption{
    \label{fig:track_res_ann}
    Tracker momentum resolution $\delta p_{\rm trk}$ evaluated at the tracker entrance
    for the reconstructed conversion electron tracks before and after the track selection cuts.
    The distributions correspond to the simulated running in high beam intensity mode and
    illustrate the critical importance of the track selection cuts for reducing
    the background due to misreconstructed tracks with large positive values of $\delta p_{\rm trk}$.
  }
\end{figure}

Accurate reconstruction of the track momentum is critical for separating the conversion electron signal from the DIO background which rapidly falls with momentum.
Especially important is to reject tracks with large positive values of $\delta p_{\rm trk} = p_{\rm reco} - p_{\rm MC}^{\rm trk}$,
%
% \todo[color=gray!20,linecolor=gray,tickmarkheight=10pt] {
%  (Michael) $\Delta p_{\rm trk}$ was used before when talking about resolution I believe, should
%they match this? Also, should MC be defined at some point or assumed obvious?}
%
where $p_{\rm reco}$ is the reconstructed track momentum and $p_{\rm MC}^{\rm trk}$
is the momentum of the Monte Carlo (MC) particle corresponding to the track,
both taken at the tracker entrance.
The track selection procedure utilizes an artificial neural network (ANN) trained to separate electron tracks
with $\delta p_{\rm trk} > 700$ \keVc\ from tracks with $|\delta p_{\rm trk}| < 250$ \keVc.
The ANN training uses tracks passing the pre-selections described above.
A detailed discussion of the approach can be found in Ref. \cite{TOOLS_2021_Edmonds_TrkQual}.
For conversion electron events with tracks passing the pre-selections described above,
the efficiency of the ANN-based track selection is 96\%.
Improvement in the quality of momentum reconstruction is clearly seen in Figure~\ref{fig:track_res_ann} --
after the track selection, the high-side tail of the $\delta p_{\rm trk}$ distribution is significantly suppressed.
The overall track selection efficiency is 81\%, so 26\% of the simulated \MuToEm\ conversion events
have well reconstructed tracks.

%%%%%%%%%%%%%%%%%1%%%%%%%%%%%%%%%%%%%%%%%%%%%%%%%%%%%%%%%%%%%%%%%%%%%%%%%%%%%%%
\subsection{Particle Identification}
\label{sec:pid}
\begin{figure}[H]
  \centering
  \includegraphics[width=1.0\linewidth]{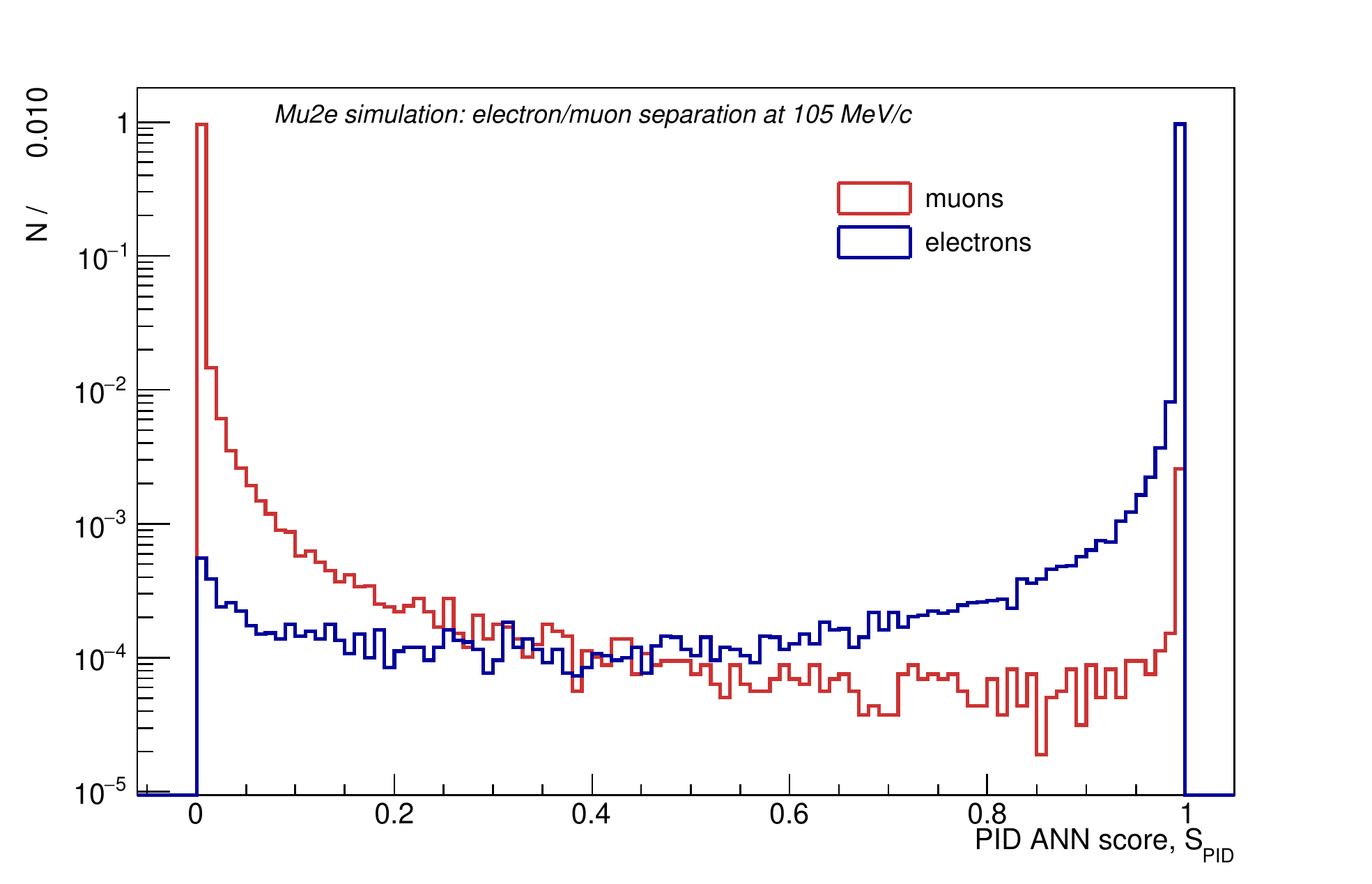}
  \caption{
    \label{fig:pid_score}
    Distributions of the PID ANN output score, $S_{\rm PID}$ for 105 \MeVc\ electrons and muons.
    The spike in the distribution of the muon PID score is due to muon decays in flight in front
    of the tracker and in the tracker volume.
  }
\end{figure}

Most cosmic ray muons entering the detector do not decay within the detector volume.
Events with reconstructed muons are discriminated from the events with reconstructed
electrons by a particle identification (PID) ANN.
The PID ANN is trained using samples of simulated 105 \MeVc\ electron and muon events
with the reconstructed tracks passing the track selection cuts described
in Section \ref{sec:track-selection_cuts_summary}.
Events with muon decays in flight are excluded from the training.
% Not every event with a reconstructed track has a calorimeter cluster produced by the same particle.
% The variables above are defined in 93\% of all events with the reconstructed tracks passing the selection cuts.
The distributions of the output score of the PID ANN, $S_{\rm PID}$, for electron and muon samples
are presented in Figure~\ref{fig:pid_score}. The requirement $S_{\rm PID}> 0.5$ identifies events
with reconstructed electrons with an efficiency of 99.3\%.
The corresponding muon misidentification rate is 0.4\%.

\section {Backgrounds}
\label{sec:background_estimates}

Optimization of the search sensitivity used in this paper is based
on finding the 2D momentum-time signal window maximizing the discovery
potential of the experiment.
As will be shown in Section \ref{sec:sensitivity_optimization},
the Mu2e Run I discovery potential is optimized for the momentum
and time window $103.6 < p < 104.9 ~\MeVc$ and $640 < T_0 < 1650$ ns.
The individual background contributions, discussed below, are integrated
over this window.

enterin%%%%%%%%%%%%%%%%%%%%%%%%%%%%%%%%%%%%%%%%%%%%%%%%%%%%%%%%%%%%%%%%%%%%%%%%%%%%%%
\subsection{Cosmic Rays}
\label{sec:cosmic_bkg}
Interactions and decays of cosmic ray particles in the DS are expected to
produce the dominant background in the \MuToEm conversion
search. Detailed simulation studies performed using the CRY event generator
\cite{2007_COSMICS_Hagmann_CRY} to simulate the cosmic rays helped
identify three distinct types of cosmic background events: (1)
cosmic ray muons passing through the CRV coverage, (2) cosmic ray
muons entering through the detector regions not covered by the CRV,
and (3) neutrally-charged cosmic ray hadrons.

\begin{figure}[H]
  % \hspace{-0.6in}
  \begin{tikzpicture}
    \node[anchor=south west,inner sep=0] at (0,0.) {
      % \node[shift={(0 cm,0.cm)},inner sep=0,rotate={90}] at (0,0) {}
      \makebox[\linewidth][c] {
        \includegraphics[width=1.0\linewidth]{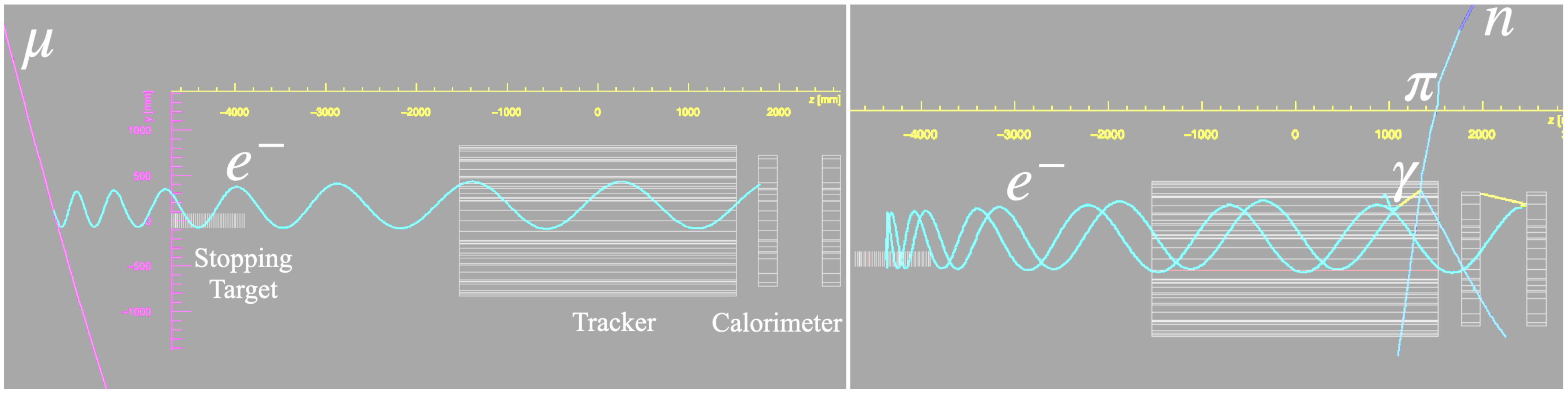}
      }
    };
%    \node [text width=2cm, text=white, scale=1.0] at (1.2,0.3) {(a)};
%    \node [text width=2cm, text=white, scale=1.0] at (8.7,0.3) {(b)};
  \end{tikzpicture}
  \caption{
    \label{fig:cosmic_event}
    Left : A background event produced by a cosmic ray muon
    that knocks out a signal-like electron in the DS.
    Reconstruction of the CRV stub allows the event to be vetoed.
    Right: A cosmic ray neutron entering the detector in the upper right corner
    of the event display interacts in the apparatus to produce an upstream-moving electron.
    The electron gets reflected by the DS magnetic mirror and passes through
    the tracker for the second time.
    This event can not vetoed by the CRV, but can be rejected based
    on the presence of the upstream track.
  }
\end{figure}

The first type of cosmic ray background events
originates from muons striking the detector, or beamline components,
and knocking out electrons
with energies close to 105 MeV, see Figure~\ref{fig:cosmic_event} (left).
Most of the potential background is due to these muons, so this background
contribution is primarily determined by the CRV veto efficiency.

The second type of events consists of cosmic ray muons entering
the detector through the uninstrumented regions. For instance, there
is a significant penetration in the CRV to permit the muon beamline to
enter the DS (see Figure~\ref{fig:crv}). Cosmic ray muons can penetrate
these regions without being vetoed and produce signal-like particles.

The third type of background contribution originates from the neutral component
of cosmic showers, predominantly neutrons, which do not generate signals
in the CRV counters.
Figure~\ref{fig:cosmic_event} (right) shows a conversion-like event resulting
from a cosmic ray neutron interaction in the detector. Cosmic ray neutrons interacting
with the material around the stopping target can produce events without
an upstream-going electron component. Current estimates suggest that
the background from the neutral  component does not impact the Run I
sensitivity. Comparison of the differential cosmic neutron flux used
by CRY to the measurements of
Ref. \cite{2004_COSMICS_NEUTRON_FLUX_GORDON} indicates that CRY may be
underestimating the neutron component of cosmic showers by a factor
of ${\sim} 1.5-2$. In Run II, the background from cosmic ray neutrons
could be reduced with additional shielding.

\begin{figure}[H]
  \centering
  \includegraphics[width=0.9\linewidth]{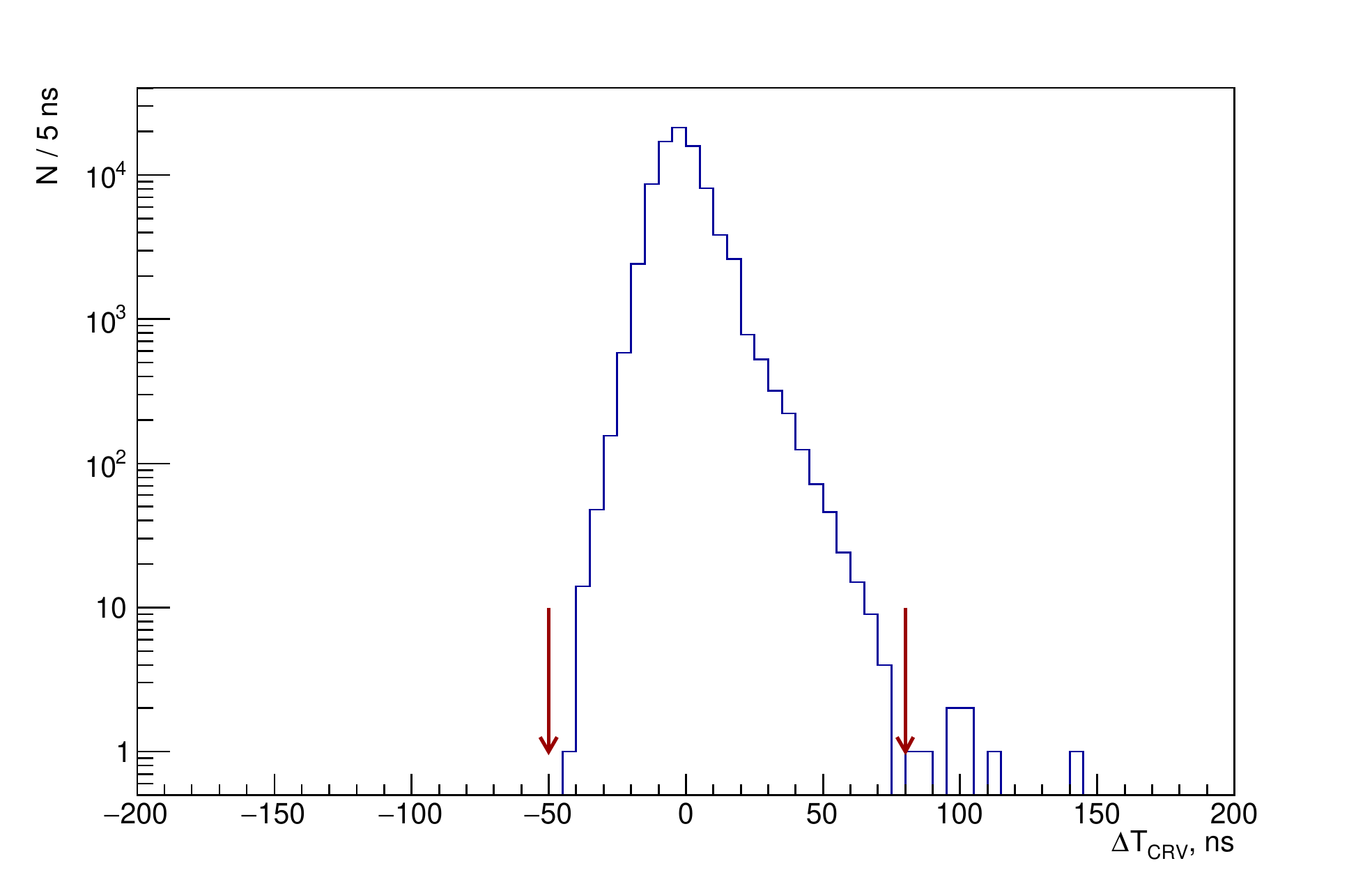}
  \caption{
    \label{fig:cosmics_dtcrv2}
    Distribution of timing residuals $\Delta T_{\rm CRV} ~=~ T_{0}-T_{\rm CRV}$
    between the reconstructed track and the CRV stub. Arrows represent
    the timing window used in the event selection.
  }
\end{figure}

Cosmic background events have the following characteristic signatures
in the Mu2e detector:

\begin{itemize}
\item
 A typical cosmic background event consists of a reconstructed
 downstream propagating electron and a CRV stub, see Figure~\ref{fig:cosmic_event} (left).
 The distribution of the timing residuals
 $\Delta T_{\rm CRV} ~=~ T_{0}-T_{\rm CRV}$ between the  reconstructed
 electron and the CRV stub  is shown in Figure~\ref{fig:cosmics_dtcrv2}.
 Cosmic event candidates are identified by
 the timing window $-50 < \Delta T_{\rm CRV} < 80$ ns.

\item
  A cosmic ray particle can also interact in the calorimeter or decay
  in the tracker volume producing a particle moving upstream,
  see Figure~\ref{fig:cosmic_event} (right).
  Both upstream and downstream
  moving electrons are reconstructed and the upstream component of the
  track can be used to reject this type of cosmic background events.
\end{itemize}

Based on the data taking plan for Run I, specified in Table \ref{tab:running_time},
we have estimated the total cosmic background of $0.046 \pm 0.010$ (stat) events.
%
% Instead, to estimate the impact of the uncertainty of the CRV aging on the expected Mu2e
% physics reach, in Section \ref{sec:sensitivity_estimate} the experimental sensitivity
% is estimated under different assumptions corresponding to significantly different CRV
% aging rates.
%
%Similar to assumptions made about performance of other subdetectors,
%we assume that the performance of the CRV extrusions is measured
%accurately enough not to dominate the systematic uncertainty of the background prediction.
%

Currently, the largest uncertainty on the cosmic background prediction comes from
the uncertainty on the CRV counter aging rate.
To simulate performance of the counters in Run I, we use results of early Mu2e
measurements which yielded an aging rate of 8.7\%/year.
% The most recent measurements of the counter aging indicate that after the first two years
% from production the aging rate significantly slows down, 
% In addition, the expected Run I sensitivity does not depend critically
% on the expected CRV aging rate.
%
% A conservative assumption of the light yield
% degrading by ${\sim} 15$\%/year, which is about two times the currently measured rate,
% corresponds to the expected cosmic background mean increase by about a factor of six, but
% reduces the expected sensitivity by only about 30\%.
%
The ongoing measurements of the counter aging will significantly reduce the associated uncertainties.
Current uncertainties of the aging model are not considered in the evaluation of the
systematic uncertainties -- see discussion in Section~\ref{sec:sensitivity_optimization}.

Out of considered sources of the systematic uncertainties the largest contribution comes
from the uncertainty on the cosmic flux normalization.
The flux of cosmic particles integrated over the data taking time depends on the latitude,
altitude, local magnetic field of Earth, etc.
In addition, the solar activity cycle, which has a period of about 11 years,
makes the integral time-dependent.
Based on the data presented in Ref. \cite{2017_COSMICS_MIYAKE}, the uncertainty
in predicting the time-dependent intensity of the cosmic particle flux does not exceed 15\%.
The simulation using a different cosmic shower generator, CORSIKA \cite{1998_COSMICS_FZKA-6019},
leads to a 5\% different yield of reconstructed electrons per cosmic muon. Added linearly, the
two sources give an overall systematic uncertainty of 20\% on the cosmic ray background estimate.
With the systematic uncertainties included, the cosmic background in Mu2e Run I is
$\rm 0.046 \pm 0.010~ (stat) \pm 0.009~(syst)$. It is worth noting that about 3/4 of the total
is due to cosmic muons entering the DS through the area not covered by the CRV.

Reconstructed cosmic event candidates are excluded from the analysis.
As the CRV will operate in a high radiation environment, accidental timing coincidences
of the reconstructed tracks with CRV hits produced by neutrons and photons
from proton beam interactions could mimic cosmic ray muons
and introduce an inefficiency in the signal selection.
The inefficiencies are estimated at 4\% and 15\% for the low and high 
intensity running modes,
% \todo[color=gray!20,linecolor=gray,tickmarkheight=10pt] { (Michael) Is it ``intensity mode?''}
respectively.

%%%%%%%%%%%%%%%%%%%%%%%%%%%%%%%%%%%%%%%%%%%%%%%%%%%%%%%%%%%%%%%%%%%%%%%%%%%%%%
\subsection{Muon Decays In Orbit }
\label{sec:dio_bkg}

\begin{figure}[h]
  \centering
  \includegraphics[width=0.51\linewidth]{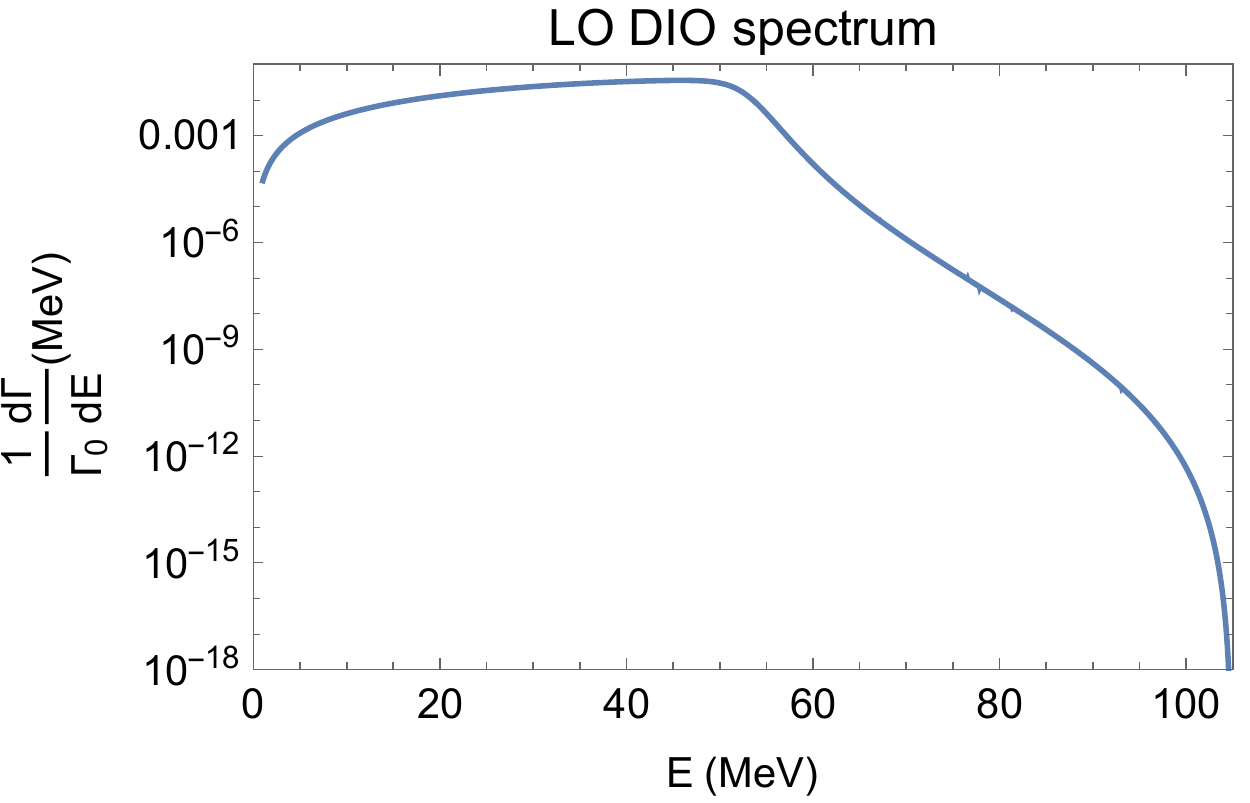}
  \includegraphics[width=0.47\linewidth]{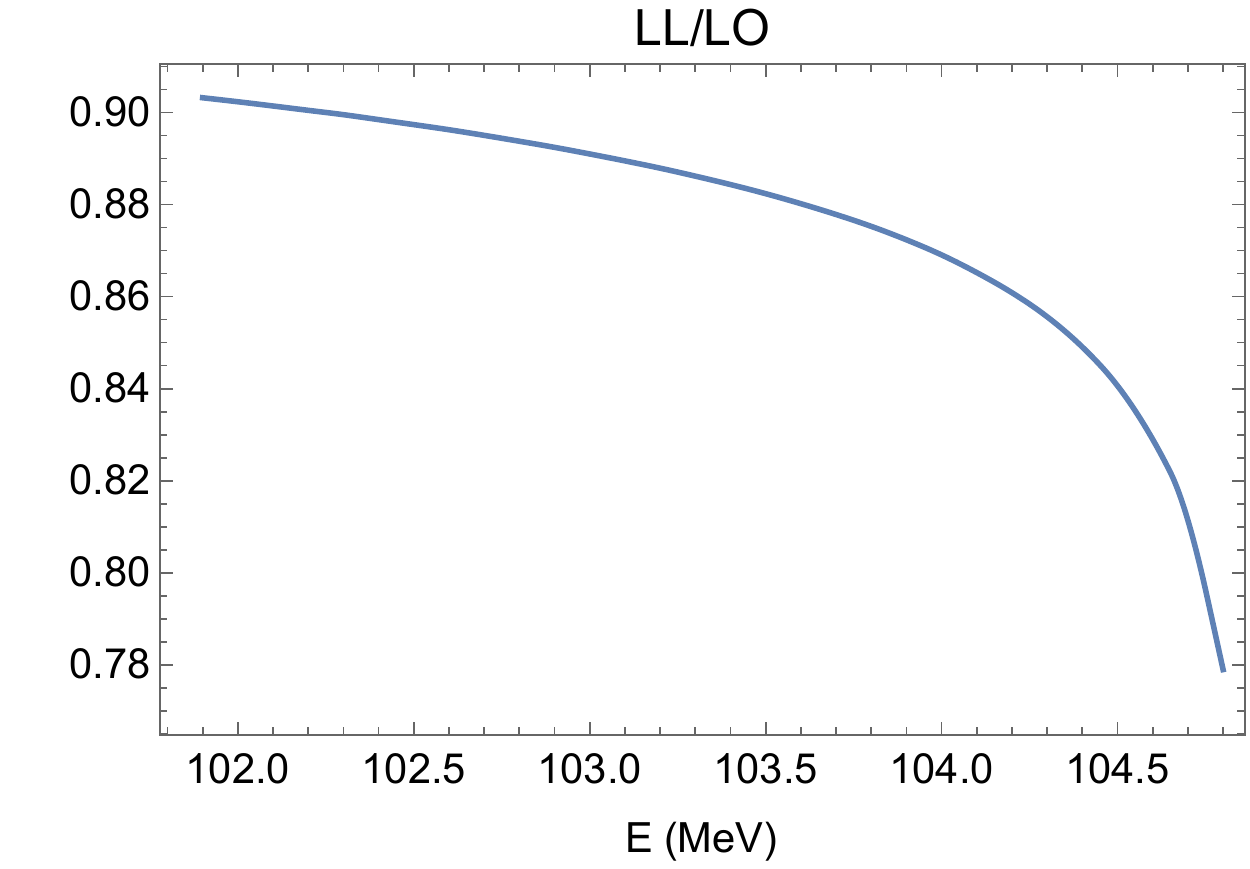}
  \caption{
    \label{fig:DIO_LL}
    Left: LO DIO spectrum on Al from Ref.~\cite{2011_DIO_Czarnecki}.
    Right: Ratio of LL and LO DIO spectra on Al for $E > 102$ MeV.
  }
\end{figure}

% \todo[color=green!20, inline] { (Bob/Sophie) Suggestion for the initial paragraph: \\
%   Muon decay in orbit is an intrinsic background to the conversion electron search. Electron energies associated with muon to electron conversion
%   are far away from those associated with free muon decay (endpoint at $\sim$ 52 MeV). However, when the decay happens in the field of a nucleus
%   the electron momentum spectrum is altered through the exchange of virtual photons. As a result DIO electrons can have energies up to ${\sim} 105$ MeV.
%   Precise calculations are necessary to understand the DIO background close to the signal region, such as those calculated to the leading order (LO) in
%   Ref.~\cite{2011_DIO_Czarnecki} and shown in Fig.~\ref{fig:DIO_LL} (left). The DIO spectrum rapidly decreases for $E>$ 100 MeV,
%   as the decay final state phase space shrinks as ($E_{CE}-E$)$^5$ as the DIO electron energy approaches the endpoint.
% }
Electrons produced in decays of free muons at rest have energies up to $m_{\mu}/2$,
well below $E_{CE}$. However, negative muons stopped in the stopping target 
get captured by the Al atoms and form muonic atoms. The energy spectrum of electrons
from decays of bound muons extends up to $E_{CE}$, making DIO one of the major background
sources to the \MuToEm\ conversion search.
Near the endpoint, the DIO spectrum falls as $(E_{CE}-E)^5$, driving requirements
on the experimental momentum resolution. The leading order (LO) 
DIO spectrum on Al calculated in Ref. ~\cite{2011_DIO_Czarnecki} 
is shown in Figure~\ref{fig:DIO_LL} (left). The leading logarithm (LL) level corrections to the DIO spectrum
have been calculated in Refs. ~\cite{2016_Szafron_DIO_LL_PhysRev,2016_Szafron_DIO_LL_PhysLettB}.
Taking into account the higher order corrections lowers the DIO background estimate and 
as shown in Figure~\ref{fig:DIO_LL} (right), the integral of the DIO spectrum calculated
at the LL level over the region [103.6, 104.9]~MeV is reduced by ${\sim} 13$\% compared to the LO
calculation. In this paper, the LL DIO spectrum is used to model the DIO background.

%Interactions with the host nucleus through the exchange of virtual photons modifiy
%the kinematics of the bound negative muon decay.
%While the maximum energy of an electron from the decay of a free muon at rest
%is $m_{\mu}c^2/2$ or $\sim 52.8$ MeV, the recoil taken by the nucleus extends the decay
%electron spectrum up to ${\sim} 105$ MeV, the same as the \MuToEm\ conversion
%electron energy.
%The DIO spectrum calculated in the leading order (LO)~\cite{2011_DIO_Czarnecki}
%is shown in Figure~\ref{fig:DIO_LL} (left). At $E>$ 100 MeV, the decay phase space shrinks
%as ($E_{CE}-E$)$^5$, and the DIO spectrum rapidly falls as the electron energy
%approaches the endpoint.
%\todo[color=gray!20,linecolor=gray,tickmarkheight=10pt] { (Michael) Is the phase space shrinking as $\Delta E^5$? Is the
 % spectrum shape here directly proportional to phase space size? Genuinely curious!}
%
%Radiative effects at the leading logarithm (LL) level~\cite{2016_Szafron_DIO_LL_PhysRev,2016_Szafron_DIO_LL_PhysLettB},
%such as soft photon emission and vacuum polarization effects,
%enhanced by the large logarithm factor $log \big (\frac{m^2_{\mu}}{m^2_e} \big )$,
%reduce the DIO background expectation.
%
%As shown in Figure~\ref{fig:DIO_LL} (right), the integral of the DIO spectrum over the region [103.6, 104.9]~MeV
%is reduced by ${\sim} 13$\%. In this paper, the LL DIO spectrum is used to model the DIO background.

%%%%%%%%%%%%%%%%%%%%%%%%%%%%%%%%%%%%%%%%%%%%%%%%%%%%%%%%%%%%%%%%%%%%%%%%%%%%%%
\subsubsection {Calibration of the Tracker Resolution and Momentum Scale}
\label{sec:dio_momentum_calibration}

\begin{figure}[h]
  \centering
  \includegraphics[width=0.35\textwidth]{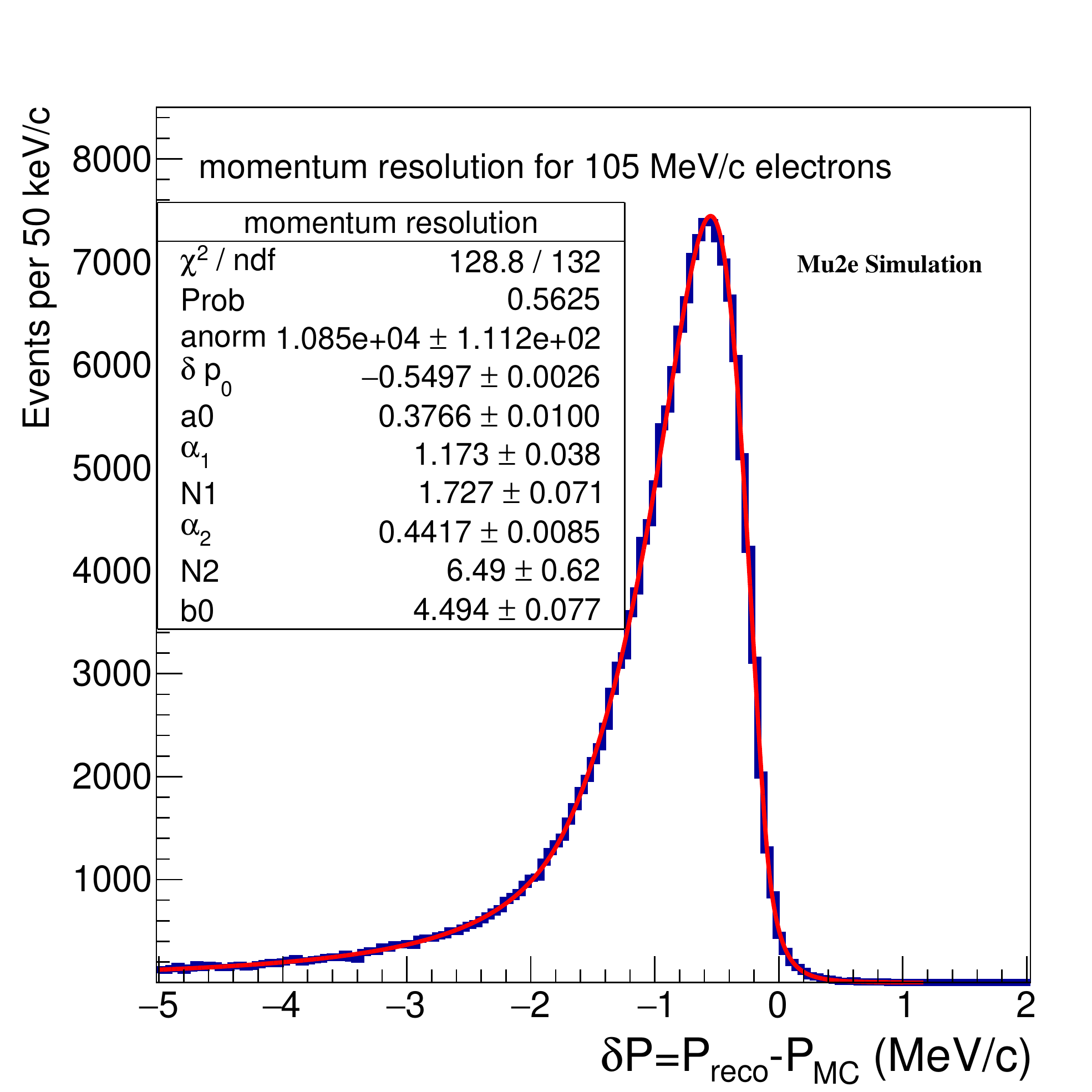}
    \includegraphics[width=0.35\textwidth]{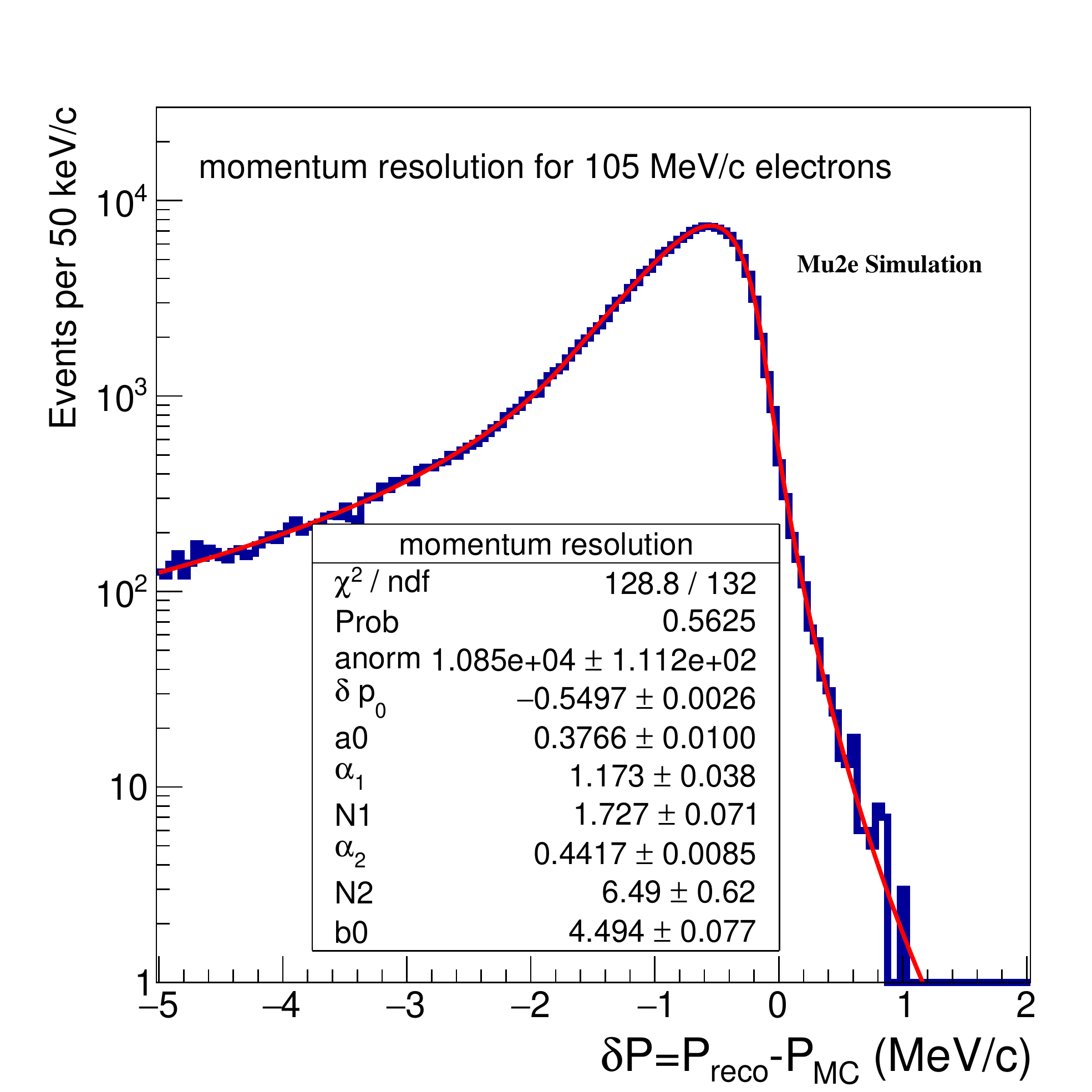}
 \caption{
  \label{fig:CE_fit}
  Left: $\delta_{p}$ distribution for 105 \MeVc\ generated electrons and its fit with the resolution
  function defined in the text. Right: The same distribution, but displayed in a log scale
  to highlight the tail and demonstrate the quality of the fit in the tail regions.
}
\end{figure}

A reliable estimate of the DIO background requires understanding of the tracker momentum scale
and resolution. Shown in Figure~\ref{fig:CE_fit} is the distribution of
$\delta{p}~=~p_{\rm reco}~-~p_{MC}$, 
the momentum resolution of the experiment, for the simulated \MuToEm\ conversion electrons.
$p_{MC}$ here is the CE momentum at the production vertex.

The most probable value of the energy losses in front of the tracker is $\sim$ 0.5 MeV, and the
% \todo[color=green!20,linecolor=green,tickmarkheight=10pt] { Sophie - "these" instead of "their"?}
fluctuations of the energy losses dominate the experimental resolution. The momentum response
is well fitted  by the following function:

\begin{equation}
  \label{eq:dio_mom_res}
  R(\delta{p}) =
 \begin{cases}
    A_1(B_1-(\delta p-\delta p_0))^{-N_1} &\delta p-\delta p_0<-\alpha_1 \\
    A_{norm} \exp(a_0(b_0(\delta p-\delta p_0)-e^{[b_0(\delta p-\delta p_0]})) &-\alpha_1<\delta p-\delta p_0<\alpha_2 \\
    A_2(B_2+\delta p-\delta p_0)^{-N_2} &\delta p-\delta p_0>\alpha_2
  \end{cases}.
\end{equation}
%\todo[color=gray!20,linecolor=gray,tickmarkheight=10pt] { (Michael) I suspect the first line should
%  have ``$B_1 - (\delta p - \delta p_0)$ to avoid 1/0 for positive $B_1,N_1$ (the A/B parameters aren't given
%  so it's unclear what values they take on)
%}

The core part of the resolution function is largely due to the energy losses,
and its parameterization is generalized from an approximation to the Landau distribution~\cite{DIO_1955_moyal},
in which $a_0$ is fixed at 1/2. Introducing $a_0$ in the parameterization allows
for an extra degree of freedom which absorbs effects of widening due to the multiple scattering
and results in a better fit.
% \todo[color=gray!20,linecolor=gray,tickmarkheight=10pt] { (Michael) Is this ``data'' or a better fit to ``simulations?''}
The tail on the low momentum side accounts for tracks with large energy losses,
while the high-side tail is due to misreconstruction of tracks.
Both tails are well described by power law functions.
Parameter $\delta p_0$, the peak position, is defined by the most probable energy losses,
$b_0$ is the inverse of the Landau scale parameter~\cite{Landau}.
Parameters $\alpha_1$ and $\alpha_2$ determine the transition points from the Landau "core"
to the tails. $A_{norm}$ is the overall normalization factor, while $A_1$, $A_2$, $B_1$ and $B_2$
are factors determined by the requirement of the continuity of the function and its first derivative.
Parameters $N_1$ and $N_2$ determine how fast the power-law tails fall,
thus the relative contribution of the tails.
The uncertainty on the DIO background resulting from the high momentum resolution tail
is dominated by the uncertainty on N$_2$.

Parameters of the momentum resolution will be measured as follows.
Calibration of the energy losses, parameter $\delta p_0$, relies on cosmic ray events
%cosmics-produced events
%with a particle, an electron or a muon,
entering the tracker in the upstream direction,
reflecting in the DS  magnetic mirror, and returning back to the tracker.
Such events have two reconstructed tracks corresponding to the same particle,
and the difference between the momenta of the upstream and downstream tracks
is defined by the total amount of material crossed by the particle.

Determination of the momentum scale and the core resolution width uses the
positive beam.
It is based on the reconstruction of the 69.8 \MeVc\ positron peak from $\pi^+ \to e^+\nu$
decays of stopped positive pions.
As described in Section~\ref{mu2e_detector}, switching the beam polarity
requires rotating
the TS3 collimator by 180 degrees, however, the polarity of the B-field 
stays the same.
An independent calibration of the momentum scale comes from the reconstruction
of the momentum spectrum of positrons from Michel decays of stopped positive
muons, which has a sharp edge at 52.8 \MeVc. Both measurements will be performed
at a reduced magnetic field to keep the track curvature the same as the curvature
of conversion electron tracks at full field.

The measurement of the positron Michel spectrum has a very low background, so
the high-momentum tail of the spectrum is dominated by misreconstructed
tracks with large $\delta p-\delta p_0>0$.
That allows the determination of the parameter $N_2$ from the fit of the
high-momentum part of the spectrum.

\begin{figure}[ht]
  \centering
  \includegraphics[width=0.35\textwidth, height=0.25\textheight]{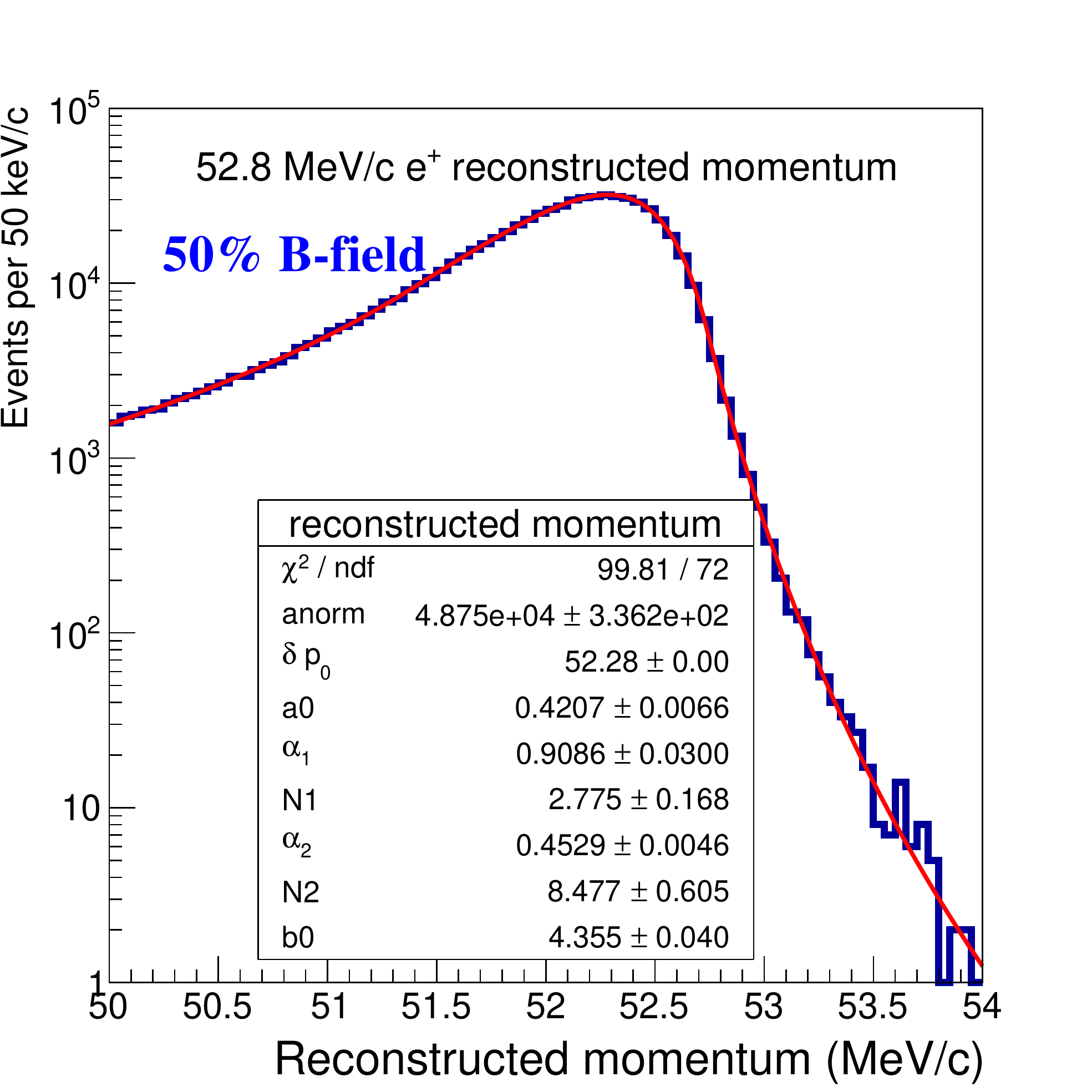}
  \includegraphics[width=0.35\textwidth, height=0.25\textheight]{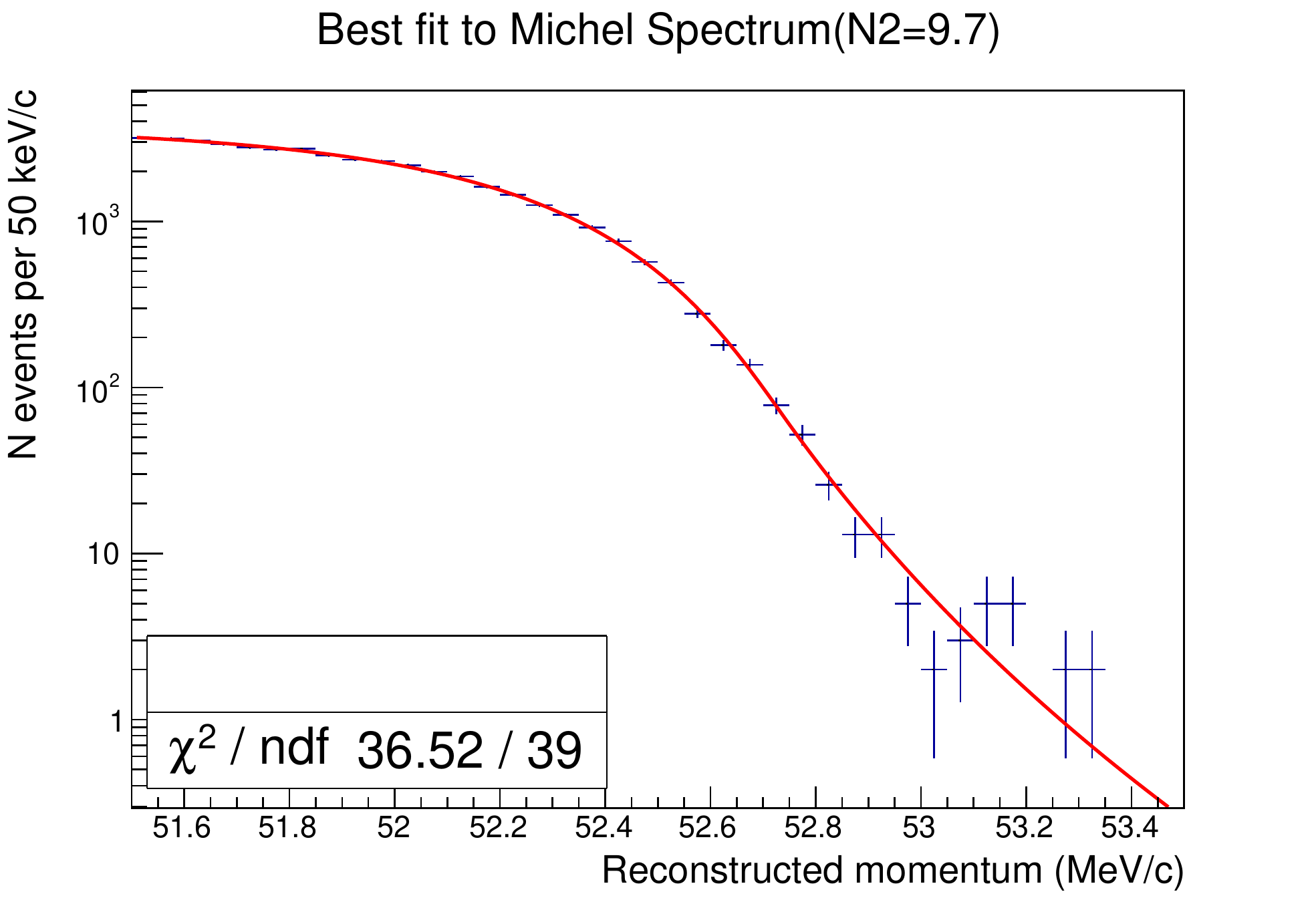}

  \caption {
    \label{fig:dio_mupl2_chisquare}
    Left: Fit of the resolution function corresponding to the monochromatic 52.8 \MeVc\ positrons
    simulated and reconstructed at B = 0.5 T. The fit yields $N_2 = 8.5 \pm 0.6$.
    Right: Fit of the momentum spectrum of positrons from Michel decays of stopped $\mu^+$'s.
    also simulated and reconstructed at B = 0.5 T.
    The best value of $N_2=9.7$ is determined using the procedure described in the text.
%     Right: Fit of the Michel spectrum at 50\% B-field, using a convolution of the the resolution function
%     with the true momentum spectrum of the reconstruced positrons. All parameters of the resolution function
%     except $N_2$ are taken from the fit of the 52.8 MeV/c $e^+$ dataset, and are fixed. Multiple fits of the
%     Michel spectrum are then performed. During each fit, the $N_2$ parameter of the resolution function takes a
%     different value and is fixed. The only free parameter is an overall normalization factor.
%     Shown is the best fit corresponding to the value of $N_2$=9.7, which minimizes the fit $\chi^2$.
%     (Upper Right) Fit log likelihood as a function of parameter N$_2$, obtained by fitting
%     the reconstructed Michel decay $e^+$ spectrum at 50\% B-field, using a convolution
%     of the theoretical Michel spectrum, tracker acceptance and the resolution line shape
%     obtained at 52.8 MeV/c. The best fit yields $N_2$ =9.7$^{+2}_{-1.4}$.
%     The N$_2$ parameter estimate from the two datasets are statistically compatible.
  }
\end{figure}

%%%%%%%%%%%%%%%%%%%%%%%%%%%%%%%%%%%%%%%%%%%%%%%%%%%%%%%%%%%%%%%%%%%%%%%%%%%%%%%%
\subsubsection {Systematic Uncertainties}
The main sources of systematic uncertainties on the DIO background are listed in Table~\ref{tab:DIO_systematic_stat}.

\begin{specialtable}[H]
  \centering
  \small
  \caption{
    \label{tab:DIO_systematic_stat}
    Breakdown of the DIO background relative systematic uncertainties.
%%    The quoted uncertainty on the high-momentum resolution tail is
%%    the estimated uncertainty of the method, not included into the total.
%    \todo[color=gray!20, inline] {
%       (Michael) Why is the resolution tail effect not included in the total?
%       So the total is momentum scale + theoretical only?
%     }
  }
   \begin{tabular}{c c c c}
\multirow{2}{*}{\textbf{Source}} & \textbf{low intensity} & \textbf{high intensity} & \multirow{2}{*}{\textbf{Run I error}}\\
                                 & \textbf{running mode } & \textbf{running mode  } & \\
       \toprule
Momentum Scale                   &  +62\%,-38\%     &  +50\%,-34\%  & +59\%,-37\%  \\
% Resolution Tail                 &  +23\%,-11\%     &   +23\%,-11\%  & +23\%,-11\% \\
     Theory                      &   $\pm$ 2.5\%   & $\pm$2.5\%   &$\pm$2.5\% \\
 \midrule
    Total                        &  +62\%,-38\%       & +50\%,-34\%  & +59\%,-37\%  \\
%    Total                       &  +66\%,-40\%       & +55\%,-36\%  & +63\%,-39\%  \\
      \bottomrule
%Track Reconstruction Efficiency &    negligible  & negligible & negligible \\
% Particle ID                     &    negligible  & negligible & negligible \\
  \end{tabular}
\end{specialtable}

\begin{enumerate}[listparindent=\parindent]
\item\textbf{ Uncertainty on the absolute momentum scale}.
  \noindent
  Currently, this is the dominant systematic uncertainty on the DIO background.
  We expect the momentum scale of the Mu2e tracker to be calibrated to an accuracy
  of better than 100 \keVc\ at $p = 100 ~\MeVc$.
  However, it is not possible to predict the exact value of the resulting systematic uncertainty,
  so a conservative estimate of 100 \keVc\ is used.
  \noindent
  Shifting the optimized momentum window by $\pm100$ \keVc\
  changes the DIO background estimate asymmetrically by [+59\%, -37\%].
  For the high beam intensity running mode, the relative uncertainty is slightly lower.
  This is expected: at higher occupancy, the momentum resolution degrades, and although
  the absolute value of the background increases, the slope of the measured DIO spectrum
  becomes less steep, reducing the relative uncertainty.
\item\textbf{Uncertainty on the momentum resolution tail.}
  The momentum resolution function shown in Figure~\ref{fig:CE_fit}
  has a non-Gaussian tail on the high-momentum side.
  As the DIO spectrum is rapidly falling towards the endpoint, the uncertainty
  on the tail may lead to a non-negligible uncertainty on the expected background.
  % In general, the resolution function depends on the track momentum.
  The resolution tail at 100 \MeVc\ can not be studied directly using the data --
  there is no physics process which could be used for that.
  We therefore plan to perform a detailed study of the detector momentum response
  using the sharp high energy ($\sim$52 MeV) edge of the positron
  spectrum measured from the decays of stopped positive muons. The magnetic field
  in the tracker will be reduced by $\sim$50\% to match the curvature of the
  reconstructed positron tracks with the curvature of the conversion electron tracks
  in the nominal magnetic field.
  Below, we outline the proposed method and demonstrate that its intrinsic uncertainty
  is small.

  \indent
  From Eq.~\ref{eq:dio_mom_res}, the uncertainty on the tail is dominated by
  the uncertainty on the parameter N$_2$.
  A direct fit of the resolution function for simulated 52.8 \MeVc\ positrons,
  shown in Figure~\ref{fig:dio_mupl2_chisquare} (left), returns $N_2 = 8.5 \pm 0.6$.
  To determine the value of $N_2$ from the analysis of the Michel spectrum,
  we assume that all parameters in Eq.~\ref{eq:dio_mom_res},
  except $N_2$, are fixed from the studies of cosmic and $\pi^+ \to e^+ \nu$ events,
  and for the present study their values are taken from the fit of the 52.8 \MeVc\
  positron dataset.
  A convolution of the theoretical Michel spectrum with the resolution function
  corresponding to different values of $N_2$ produces multiple templates.
  Each template is used to fit the spectrum of Michel positrons simulated
  and reconstructed in B = 0.5 T, with the only floating parameter in the fit
  being the overall normalization.
  The analysis of the $\chi^2$ distribution dependence on $N_2$ yields the
  best value of $N_2 = 9.7 ^{+2.1}_{-1.4}$. The best fit is shown
  in Figure~\ref{fig:dio_mupl2_chisquare} (right).
%
%  \todo[color=gray!20,linecolor=gray,tickmarkheight=10pt] { (Michael) I don't think we need this level of detail of the fit, perhaps:\\
%    ``...taken from the fit of the 52.8 \MeVc\ positron dataset. The tail parameter $N_{2}$ is then fit to the simulated Michel
%    positron dataset in a B = 0.5 T field, which yields the best value of $N_{2} = 9.7^{+2.1}_{1.4}$. The best fit...}
%
  The two results are statistically consistent, and their relative difference of 14\%
  can be used to estimate the systematic uncertainty of the method.
%
%  \todo[color=gray!20,linecolor=gray,tickmarkheight=10pt] { (Michael) How does this quantify the method uncertainty if they're consistent? And this is not
%  added to the total, as described above, so why assume this? Perhaps just say $\sigma < 14\%$ and then note the effects at this limit?
%  }
  Assuming the relative uncertainty scales with the track curvature,
  the resolution function for 100 \MeVc\ electrons reconstructed at B = 1 T
  should have the same relative uncertainty on $N_2$.
%  The fit to the momentum resolution of 105 MeV/c electron tracks,
%  simulated and reconstructed in full field, results in the value of $N_2 = 6.5 \pm 0.6$,
%  as shown in Figure~\ref{fig:CE_fit}.
  %
  Under this assumption, convolving the momentum resolution function at 105 \MeVc\ from Figure ~\ref{fig:CE_fit} with the DIO spectrum
  % and using the relative uncertainty on N$_2$ of 14\%
  results in the relative uncertainty on the DIO background of $[+23\%,-11\%]$. 
  This uncertainty, contributed to by the experimental procedure, is already small
  compared to the uncertainty due to the momentum scale and can be further
  reduced in the future.
%  We note, that the statistics of $10^{10}$ Michel decays of positive muons could be collected
%  in a calibration run within one hour, so the statistical uncertainties are expected to be small
%  compared to the systematic ones.
%
%  \todo[color=gray!20,linecolor=gray,tickmarkheight=10pt] {
%    (Michael) Perhaps add ``, where future studies will continue to understand these uncertainties'' since
%    you don't measure any non-statistical effect here
%  }
\item \textbf{Theoretical uncertainty}
  on the DIO spectrum \cite{2016_Szafron_DIO_LL_PhysRev,2016_Szafron_DIO_LL_PhysLettB} is already small,
  at less than $\pm$2.5\%.
  The largest uncertainty comes from the uncertainty in the nuclear charge distribution ($\pm$2\%).
  %The radiative effects has been included, while the remaining higher order radiative effects are
\end{enumerate}

%%%%%%%%%%%%%%%%%%%%%%%%%%%%%%%%%%%%%%%%%%%%%%%%%%%%%%%%%%%%%%%%%%%%%%%%%%%%%%
\subsubsection{Expected Yield of the DIO Electrons}

\begin{figure}[ht]
\centering
\begin{tikzpicture}
  \node[anchor=south west,inner sep=0] at (0,0.) {
    % \node[shift={(0 cm,0.cm)},inner sep=0,rotate={90}] at (0,0) {}
    % \makebox[\textwidth][c] {
      \includegraphics[width=0.5\textwidth]{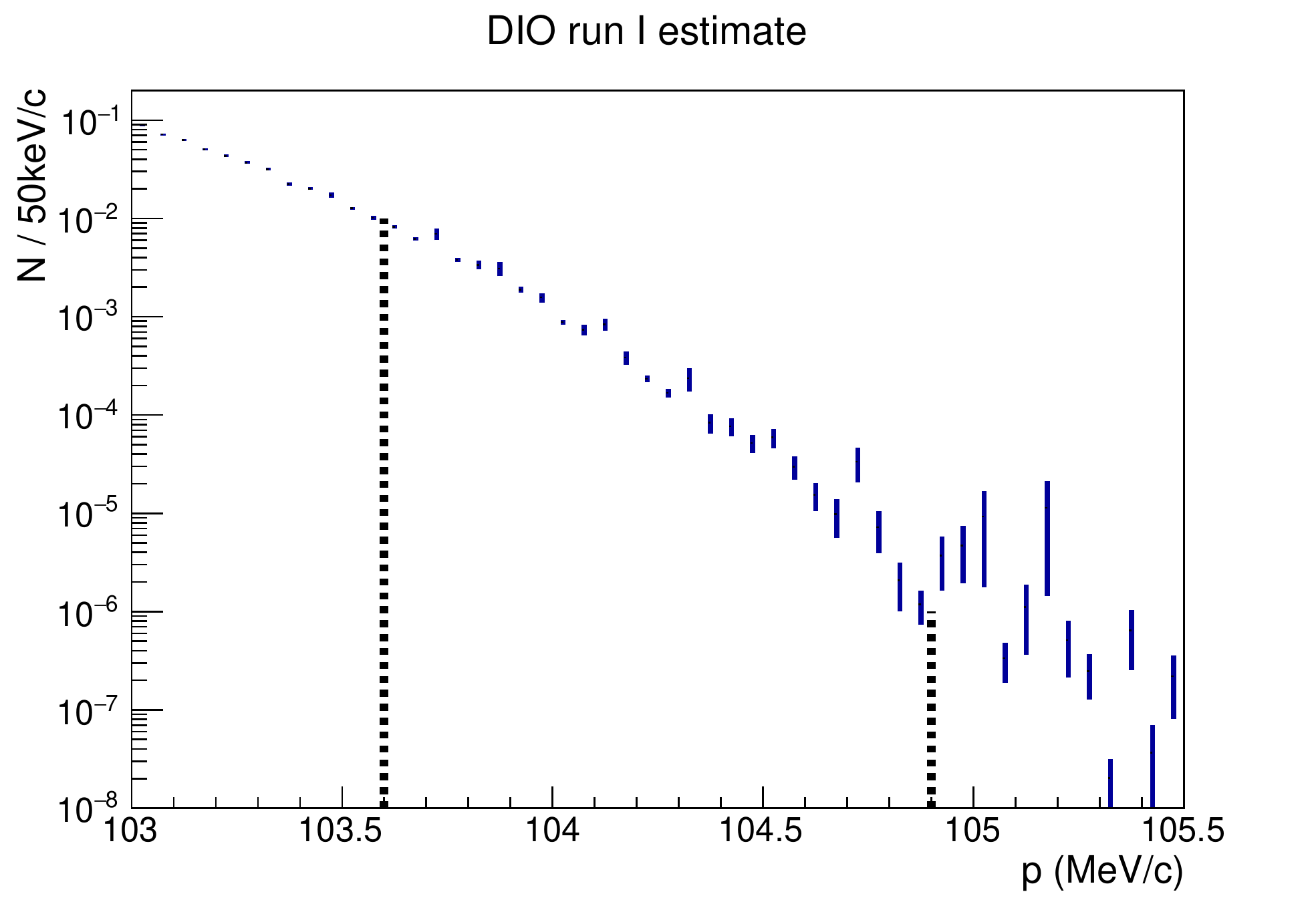}
    %}
  };
  % \node [text width=6cm, scale=0.8] at (4.5,6.4) {mu2e-18894 by Kevin Lynch and Jim Popp};
\end{tikzpicture}

\caption{
  \label{fig:fele_run1_DIO}
  DIO electron spectrum normalized to Mu2e Run I scenario, 6$\times10^{16}$ stopped muons.
  The DIO background integral over the optimized signal region, shown with the dashed lines,
  is N$_{DIO}$=0.038$\pm$0.002~(stat)$^{+0.025}_{-0.015}$(syst).
}
\end{figure}

The DIO background normalized to the stopped muon flux of Run I is shown in Figure~\ref{fig:fele_run1_DIO}.
The estimated DIO background for Mu2e Run I is 0.038 $\pm$ 0.002(stat)$^{+0.025}_{-0.015}$ (syst).
\noindent

%%%%%%%%%%%%%%%%%%%%%%%%%%%%%%%%%%%%%%%%%%%%%%%%%%%%%%%%%%%%%%%%%%%%%%%%%%%%%%%
\subsection{Radiative Pion Capture}
\label{sec:rpc_bkg}

RPC 
% \todo[color=gray!20,linecolor=gray,tickmarkheight=10pt] { (Michael) ``Radiative pion capture occurs...'' -- RPC is already defined}
occurs when pions contaminate the muon beam and stop within the stopping target. The stopped pions undergo the process
$\pi^{-} + N(A,Z) \rightarrow \gamma^{(*)} + N(A, Z-1)$ , followed by an asymmetric $\gamma \ra e^+e^-$ conversion producing electrons
with an energy spectrum extending above 130 MeV. This is one of the main background sources
to the \MuToEmConv\ search. Emission of virtual photons with $q^2 > (2 m_e)^2$ is a direct source of $e^+e^-$ pairs.
Following Refs. \cite{RPC_1955_Kroll_Wada_PhysRev.98.1355, RPC_1960_Joseph_IlNuovoCimento.16.6.997}, this process is referred to as
internal conversion. By extension, the conversion of on-shell photons in the detector material is referred to as the process of external
conversion. Compton scattering of on-shell RPC photons in the detector also produces background electrons.
This causes an increase in the RPC background electron yield for external conversions and makes the spectra of electrons
and positrons differ.

The internal conversion fraction ($\rho$),
% \todo[color=gray!20,linecolor=green,tickmarkheight=10pt] { (Michael) Perhaps ``...fraction ($\rho$), the...''}
the ratio of the off-shell and on-shell photon emission rates,
has been calculated in Refs. \cite{RPC_1955_Kroll_Wada_PhysRev.98.1355, RPC_1960_Joseph_IlNuovoCimento.16.6.997}.
In this analysis, the internal conversion fraction is assumed to be independent of the photon energy,
and the value of $\rho~=~0.0069~\pm~0.0003$, measured in Ref. \cite{samios}, is used.

The RPC background modeling relies on the RPC measurements on nuclei published in Ref. \cite{Bistrilich}.
As there is no published data on Al,  the spectrum of RPC photons measured on a Mg target is used.
According to Ref. \cite{Bistrilich}, for nuclei with the nuclear charge Z in the range $6 < Z < 20$,
the measured RPC branching ratio varies by ${\sim} 10\%$. Although the measured spectra are not exactly the same,
the difference between Al and Mg should not introduce a significant additional systematic uncertainty.

\subsubsection{RPC Sources}

A pulsed timing structure of the proton beam leads to two distinct components of RPC background:

\begin{enumerate}
\item   \textbf{ In-time RPC:} radiative capture of pions produced by protons arriving in the beam pulse.
The rate of in-time RPC rapidly decreases with time roughly
    following the negative pion lifetime, and the corresponding
    background can be minimized by sufficiently delaying the live-time
    search window with respect to the beam pulse.
\item \textbf{Out-of-time RPC:} radiative capture of pions produced by out-of-time protons.
  A delayed live-time window cannot eliminate such pions, only extinction of out-of-time protons can do this.
\end{enumerate}

A third source of delayed RPC background results from antiproton annihilation in the transport solenoid
and is described in Section~\ref{sec:pbar_bkg}.

\subsubsection{Momentum and Time Distributions}

\begin{figure}[ht]
  \centering
  \includegraphics[width=4in]{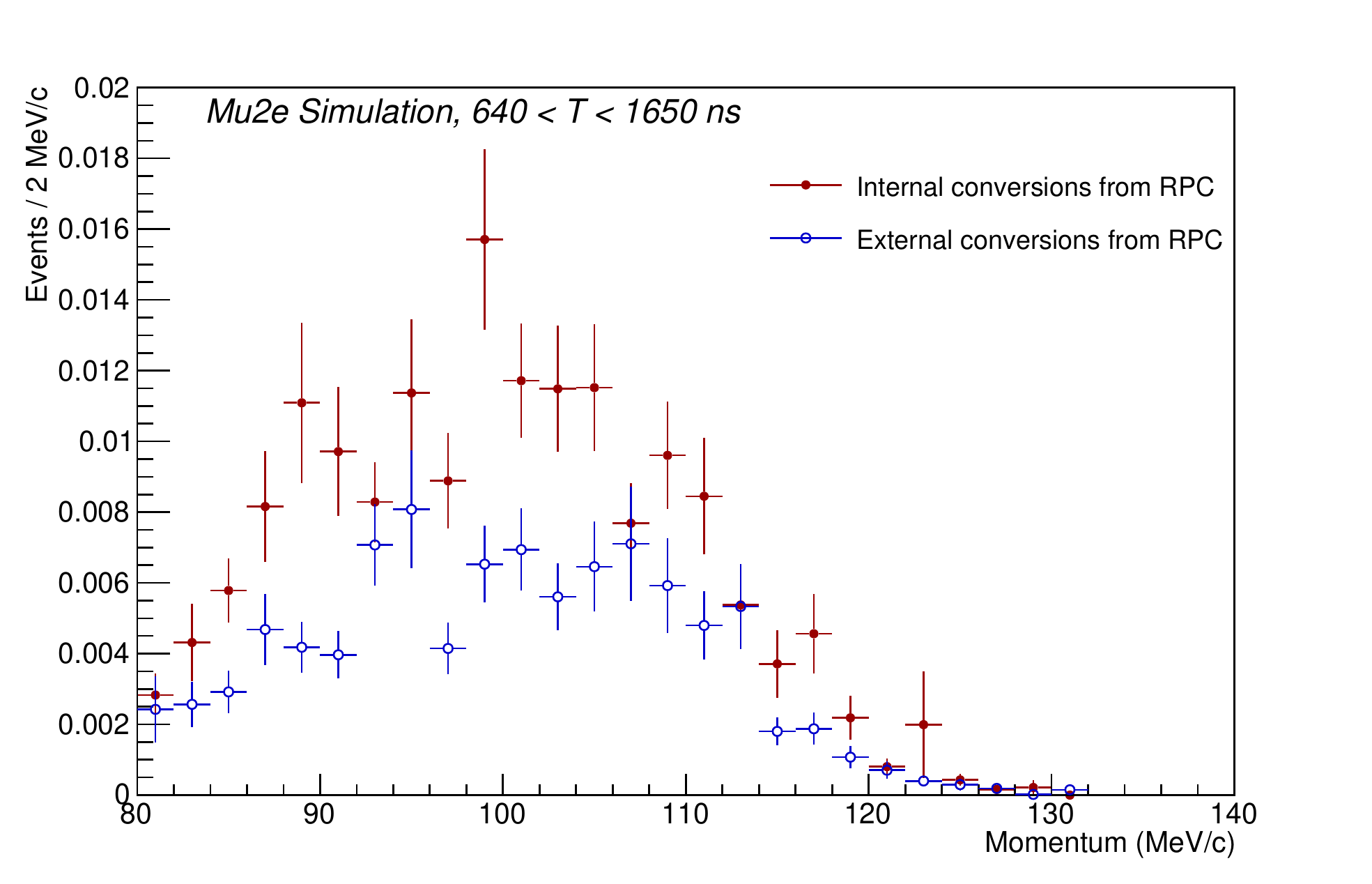}
  \includegraphics[width=4in]{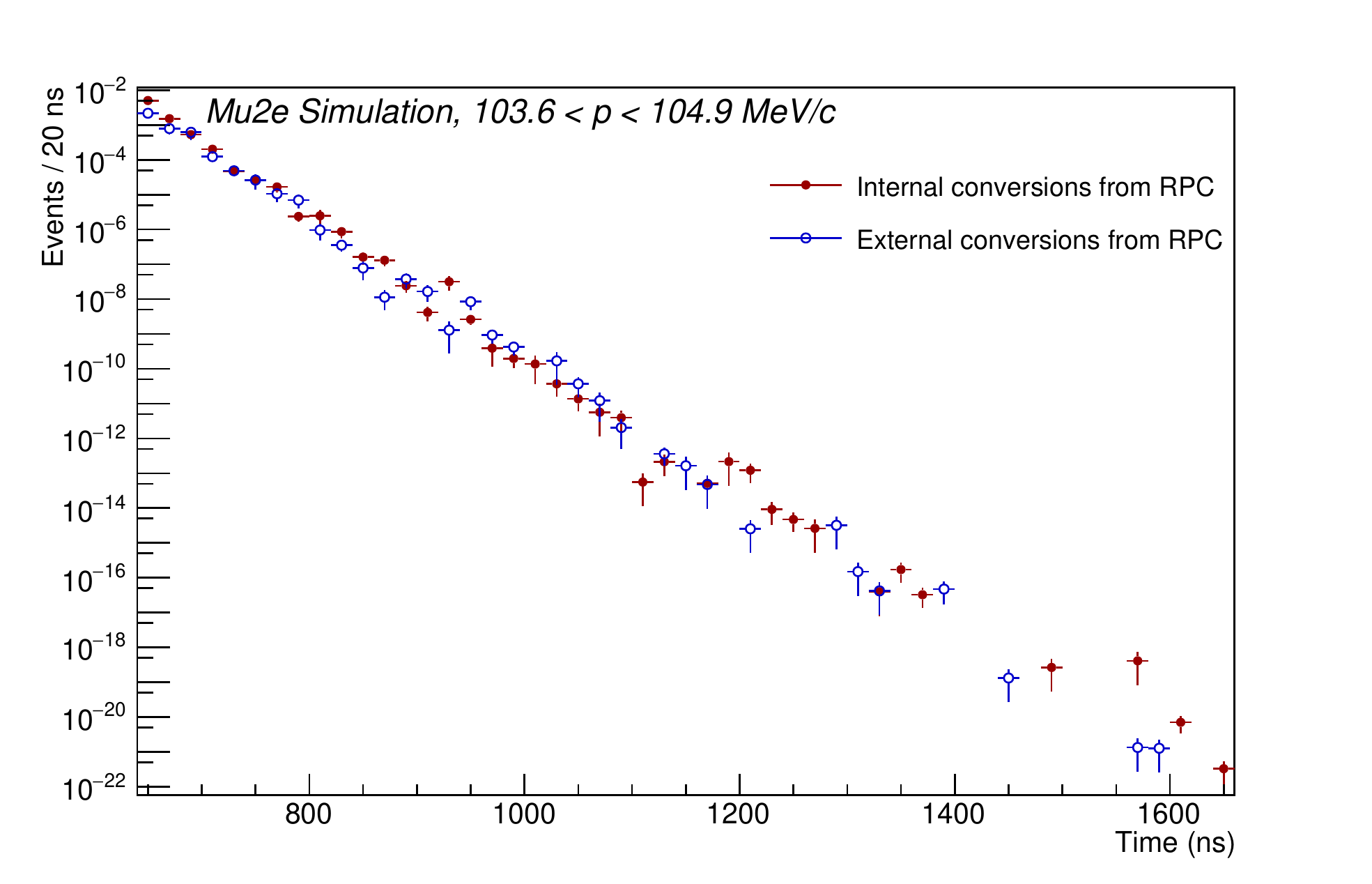}
  \caption{
  \label{fig:Both_Mom_and_Time}
  Momentum and time distributions for electrons from the in-time RPC background. All track selections
  except momentum and timing cuts are applied in both cases. In addition, the momentum distribution
  includes a cut on the reconstructed electron track time, $640~<~T_0~<~1650$ ns, the timing distribution
  is plotted for events with the reconstructed electron track momentum $103.6~<~p~<~104.9$ \MeVc.
  The plots are normalized to represent the expected Run I background.
  % \todo[color=gray!20, inline] { (Michael)  Perhaps ``expected Run I background?''}
  }
\end{figure}

Fig.~\ref{fig:Both_Mom_and_Time} shows the distributions of the reconstructed track momentum and time
for in-time RPC electrons. All track selection criteria are enforced except for momentum and time cuts. 
The plots are normalized to represent the number of protons on target expected in Run I.
The RPC photon spectrum with the endpoint at $\sim$134 \MeVc\ defines the maximal momentum
of the reconstructed electrons, and below $\sim$80 \MeVc\ the reconstruction is limited by the
tracker acceptance. RPC photons contributing to the background predominantly convert in the same
stopping target foil in which they were produced. Due to the small thickness of the stopping target foils,
the contribution of external conversions is about 50\% lower than the contribution of internal conversions.
The time distribution displays a characteristic exponential slope.
Pions produced by out-of-time protons can arrive at the stopping target at any point within the event
and, consequently, the time distribution for out-of-time electrons is assumed to be flat.

\sloppy The estimated contribution of the in-time RPC is $0.010 \pm 0.002 ({\rm stat})$ events.
The contribution of the out-of-time RPC, proportional to the proton beam extinction,
is $(1.2\pm0.1({\rm stat}))10^{-3}\times(\zeta/10^{-10})$ events.

%%%%%%%%%%%%%%%%%%%%%%%%%%%%%%%%%%%%%%%%%%%%%%%%%%%%%%%%%%%%%%%%%%%%%%%%%%%%%%
\subsubsection{Systematic Uncertainties}
\begin{itemize}
\item \textbf{RPC Photon Spectrum}

  A RPC branching rate of  $BR_{RPC} = (2.15 \pm 0.2) \%$, taken from Ref.~\cite{Bistrilich},
  is used in this study. A relative uncertainty of 9.3$\%$ on this measured rate  is assigned
  as the corresponding systematic uncertainty on $BR_{RPC}$ for Al.

\item \textbf{Internal Conversion Fraction}
  
  The internal conversion fraction measured in Ref.~\cite{samios},  $\rho=0.0069\pm0.00031$, is used.
  Its value is assumed to be independent of the photon energy.
  The measurement presented in Ref.~\cite{samios} was performed using hydrogen, where $E_{\gamma} = 129.4$ MeV.
  As the energy region of interest for the \MuToEm\ conversion search is around 105 MeV,
  and the theory predicts a decrease of $\rho$ as the photon energy goes down, this assumption
  is conservative.

\item \textbf{Proton Pulse Shape}

  The variation in the pion-capture background due to uncertainty in the simulated
  shape of the incoming proton beam time structure was found to be negligible.

\item \textbf{Pion Production Cross Section}

   The Run I data taking plan assumes collection of $6\times 10^{16}$ stopped negative muons (see Table \ref{tab:running_time}).
    As muons are primarily produced in pion decays, one might think that the ratio
    of the number of stopped negative pions to the number of stopped negative muons, $N_{\rm stopped}^{\pi^-}/N_{\rm stopped}^{\mu^-}$, is constant,
    and that, for a fixed number of stopped muons, the RPC background would not depend on the pion production cross section.
    However, the pions which stop in the stopping target have momenta significantly lower than the pions
    producing stopped muons, so the ratio $N_{\rm stopped}^{\pi^-}/N_{\rm stopped}^{\mu^-}$
    depends on the energy spectrum of the produced pions.
    As there is no experimental data on production of charged pions with momenta below 100 \MeVc,
    model-dependent predictions have to be used. For a fixed number of stopped negative muons,
    different hadro-production models implemented in Geant4 predict variations of the RPC background. The relative change in the RPC background yield depends on the model used, and results in an asymmetric systematic,
    shown in Table~\ref{table:rpc_sys}.
  \end{itemize}

\subsubsection{Summary of Systematic Uncertainties on the RPC yield}

Table~\ref{table:rpc_sys} lists all the systematic uncertainties discussed.
For each column the contributions are added in quadrature to provide total uncertainties. It must be noted that the major systematic uncertainties in this result come from assumptions made within our modeling and can be reduced through using a data-driven estimate.The RPC yield could potentially be estimated through measurements of electrons from pions arriving early at the stopping target (before any conversion electron is expected). They  could be fitted with an exponential expression and the yield in the signal region could be extrapolated from that fit. It is important to note that data from Run I can be used to measure the RPC photon spectrum and RPC branching fraction in aluminum, and also help validate our pion production cross section model, thus reducing systematic uncertainties in future physics runs.

\begin{specialtable}[ht]
  \centering
  \small
  \caption{
    \label{table:rpc_sys}
   List of systematic uncertainties and their relative contributions to the RPC yield.  }

   \begin{tabular}{c c  c}
\toprule
 \textbf{Systematic Contribs.}                & Internal Conv.    & External Conv.    \\ [0.5ex]
\midrule
RPC fraction \cite{Bistrilich}                & 9.3$\%$           &  $9.3 \%$          \\
Internal conversion coefficient \cite{samios} & 4.5$\%$           &                    \\
Pion Production Model                         & $(+9, -27)\%$     &  $(+9, -27)\%$     \\
\midrule
\textbf{Total Sys.}                           & $(+13.7,-28.9)\%$ &  $(+12.9,-28.5\%$  \\   \hline
\bottomrule
 \end{tabular}
\end{specialtable}

With the systematic uncertainties included, the expected background contributions
of the in-time and out-of-time RPC are $0.010 \pm 0.002 ({\rm stat}) ^{+0.001}_{-0.003} ({\rm syst})$  and 
$(1.2 \pm 0.1 ({\rm stat}) ^{+0.1}_{-0.3}({\rm syst})) 10^{-3} \times (\zeta/10^{-10})$, respectively.

%Main author: Michael MacKenzie

\subsection{Radiative Muon Capture}
\label{sec:rmc_bkg}

The process of radiative muon capture, $\mu^{-} + N(A,Z) \rightarrow \gamma^{(*)} + \nu_\mu + N(A, Z-1)$, in many aspects
is similar to RPC.
The theoretical framework
developed to describe internal pair production in nuclear RPC \cite{RPC_1955_Kroll_Wada_PhysRev.98.1355}
is general enough to include nuclear RMC,
and the probability of internal RMC conversion is defined by a very similar
calculation \cite{RMC_2020_Plestid_Hill_arXiv_2010.09509}.

However, there are also important differences.
The maximal energy of the RMC photon, defined by the muon mass, is about 34 MeV lower than the maximal
energy of the RPC photon, which is defined by the charged pion mass.
For $^{27}$Al, the maximal energy of the RMC photon is ${\sim} 101.9$ MeV,
about 3 MeV below the expected $\MuToEm$ conversion signal.
The timing dependence of the RMC electron rate is defined by the lifetime of the muonic aluminum atom,
common for all processes which proceed through muon capture.

The energy spectrum of the RMC photons is also very different from the spectrum of RPC photons.
General features of the RMC spectra are well described within the closure approximation,
which replaces the sum over transitions into multiple final nuclear states with a transition into
a single state with the mean excitation energy \cite{RMC_1980_CHRISTILLIN}.
Within the closure approximation, the RMC photon spectrum is fully defined by one parameter --
the endpoint of the photon spectrum, $\kmax$:

\begin{equation}
R(x) =  \frac{e^2}{\pi} \frac{\kmax^2} {m^2_\mu} (1-\alpha) (1-2x+2x^2) x (1-x)^2\ ,
  \label{eq:closure}
\end{equation}

\noindent
where $x = E_\gamma/\kmax$ and $\alpha = \frac{N-Z}{A}$.
% \todo[color=gray!20,linecolor=gray,tickmarkheight=10pt] { (Michael) ``and $\alpha = \frac{N-Z}{A}$''}
\cite{RMC_1980_CHRISTILLIN}.
The closure approximation captures reasonably well the total RMC rate and the shape of the RMC photon spectra,
however, as $\kmax$ is a model parameter, it can not be relied upon to determine the spectrum endpoint.
Typically, the closure approximation fits return \kmax\ values 5-10 MeV below the kinematic limit.
For example, for a $\mu ~^{27}Al ~\ra ~\gamma~\nu~^{27}Mg $ RMC transition, the maximal kinematically allowed photon energy
is ${\sim} 101.9$ MeV, while fits to the experimental data return $\kmax = 90.1 \pm 1.8$ MeV \cite{RMC_1999_PhysRevC.59.2853}.

As the $\MuToEm$ conversion electron energy is ${\sim} 105$ MeV and the Mu2e momentum resolution
$\Delta{p} \lesssim 1$ \MeVc\ FWHM, the background from RMC, estimated using the closure approximation
spectrum with the endpoint of $\kmax = 90.1$ \MeVc, is negligible.
As there is nothing that explicitly forbids RMC photons up to the kinematic limit,
it is reasonable to assume that the spectrum has a tail up to this limit 
with an event rate too low to have been measured by the performed experiments.
To test the sensitivity of the $\MuToEm$ conversion search to this assumption,
the RMC photon spectrum on aluminum described by Eq.~\ref{eq:closure} with \kmax = 90.1 MeV
is modified by adding to it a tail extending up to the kinematic limit.
Two parameterizations of the tail are considered:
1) a closure approximation spectrum with \kmax = 101.9 MeV and
2) a flat distribution.
%% \del{The first spectrum is similar to what was reported by the 1988 TRIUMF search for muon to positron conversion
%% on titanium \cite{1988_CLFV_TRIUMF_TITANIUM}, while the latter spectrum represents the worst case under the assumption
%% that the photon energy spectrum should be falling.}

\begin{figure}
  \centering
  \includegraphics[width = 0.99\linewidth]{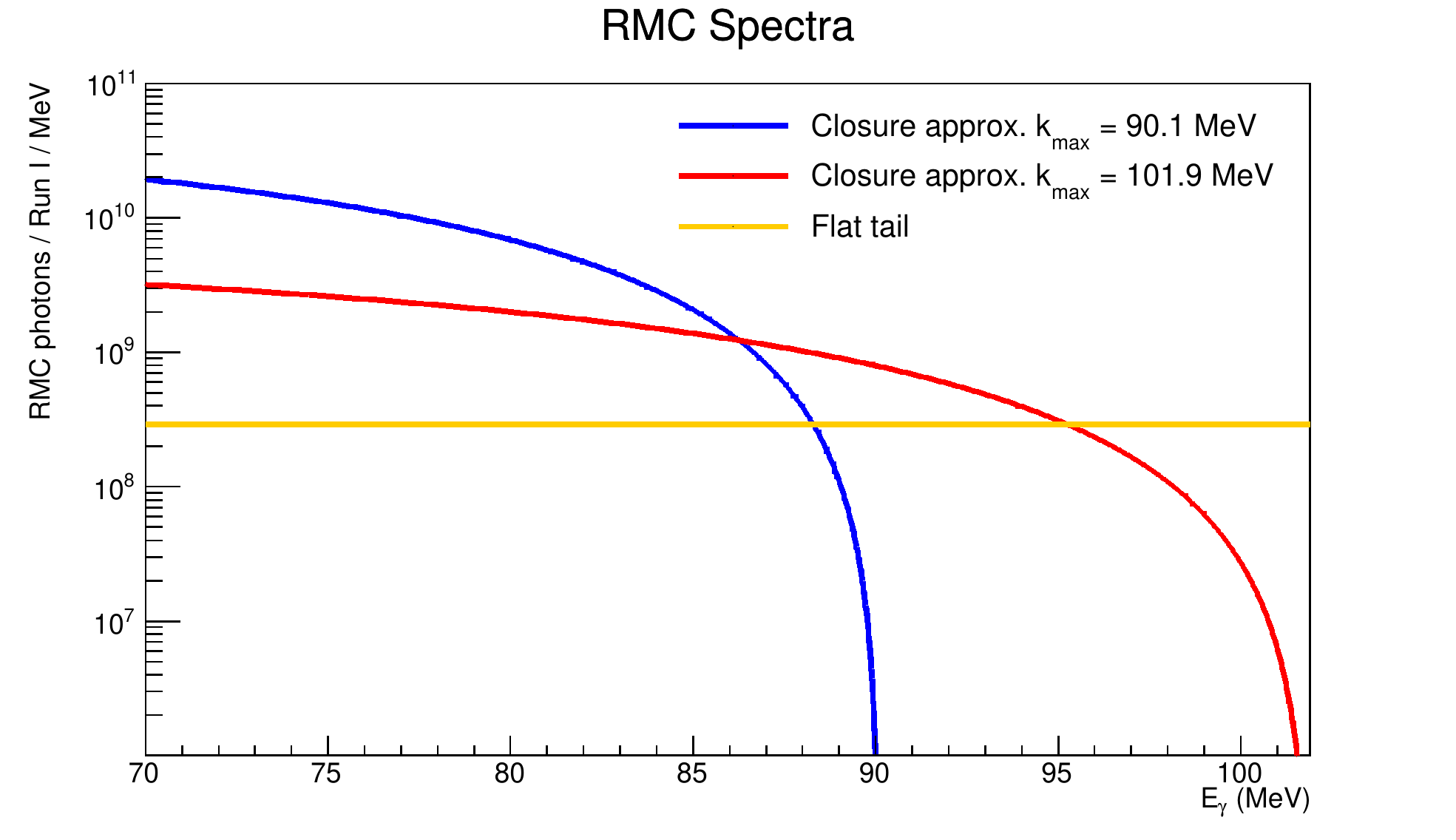}
  \caption{
    \label{fig:rmc_spectra}
    The RMC photon energy spectrum corresponding to the closure approximation with $\kmax = 90.1$ MeV,
    shown in blue, together with the two parameterizations of the tail, described in the text.
    All three spectra are normalized to the Run I expectations.
  }
\end{figure}

The first choice is similar to that used in Ref. \cite{1988_CLFV_TRIUMF_TITANIUM}, while the second choice ignores
the phase space reduction and should result in an overly-conservative background estimate.
In each case, the tail is normalized to 3 events above 90 MeV in the previous measurement, which is close to the sensitivity
limit of Ref. \cite{RMC_1999_PhysRevC.59.2853}.
The chosen normalization corresponds to a rate of
$R_{\rm RMC}(E_{\gamma} > 90~{\rm MeV}) = \frac{3}{3,051} \times R_{\rm RMC}(E_{\gamma} > 57~{\rm MeV}) \approx 1.6 \times 10^{-8}$.
The two parameterizations of the RMC photon tail are shown in Figure~\ref{fig:rmc_spectra}
along with the closure approximation fit of Ref. \cite{RMC_1999_PhysRevC.59.2853},
normalized to the number of stopped muons expected in Run I.

Table \ref{tab:rmc_background} gives the background estimates for
both considered parameterizations of the tail
in the optimized signal window introduced in Section \ref{sec:background_estimates}.
%% \del{The on-shell contribution is the dominant background source due to the higher endpoint
%%   of Compton scattered electrons compared to (internally) pair-produced electrons.
%%   The flat tail extension model only begins to be a relevant background, but this represents a very conservative
%%   background shape at the upper limit of the allowed branching fraction of RMC above 90 MeV.
%%   Therefore, the RMC background is considered negligibly small for the $\MuToEm$ conversion search.
%% }
The dominant contribution comes from the on-shell photons: for the same photon energy,
Compton scattering produces electrons with a momentum spectrum that extends higher than the spectrum
of pair-produced electrons. Table \ref{tab:rmc_background} shows that even under an overly-conservative
assumption the RMC background to the $\MuToEm$ conversion search is negligibly small.
However, the high energy tail of the RMC photon spectrum may modify the total electron spectrum
around 100 \MeVc\ and impact measurements of the high-momentum end of the DIO spectrum.

\begin{specialtable}[H]
  \centering
  \small
  \caption{
    \label{tab:rmc_background}
    RMC background estimates using the altered closure approximations. These estimates use
    the optimized signal window introduced in Section \ref{sec:background_estimates}, $640 < T_{0} < 1650$ ns and $103.6 < p < 104.9$ \MeVc.
    The estimates have a statistical accuracy of ${\sim} 50\%$.
  }
  \begin{tabular}{lcc}
    \toprule
    RMC tail parameterization     & Production mechanism & Run I background \\
    \midrule
    Closure approx., \kmax=101.9 MeV & On-shell             & $1.2\times 10^{-5}$ \\
    Closure approx., \kmax=101.9 MeV & Off-shell            & $1.5\times 10^{-7}$ \\
    \midrule
    Flat                          & On-shell             & $2.4\times 10^{-3}$ \\
    Flat                          & Off-shell            & $5.5\times 10^{-5}$ \\
    \bottomrule
  \end{tabular}
\end{specialtable}

The value of $2.4\times 10^{-3}$ events is used as a conservative upper limit on the expected RMC background.

\subsection{Antiprotons}
\label{sec:pbar_bkg}
Another potentially significant source of background is due to the annihilation
of antiprotons produced in the interactions of the $\Ekin = 8$ GeV proton beam
at the tungsten target and entering the TS. Such antiprotons can pass through the
TS, enter the DS, and annihilate in the stopping target producing signal-like electrons.
In addition, radiative capture of negative pions produced in the antiproton annihilation
along the beamline and reaching the stopping target increases the overall RPC background,
adding a component with a time dependence very different from those discussed in Section ~\ref{sec:rpc_bkg}.

The background induced by antiprotons cannot be efficiently suppressed by the
time window cut used to reduce the prompt background because the antiprotons
are significantly slower than the other beam particles and their secondary products
are delayed with respect to the beam. The only way to suppress the antiproton background
is to use additional absorber elements, located at the entrance and at the center
of the TS.
The antiproton background estimate is mostly affected by the uncertainty
on the antiproton production cross section that has never been measured
at such low energies.

%% To reduce
%% the impact of this background, additional absorber elements are added to the entrance of the transport
%% solenoid and the central collimator in the transport solenoid. These absorber elements are optimized
%% to reduce the antiproton acceptance in the beamline, without significant reductions in the muon yield,
%% while also not introducing more delayed pion production due to antiproton interactions in these absorber
%% elements. These delayed pions are particularly dangerous as they are more likely to arrive past the
%% delayed livegate start meant to reduce backgrounds from beam pions.

\subsubsection{Antiproton Production Cross Section}

%% The transport solenoid is designed to only accept low momentum particles, and as such low momentum antiprotons
%% have the largest geometric acceptance, especially those produced backwards, away from the incoming proton
%% direction and towards the transport solenoid. Unfortunately, there is little data on the cross section for
%% antiproton production in this region. There is significant data in the regime with primary proton momenta
%% above 100 \GeVc, but the Mu2e primary proton beam has a momentum of ${\sim} 8.9$ \GeVc.
%% To parametize the antiproton production at Mu2e, we utilize the data shown in Table \ref{tab:pbar_data}.

The Mu2e primary proton beam has a momentum of ${\sim} 8.9$ \GeVc, but the lowest proton momentum at
which cross section experimental data are available is 10 \GeVc\ (Table \ref{tab:pbar_data}).
\begin{specialtable}[H]
  \centering
  \small
  \caption{
    Available data for antiproton production from proton interactions on different heavy nuclei.
    The antiproton momentum column ($p_{\bar{p}}$) indicates the minimum and maximum measured momentum;
    when these are separated by a $\div$ more than 2 points have been measured.
    \label{tab:pbar_data}
  }
  \begin{tabular}{ccccl}
    \toprule
    $N_{points}$ & $p_{\rm proton}$ (\GeVc) & $\theta_{\bar{p}}$ ($\sigma$, $^o$) & $p_{\bar{p}}$ (\GeVc) & Nuclear target, reference \\
    \midrule
    2          & 10            & 0                                 & $1.06, 1.40$      & Tungsten, Anmann et al.    \cite{pbar_0deg}        \\
    13         & 10            & 3.5                               & $1.25 \div 4.50$  & Tantalum, Sibirtsev et al. \cite{1991_pbar_3.5deg} \\
    5          & 10            & 10.5                              & $0.73 \div 2.47$  & Tantalum, Kiselev et al.   \cite{2012_pbar_10.5-59deg} \\
    8          & 10            & 10.8                              & $0.72 \div 1.87$  & Gold,     Barabash et al.  \cite{pbar_10.8deg} \\
    8          & 10            & 59                                & $0.58 \div 1.35$  & Tantalum, Kiselev et al.   \cite{2012_pbar_10.5-59deg} \\
    4          & 10            & 97                                & $0.60 \div 1.05$  & Tantalum, Boyarinov et al. \cite{pbar_97-119deg} \\
    2          & 10            & 119                               & $0.59, 0.66$      & Tantalum, Boyarinov et al. \cite{pbar_97-119deg} \\
    \bottomrule
  \end{tabular}
\end{specialtable}

To generate antiprotons from protons of any momentum the invariant differential cross section ($Ed^3\sigma/dp^3$) has been parametrized as a function of the antiproton momentum  ($p^*$) in the center of mass system (c.m.).

In the simple case of a p+p interaction,

\begin{equation}
  \label{eq:pp_pbar_process}
  p + p \ra (p + \bar{p}) + p + p
\end{equation}

\noindent
the maximum $p^*$ ($p^*_{max}$) corresponds to the case in which the three protons in the final state act as a single body and recoil in the direction opposite to the $ \bar p $:

\begin{equation}\label{pstarmax}
  p^{*}_{max} =  \sqrt{\bigg(\dfrac{s - (3\,m_{p})^2 + m_{p}}{2\sqrt{s}}\bigg)^2 - m_{p}^2}
\end{equation}

\noindent
where $s$ is the Mandelstam invariant variable and $m_p$ is the proton mass.

When the nucleus, tungsten in the case of Mu2e, is considered as the target,

\begin{equation}
  \label{eq:pbar_reaction}
  p + W \ra (W^* + \bar{p}) + X
\end{equation}

\noindent
more nucleons can be involved in the interaction and the antiproton momentum in the c.m. can be larger than $p^*_{max}$. The ratio $p^*/p^*_{max}$ is then correlated to the multi-nucleon state participating in the interaction.
The concept of the fraction of maximum momentum in the c.m. can be improved using the variable

\begin{equation}\label{xcm}
  x_{cm}=\frac{p^*}{p^*_{max}}\left(1-\frac{2}{1+e^{\frac{cos\theta^*}{\lambda_F}}}\right)
\end{equation}

\noindent
where the dependence on the antiproton angle in the c.m. system with
respect to the incident proton direction ($\theta^*$) takes into
account the different matter density seen by the particle in case of
forward or backward scattering and $\lambda_F=0.06$ is a parameter
that ensures a smooth transition between the forward and the backward
region. The value of $\lambda_F$ is obtained by fitting the data.
%\todo[color=green!10, inline] { (Michael) Should we specify what this angle is with respect to in the system?}

The parametrization of the invariant cross section as function of $x_{cm}$ is given by Ref. \cite{CLFV_2020_DeFelice_osti_1763411}:

\begin{equation}\label{defbob}
  E\, \dfrac{d^{3}\sigma}{dp^{3}}(x_{cm}) =   \begin{cases}
    N_{G}\,\dfrac{1}{\sqrt{2\pi\sigma^{2}_{G}}}\,e^{-\frac{(x_{cm} -\mu_{G})^2}{2\sigma^{2}_{G}}} & \text{for} \,\,|x_{cm}|\leq 1 \\
    \vspace{0.1in}
    N_{E}\,e^{\frac{\sqrt{1+(\beta^*_{max})^2(x_{cm}^2-1)}-\sqrt{1-(\beta^*_{max})^2}}{\lambda_E}} & \text{for} \,\,x_{cm} <-1\\
   0 & \text{for} \,\, x_{cm} >1\\
  \end{cases}
\end{equation}
where $\beta^*_{max}=p^*_{max}/\sqrt{(p^*_{max})^2+m_p^2}$ and the parameters obtained by fitting the data are:
\begin{eqnarray*}
 N_{G}          & : & \text{Normalization of the Gaussian term}\\
 \sigma_{G}     & : & \text{Sigma of the Gaussian}\\
 \mu_{G}        & : & \text{Mean of the Gaussian}\\
 N_E            & : & \text{Normalization of the exponential term}\\
 \lambda_{E}    & : & \text{Slope of the exponential}
\end{eqnarray*}

\begin{figure}[H]
  \includegraphics[height=0.16\textheight]{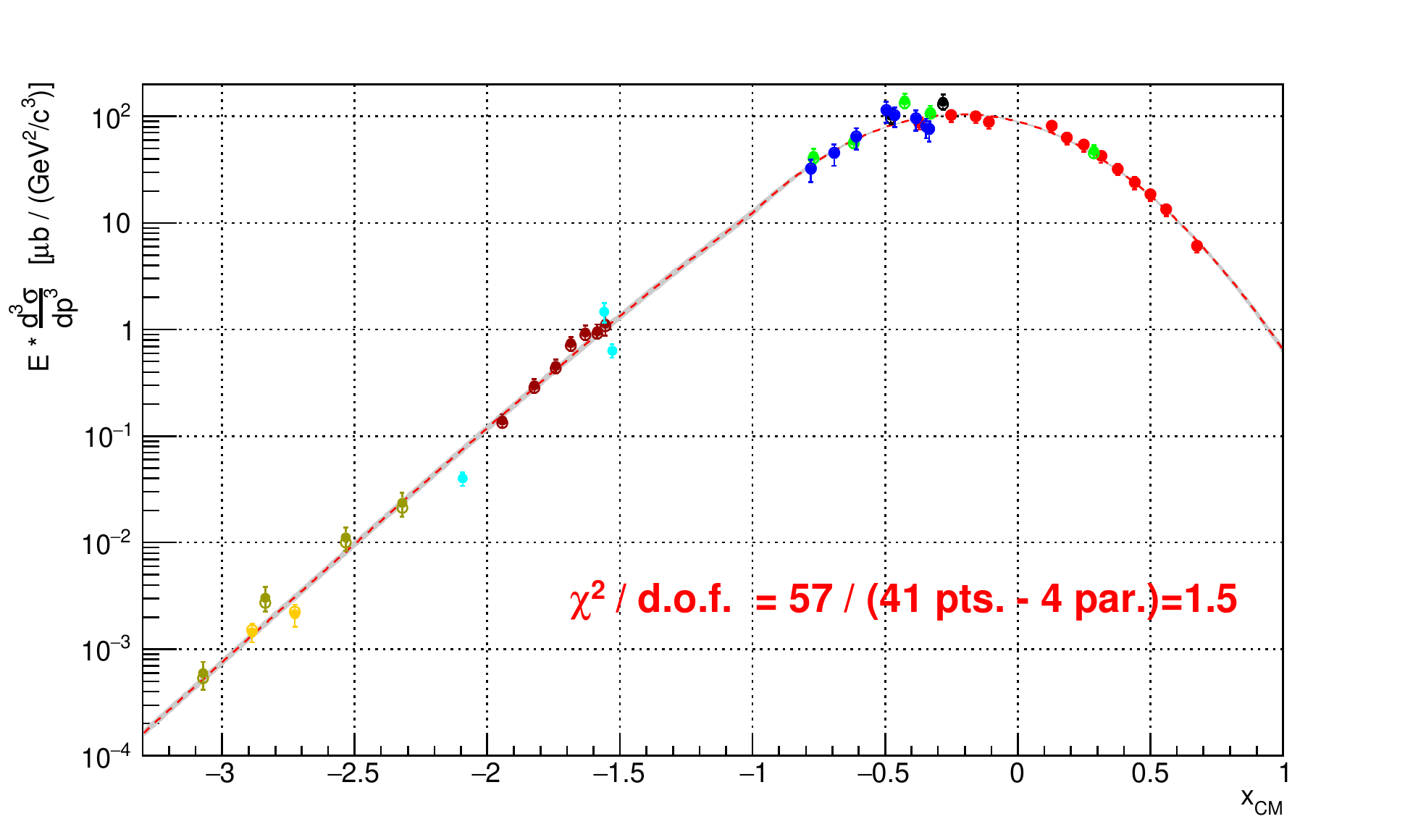}
  \includegraphics[height=0.16\textheight]{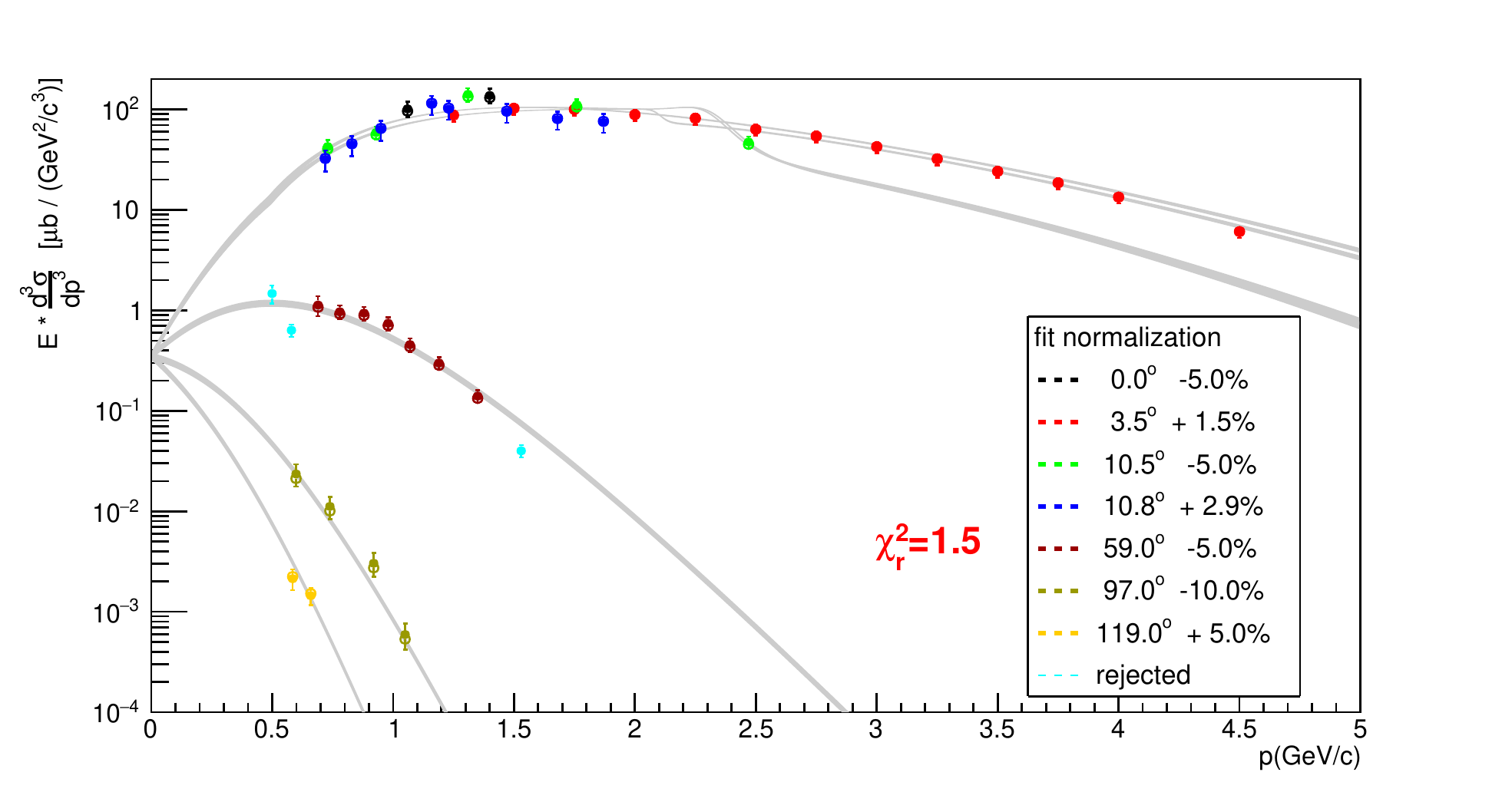}
  \caption{
    \label{fig:fitxcm}
    Invariant cross section as a function of $ x_{cm} $ (left) and $\bar{p}$ momentum (right) for all data points fit using the cross section model. Points in cyan have been excluded from the fit, as they are not consistent
    with the rest of the 59$^o$ data.
    }
\end{figure}

Figure~\ref{fig:fitxcm} shows the fit to the data in the c.m. system and in the laboratory system.
The normalization of the exponential term in Eq.~\ref{defbob} is fixed by the continuity requirement
at $x_{cm}=-1$.
The normalization of the measurements at a given angle, 
that come from the same publication, has also been used as a fit parameter.
The relative change in the normalization for each input dataset is shown in the legend.
%% \todo[color=green!10, inline] { (Michael) What does variation mean here? Difference from the expected?}
%% \todo[color=magenta!10, inline] { (Stefano) The nominal value is 1.Essentially we allow the normalization of the group of the data at the same angle to change together.}

The fit to the data at 10 \GeVc\ (Figure~\ref{fig:fitxcm}) is quite good, and
the corresponding total cross section is 282.4 $\mu$b.
Using an incident proton momentum of 8.9 \GeVc, that is the Mu2e beam momentum,
the total cross section goes down to 213.2 $\mu$b, that is 75\% of the cross section at 10 \GeVc.
%\todo[color=green!10, inline] { (Michael) It should be $\mu$b, without the space I believe}
%\todo[color=green!10, inline] {(Sophie) do we use ``@" or just say "at"?.}

This result can be compared with the one obtained using the simple model proposed in Ref. \cite{pbar_danielewicz},
where the total cross section has been parametrized as a function of the Mandelstam invariant variable $s$,
neglecting the interaction with multinucleon states:

\begin{equation}
  \label{daniel}
  \sigma^{\bar p}_{NN}\propto (\sqrt{s}-4m_p)^\frac{7}{2}
\end{equation}

\noindent
from which one gets:

\begin{equation}
  \frac{\sigma_{8.9}}{\sigma_{10}} = 29\%
\end{equation}
\noindent
where $\sigma_{8.9}$ and $\sigma_{10}$ are the total antiproton production cross sections
for the proton beam momenta of 8.9 \GeVc\ and 10 \GeVc\ 
respectively. As shown in the same paper, the effect of the interaction with more nucleons
is expected to become larger and larger when approaching the antiproton production threshold,
so that Eq. \ref{daniel} becomes less and less valid.

The discrepancy between the results of the two parametrizations reflects the uncertainty in the cross section extrapolation to lower energies
where no experimental data are available. At this point the only statement that can be made is that the cross section at
8.9 \GeVc\ must be lower than the one at 10 \GeVc.
%
% \todo[color=gray!20,linecolor=gray,tickmarkheight=10pt] {
%   (Michael) Should the uncertainty not be Gaussian with $\pm50\%$ then, since it will often
%   be above 100\%? Perhaps it should be flat or a truncated Gaussian between 0 and 100\%?
% }
According to this quite conservative assumption the antiproton production cross section at 8.9 \GeVc\ can be taken as:

\begin{equation}
  \label{sigma89}
  \left(E\, \frac{d^3\sigma}{dp^3}\right)_{8.9}= \left(E\, \frac{d^3\sigma}{dp^3}\right)_{10}\ \times (0.5\pm0.5)
\end{equation}

%%%%%%%%%%%%%%%%%%%%%%%%%%%%%%%%%%%%%%%%%%%%%%%%%%%%%%%%%%%%%%%%%%%%%%%%%%%%%%
\subsubsection{Antiproton Simulation}

The antiproton simulation has been performed in several steps.  % using the Geant4 version 10.05.p01\cite{geant4_physicsguide}.
First, vertices of inelastic proton beam interactions in the production target were simulated and stored.
% \todo[color=gray!20,linecolor=gray,tickmarkheight=10pt] { (Michael) Perhaps ``fired at'' instead ``sent to?''
% }
% the tungsten production target and their inelastic interaction vertices were stored.
The number of antiprotons produced in the production target ($N_{\bar p}^{PT}$) per POT is given by:
%%% \todo[color=gray!20,linecolor=gray,tickmarkheight=10pt] { (Michael) I don't think PT is used again, so perhaps remove this and add
%%%  ``$N_{\bar p}^{PT}$ is the number of antiprotons produced in the production target'' to the following line?}

\begin{equation}
  \label{s0yield}
  \frac{N_{\bar p}^{PT}}{POT}=\frac{\sigma_{\bar p}}{\sigma_{inelastic}}\frac{N_{inelastic}}{N_{POT}}=\frac{0.5 \times 0.2824\ mb}{1710\ mb}0.792=6.5 \times 10^{-5}
\end{equation}
%\todo[color=green!10, inline] { (Michael) ``PT'' is not defined, so perhaps just use ``production target''?}
where $\sigma_{\bar p}$ is the total antiproton production cross section obtained integrating the differential
cross section in Eq. \ref{sigma89}, $N_{inelastic}/N_{POT}=0.792$ is the probability, obtained by Monte Carlo, that
a proton in the beam produces an inelastic interaction in the tungsten target, and $\sigma_{inelastic}=1710$ mb is
taken from Ref. \cite{pbar_totinelastic}.
This value for the total proton inelastic cross section on tungsten is ${\sim} 11\%$ higher
than the value of 1517 mb obtained with MCNP \cite{MCNP61}, but this discrepancy can be neglected
with respect to the 100\% error quoted for the cross section extrapolation at threshold.

In the second step of the simulation the proton inelastic vertices were used as production vertices of antiprotons generated with the momentum
distribution flat between 0 and 5 \GeVc\ and isotropic in direction.
The generated antiprotons were propagated to the TS entrance to determine the TS acceptance
as a function of the antiproton momentum and emission angle.
The calculated TS acceptance has been used to build a significantly more efficient generation model,
where the probability to generate an antiproton with a given momentum and a polar angle was proportional
to the antiproton production cross section used by Geant4 and the square root of the TS acceptance.
To avoid reliance on the Geant4 modeling of the antiproton production, the weights of the
generated antiproton events have been corrected by the ratio of the parametrized
invariant cross section of Eq. \ref{defbob} and the inclusive cross section used by Geant4.

\begin{figure}[H]
  \includegraphics[height=0.155\textheight]{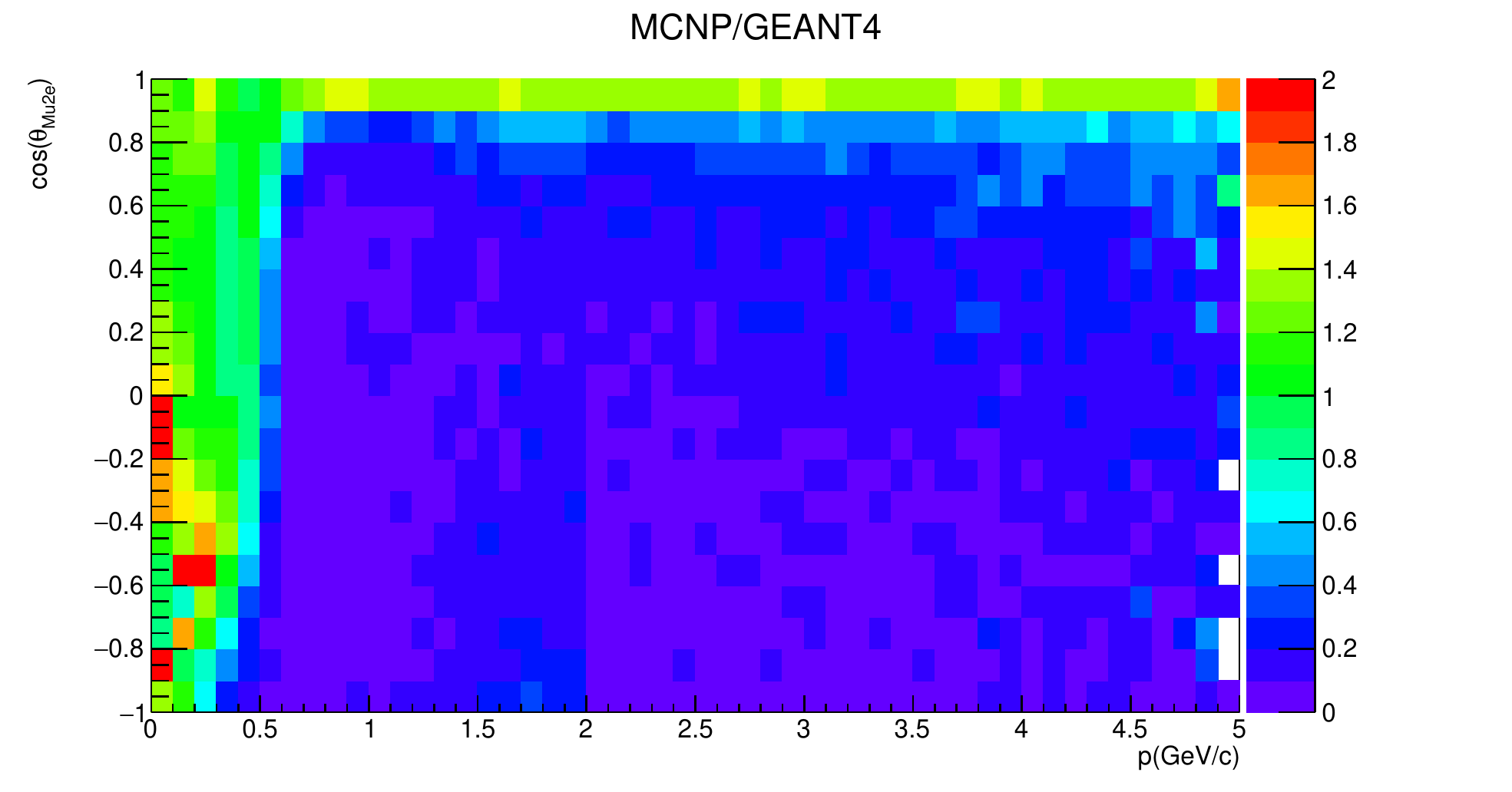}
  \includegraphics[height=0.16\textheight]{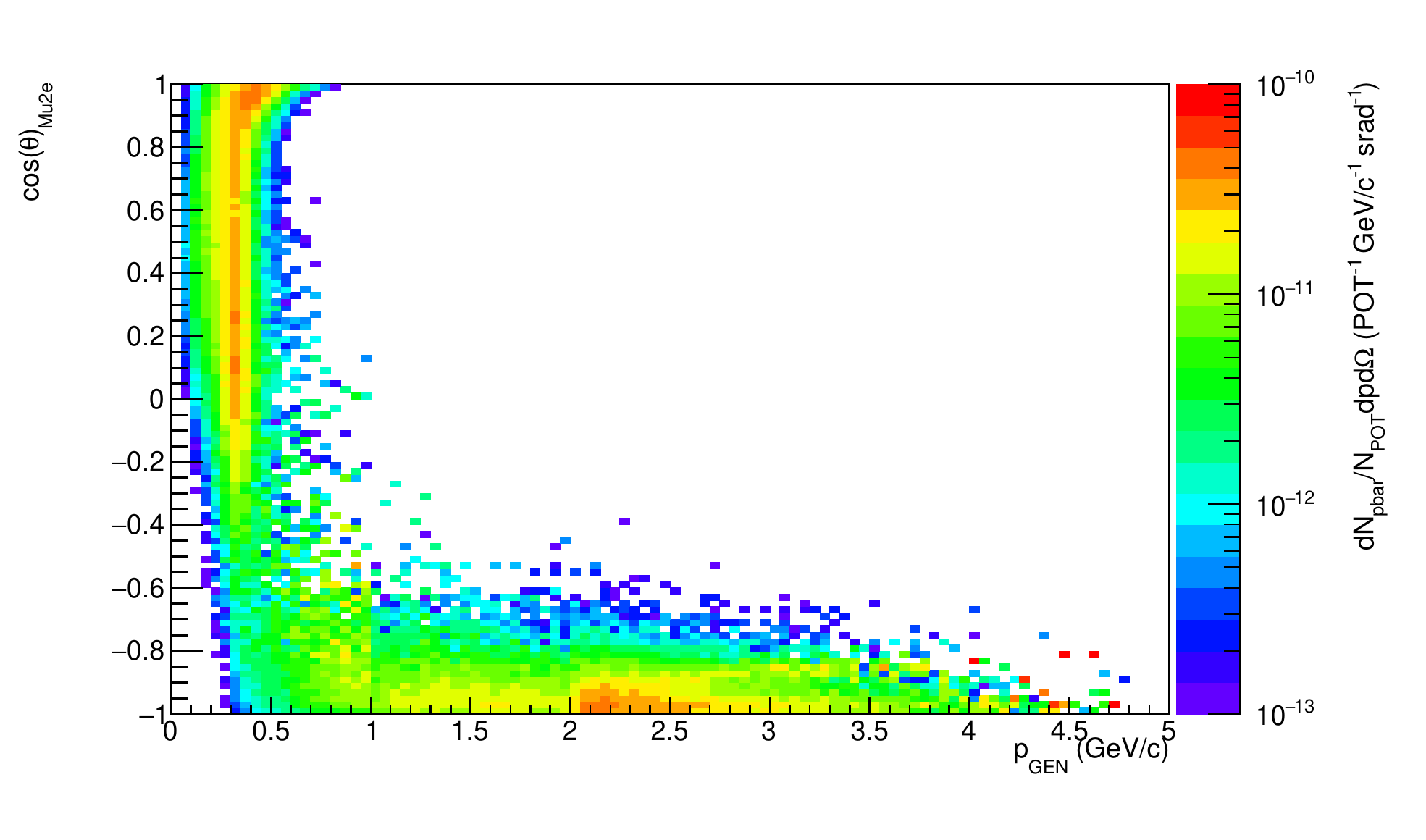}
  \caption{
    \label{fig:pbar_tsacc}
    Left: The ratio of the TS acceptances calculated using MCNP and Geant4 as a function
    of the generated antiproton momentum and $cos(\theta)$ in the Mu2e reference frame.
    The Mu2e reference frame is defined in Figure~\ref{fig:mu2e_layout}.
    Right: The number of antiprotons reaching the TS per POT, per unit momentum and solid angle.
    This includes the antiproton production cross section weights and the TS acceptance correction weights.
%    \todo[color=green!10, inline] { (Michael) The x-axis label on the left plot should have $p_{gen}$ to match $\theta_{gen}$ if they're both gen     values. Can this plot be made with MCNP/Geant4 to directly show the weights? Also, the first sentence should specify the left figure.      The last sentence isn't very clear to me and maybe should be broken up. Maybe:       ``The right figure shows the number of antiprotons reaching the TS per POT, per unit momentum and solid angle. This includes the antiproton       production cross section weights and the TS acceptance correction weights.''?}
    }
\end{figure}

The TS acceptance calculation by Geant4 was cross-checked against simulations based
on FLUKA \cite{Battistoni}, MARS \cite{Mashnik2009}, and MCNP.
% \todo[color=gray!20,linecolor=gray,tickmarkheight=10pt] {
%   (Michael) MCNP was already mentioned above, perhaps move this citation to there?
% }
Compared to Geant4, all three MC codes produced a much higher fraction of back-scattered antiprotons.
For this reason, the TS acceptance has been corrected by introducing an additional event weight
defined by the ratio of the MCNP and Geant4 acceptances -- see Figure~\ref{fig:pbar_tsacc} (left).

Figure~\ref{fig:pbar_tsacc} (right) shows the two-dimensional distribution of $cos\theta_{Mu2e}$ vs $p$
for antiprotons reaching the TS, where $p$ and $\theta_{Mu2e}$ are the momentum and the polar angle of
the generated antiproton at its production vertex.

An antiproton reaching the TS can be produced by the interaction of the generated antiproton.
This is usually the case for the forward produced antiprotons.
The antiprotons emitted in the direction of the TS ($cos\theta_{Mu2e} {\sim} 1$) can in principle have any momentum
but because of the cross section have essentially $p<1$ \GeVc. The ones generated in the direction opposite to the TS
($cos\theta_{Mu2e} {\sim} -1$) are much more enhanced by the cross section
and a small but relevant fraction of them undergo secondary interactions in the production target
producing a secondary antiproton reaching the TS.

To optimize the simulation time, each antiproton reaching the TS entrance is resampled $10^5$ times.
It has been verified that, given the large amount of material crossed
by the antiprotons from the TS entrance to the stopping target,
this resampling factor does not significantly affect the final statistical error.
A set of optimized absorbers has been added at the entrance and the center of the TS
to suppress antiproton backgrounds while minimizing the introduced delayed RPC backgrounds
and not significantly affecting the number of muons stopped in the stopping target.
The expected number of antiprotons stopped in the stopping target in Run I is

\begin{equation}
  \label{s4yield}
  % \frac{N_{\bar p}^{STOPPED}}=(4.8\pm 0.4\ (stat)^{+4.8}_{-1.9}\ (syst)\ ) \times 10^{-18}
  N_{\bar p}^{STOPPED}=180\pm 15\ (stat)\pm 180\ (syst)\ \ % \rm{(Mu2e\ Run\ I)}
\end{equation}
where the systematic error is dominated by the uncertainty on the production cross section (Eq. \ref{sigma89}).

\begin{figure}[h]
  \includegraphics[width=0.49\linewidth]{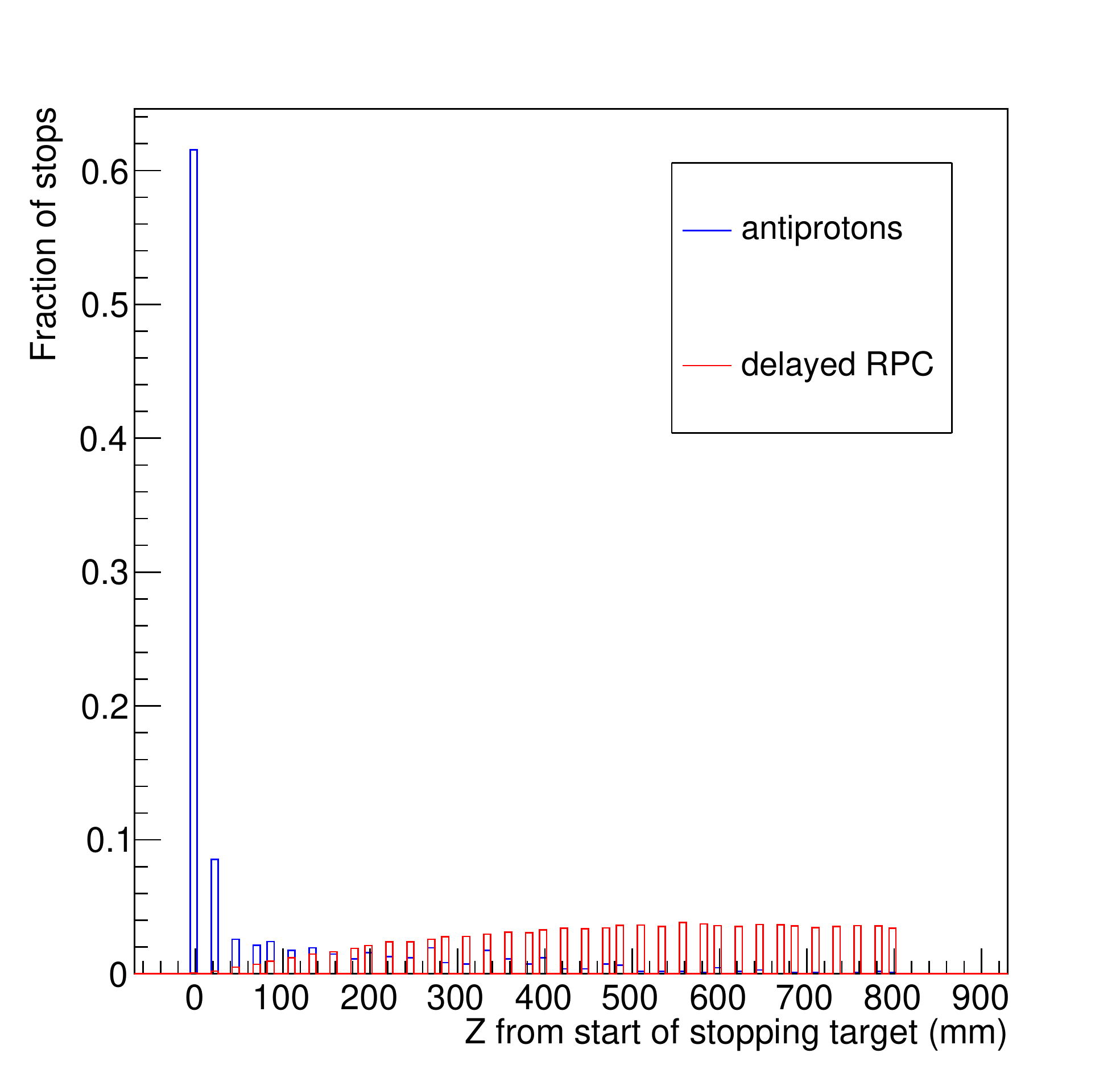}
  \includegraphics[width=0.49\linewidth]{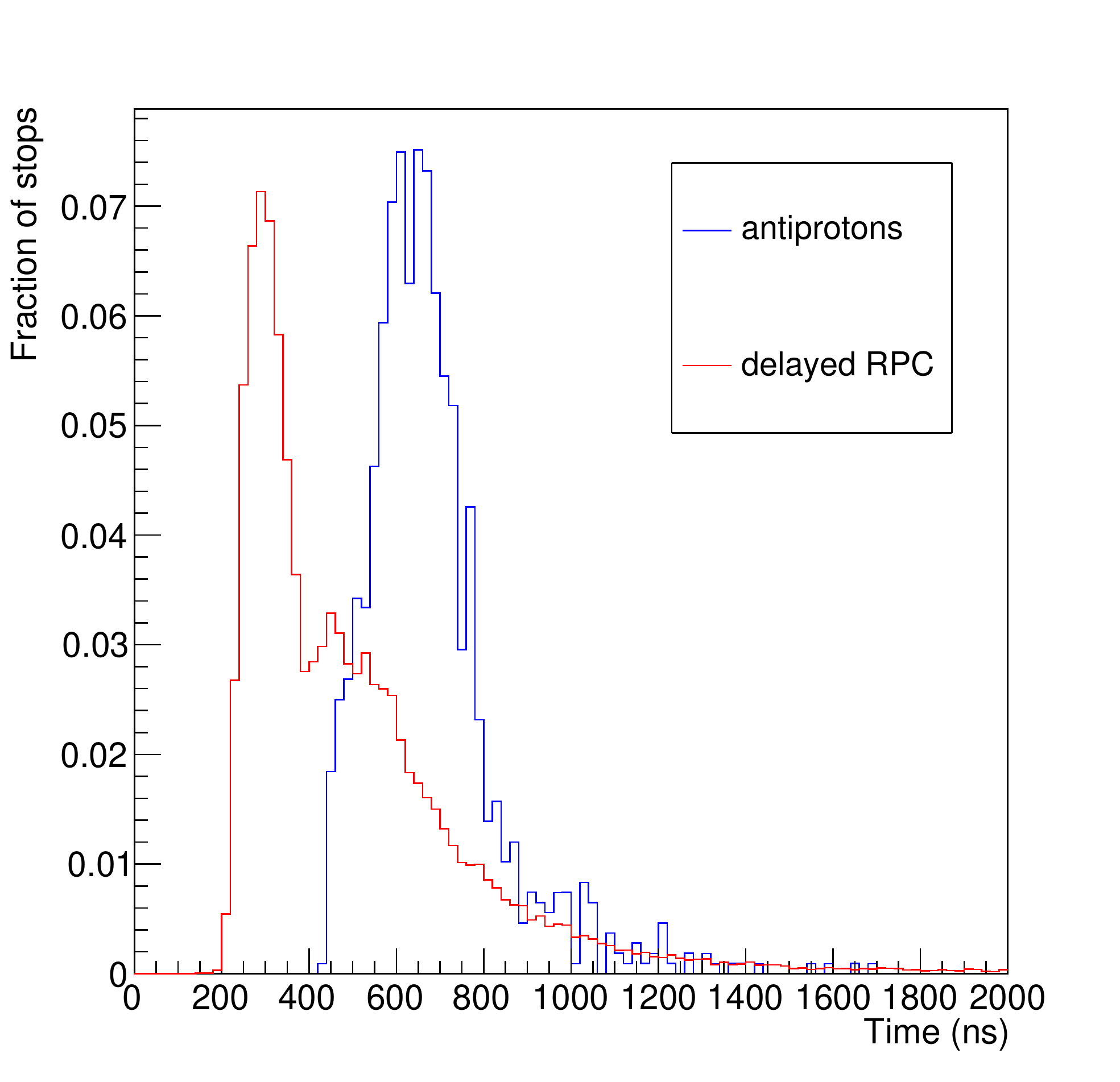}
  \caption{
    \label{fig:pbarstop}
    Longitudinal position (left) and time (right) of $\bar p$ annihilations (blue) and delayed RPC stops (red) in the stopping target.
  }
\end{figure}

The space and time distribution of the stopped $\bar p$ is shown in Figure~\ref{fig:pbarstop}.
Most of the antiprotons stop in the first aluminum foil of the stopping target.
The stopping time can be within the conversion electron search window.

%
% \noindent
% \todo[color=gray!20,linecolor=gray,tickmarkheight=10pt] {
% (Michael) Perhaps remove this ``noindent'' or move the text to paragraph above
% }

Antiproton annihilations in the stopping target are simulated using the position and time of
the stopped antiprotons. The background electrons produced in these annihilations are due
to $\pi^0 \to \gamma\gamma$ decays followed by the photon conversions and $\pi^- \to \mu^- \bar{\nu}$
decays followed by the negative muon decays.
The background due to antiproton annihilations in the signal momentum and time window for Run I
is $N_{\bar p}^{BKG}=(8.1\pm 0.7(stat)\pm 8.1(syst)) \times 10^{-3}$.
%When CE selection is applied, in the Mu2e Run-I optimized signal momentum window, $103.6<p< 104.9$ MeV/c, and time window, $640 < T < 1650$ ns, the number of background electrons per POT produced by $\bar p$ annihilations is shown in Tab. \ref{tab:PBAR_BKG_per_POT}.

%% \begin{table}[h!]
%% \begin{center}
%% \begin{tabular}{l|c|c}
%% \textbf{In Optimized Signal Window} & \textbf{first running period} & \textbf{second running period} \\
%% \hline
%% $\bar p$ BKG per POT & 2.21$\times10^{-22}$        &  $1.93\times10^{-22}$    \\
%% \end{tabular}
%% \end{center}
%% \caption{
%% \label{tab:PBAR_BKG_per_POT}
%% Antiproton background per proton on target in optimized signal window, estimated assuming first and second running periods.
%% }
%% \end{table}
%% \todo[color=green!10, inline] { (Michael) This should use the special table suggested, but it also doesn't seem important to give the
%%   rates for each running period and this could just be the background estimate for Run I}

\subsubsection{Delayed RPC Simulations}
The background due to the pions produced before the TS is considered as standard RPC background,
whether the pions come from
% \todo[color=gray!20,linecolor=gray,tickmarkheight=10pt] { (Michael) Perhaps ``...whether or not the pions come from...''}
a proton interaction or from an antiproton annihilation.
An additional antiproton-induced background comes from pions produced by antiproton interactions inside the TS.
%These events are more dangerous than the `traditional’ RPC background because they arrive later
%and are more likely to enter the time window used to select the CE signal.
These pions arrive at the stopping target later, and electrons resulting from their captures
are more likely to pass the timing cuts used to select the CE signal.
The first stages of the delayed RPC simulation are the same as used for the antiprotons.
Starting from the TS, the pions produced by antiproton annihilations are traced down
to the stopping target: they can decay along the way or, eventually, reach the stopping target and stop there.
Figure~\ref{fig:pbarstop} shows the time and position of pion stops in the stopping target.
A peak in the timing distribution around 300 ns corresponds to pions produced in the first antiproton absorber
positioned in front of the TS. A broad distribution with the maximum around 500-600 ns is due to pions
produced in antiproton annihilations in the second absorber located in the middle of the TS.
% Many pions are not delayed much because they are produced early, at the beginning of TS,
% so the $\bar p$ time of flight does not contribute too much.

The times and positions of the pion stops are used to produce RPC events.
As with the standard RPC background, the background contribution of both
the virtual (internal conversions) and the real (external conversions) photons
have been estimated separately and added up.
Assuming the proton extinction is better than $10^{-10}$, the contribution of
the out-of-time protons is negligibly small, and the background due to the delayed RPCs
in the signal momentum and  time window for Run I
is $N_{delRPC}^{BKG}=(2.3\pm 0.2(stat)\pm 2.3 (syst)) \times 10^{-3}$.
%The background yield per POT corresponding to the Mu2e Run-I optimized signal momentum and time window is shown in Tab. \ref{tab:PBARRPC_BKG_per_POT}.
%% \begin{table}[h!]
%% \begin{center}
%% \begin{tabular}{l|c|c}
%% \textbf{In Optimized Signal Window} & \textbf{first running period} & \textbf{second running period} \\
%% \hline
%% delayed RPC BKG per POT & 6.20$\times10^{-23}$        &  $5.42\times10^{-23}$    \\
%% \end{tabular}
%% \end{center}
%% \caption{
%% \label{tab:PBARRPC_BKG_per_POT}
%% Delayed RPC background per proton on target produced by $\bar p$ annihilations in the TS. The Mu2e Run I optimized signal window has been used for both the running periods.
%% \todo[color=green!10, inline] { (Michael) This should use the special table, and ``for both of the running periods''}
%% }
%% \end{table}
%
The delayed RPC background is a significant component of the antiproton background,
constituting 22\% of the total. As for the $\bar p$ annihilations,
the dominant systematic error for the estimate of this background is given
by the uncertainty on the antiproton production cross section (Eq. \ref{sigma89}).

%%%%%%%%%%%%%%%%%%%%%%%%%%%%%%%%%%%%%%%%%%%%%%%%%%%%%%%%%%%%%%%%%%%%%%%%%%%%%%
\subsection{Other Background Sources}
\label{sec:beam_bkg}

Several small beam-related background contributions are due to particles not stopping
in the stopping target. All of them originate from protons arriving at the production
target between the proton pulses and are suppressed by the proton beam extinction.
\begin{itemize}
\item
  Beam electrons with momentum around 105 \MeVc\ that arrive at the stopping target
  and scatter there could get reconstructed in the detector and fake the signal.
  The main source of such electrons are muons decaying in the downstream half
  of the TS and in the DS, in front of the stopping target.
  The small, ${\sim} 10 \time 10^{-6}$, probability of a large angle scattering
  in the stopping target
  combined with the beam extinction of $10^{-10}$ reduces the expected
  contribution from beam electrons to a level below $1 \times 10^{-3}$ events;
\item
  Negative muons and pions that enter the DS and decay in flight there,
  producing electrons with momenta above 100 \MeVc.
  The electrons could get reconstructed without scattering in the stopping target
  and mimic the \MuToEm\ conversion signal.
  The estimated contribution from decays in flight is below $ 2 \times 10^{-3}$ events.
\item
  The expected background from the DIO of muons stopped in the TS is negligibly small.
\end{itemize}
\noindent
Because of their small expected values, the backgrounds described in this section
are not considered in the sensitivity optimization procedure.

%%% Local Variables:
%%% mode: latex
%%% TeX-master: t
%%% End:

%Main author: Michael MacKenzie

\section {Sensitivity Optimization}
\label{sec:sensitivity_optimization}

%%%%%%%%%%%%%%%%%%%%%%%%%%%%%%%%%%%%%%%%%%%%%%%%%%%%%%%%%%%%%%%%%%%%%%%%%%%%%%
\subsection {Optimization Strategy}
The experimental sensitivity estimate in this analysis is based on simple event counting.
The event counting is performed in a two-dimensional momentum and time signal window,
so the optimization of the experiment's sensitivity to discovery is reduced to the
optimization of the signal window limits.

% \noindent
A standard measure of an experiment's ability to make a discovery is its "median discovery potential"
characterized by the minimal signal strength for which, given the mean background expectation $\mu_B$,
the probability to satisfy the discovery criterion would be at least 50\% .
Standard for HEP, a discovery is defined as a measurement yielding a significant, "5$\sigma$",
deviation from the expected background with the probability
% for the background-only hypothesis to result in a measured or a larger deviation

$$
% P_{\rm obs} < \int_{5}^{\infty} \! e^{-x^2/2} \, \mathrm{d}x/{\sqrt{2 \pi}}  = 2.87 \times 10^{-7}
P < \int_{5}^{\infty} \! e^{-x^2/2} \, \mathrm{d}x/{\sqrt{2 \pi}}  = 2.87 \times 10^{-7}
$$

\noindent
While this definition is very clear, it may not provide the best figure of merit for the sensitivity optimization.
Due to the discrete nature of the measurement, the same number of events is needed to claim a discovery for a range
of $\mu_B$ values. In this case, higher background values correspond to better sensitivities,
which is rather counter-intuitive. A better figure of merit is the average discovery potential,
defined as the signal strength that corresponds to an average $5\sigma$
deviation from the background-only hypothesis. Using the average discovery potential avoids
the known pathologies of the median discovery potential -- see the discussion by Bhattiprolu et al.
comparing these and other methods of quoting the discovery potential \cite{STATISTICS_2021_Bhattiprolu}.
It is also similar to the method proposed by Feldman and Cousins (FC), where the average
of the distribution of upper limits from pseudo-experiments, as opposed to the median expectation,
is used to quantify the experimental sensitivity \cite{STATISTICS_1998_FELDMAN_COUSINS}.
To combine the best of both approaches -- avoid numerical pitfalls in the optimization
procedure and have a clear definition of the discovery potential --
the sensitivity optimization is performed in two steps.
First, the sensitivity is optimized using the "mean" definition of the signal strength
as the figure of merit, and the position and size of the two-dimensional signal window
are determined.
Next, the "median" signal strength is calculated for the optimized selection
and used to quote the $5\sigma$ discovery sensitivity and the expected upper limits.

%%%%%%%%%%%%%%%%%%%%%%%%%%%%%%%%%%%%%%%%%%%%%%%%%%%%%%%%%%%%%%%%%%%%%%%%%%%%%%
\subsection {Optimization of the Momentum and Time Signal Windows}

The upper and lower edges of the momentum and time windows are optimized using
the mean discovery potential described above. The rapid rise of the DIO momentum
distribution prevents the optimization from moving the lower edge of the momentum
window significantly below ${\sim} 103.5 ~\MeVc$.
Similarly, extending the window above 105 \MeVc\ does not improve the signal acceptance,
adding only the background.
The lower edge of the timing window is constrained by the RPC background,
the contribution of which becomes large, on a scale of 0.01 events, for $T_{0}$ below 650 ns.
%% Antiproton-induced backgrounds reaches its maximum at around 700 ns, but starts decreasing at earlier times.
To avoid background from the flash from the next proton pulse, the maximal value of T$_{0}$ is set to 1650 ns.

The momentum and time windows are optimized using a grid search in steps of 50 \keVc\ in momentum
and 10 ns in time for both the upper and lower edges of the windows.
The optimized momentum window is $103.60 < p < 104.90$ \MeVc\ and
the optimized time window is $640 < T_{0} < 1650$ ns, as introduced in Section \ref{sec:background_estimates}.
%% This leads to a fairly balanced contribution of DIO, Cosmic, antiproton, and RPC backgrounds.
One of the parameters characterizing the sensitivity of an experiment to a process of interest
is its single event sensitivity (SES), defined as the signal strength corresponding to a
mean expectation of one observed signal event. The optimized Mu2e signal window corresponds
to a SES of $2.3\times 10^{-16}$ and a total signal selection efficiency of $11.7\%$.

%%%%%%%%%%%%%%%%%%%%%%%%%%%%%%%%%%%%%%%%%%%%%%%%%%%%%%%%%%%%%%%%%%%%%%%%%%%%%%
\subsection {Including Systematic Uncertainties}
\label{sec:including_systematic_uncertainties}

The signal window optimization is performed without taking systematic uncertainties
into account. After the optimal signal window is determined, the expected sensitivity
is recalculated with the systematic uncertainties included.
The expected sensitivity is optimized assuming a fixed number of stopped muons,
$6\times 10^{16}$, defined in Table \ref{tab:running_time}.
The included uncertainties represent the current best estimate of what
they will be at the time the analysis is performed.
The systematic uncertainties are treated as nuisance parameters with specified
probability density functions (PDF). % and are marginalized over.
Uncertainties associated with the current predictions of the detector 
performance are not used at this step. % in the marginalization.
An example of such uncertainty is an uncertainty of predicting
the CRV light yield during the data-taking.
The construction of the FC confidence belts in the presence of systematic uncertainties
follows the method described in Ref. \cite{STATISTICS_2003_CONRAD},
with numerical approximations made to speed up the execution.

Table \ref{tab:systematic_uncertainties} lists the systematic uncertainties.
Uncertainties on the PID and the track reconstruction efficiency are expected to be significantly
smaller than 5\%, so Table \ref{tab:systematic_uncertainties} does not include them.

In the sensitivity calculation, the uncertainties are implemented using log-normal PDFs.
In case of asymmetric errors, the larger uncertainty value has been used
to parameterize the PDF. The choice of log-normal representation of PDFs avoids
negative background expectations. In addition, compared to the choice of Gaussian
representation, it results in more conservative sensitivity estimates.
%\todo[color=gray!20,linecolor=gray,tickmarkheight=10pt] {
%  (Michael) The DIO and RPC uncertainties don't match the total in their sections
%}

\begin{specialtable}[H]
  \centering
  \small
  \caption{
    \label{tab:systematic_uncertainties}
    Systematic uncertainties used in the sensitivity optimization procedure.
    The muon flux uncertainty is correlated between the signal and the DIO
    and RPC backgrounds.
  }
  \begin{tabular}{lcl}
    \toprule
    Parameter              & Total relative uncertainty  &  Dominant contribution   \\
    \midrule
    Signal acceptance      &    4\%                      & Momentum scale \\
    Antiproton background  &  100\%                      & $\bar{p}$ production cross section   \\
    Cosmic background      &   20\%                      & Cosmic flux normalization       \\
    DIO background         &   59\%                      & Momentum scale                  \\
    RPC background         &   29\%                      & Pion production cross section   \\
    Muon flux              &   10\%                      & Flux measurement                \\
    \bottomrule
  \end{tabular}
\end{specialtable}

%%%%%%%%%%%%%%%%%%%%%%%%%%%%%%%%%%%%%%%%%%%%%%%%%%%%%%%%%%%%%%%%%%%%%%%%%%%%%%
\subsection{$\MuToEm$ Sensitivity Estimate}
\label{sec:sensitivity_estimate}

\begin{figure}[H]
  \centering
  \includegraphics[width=1.0\linewidth]{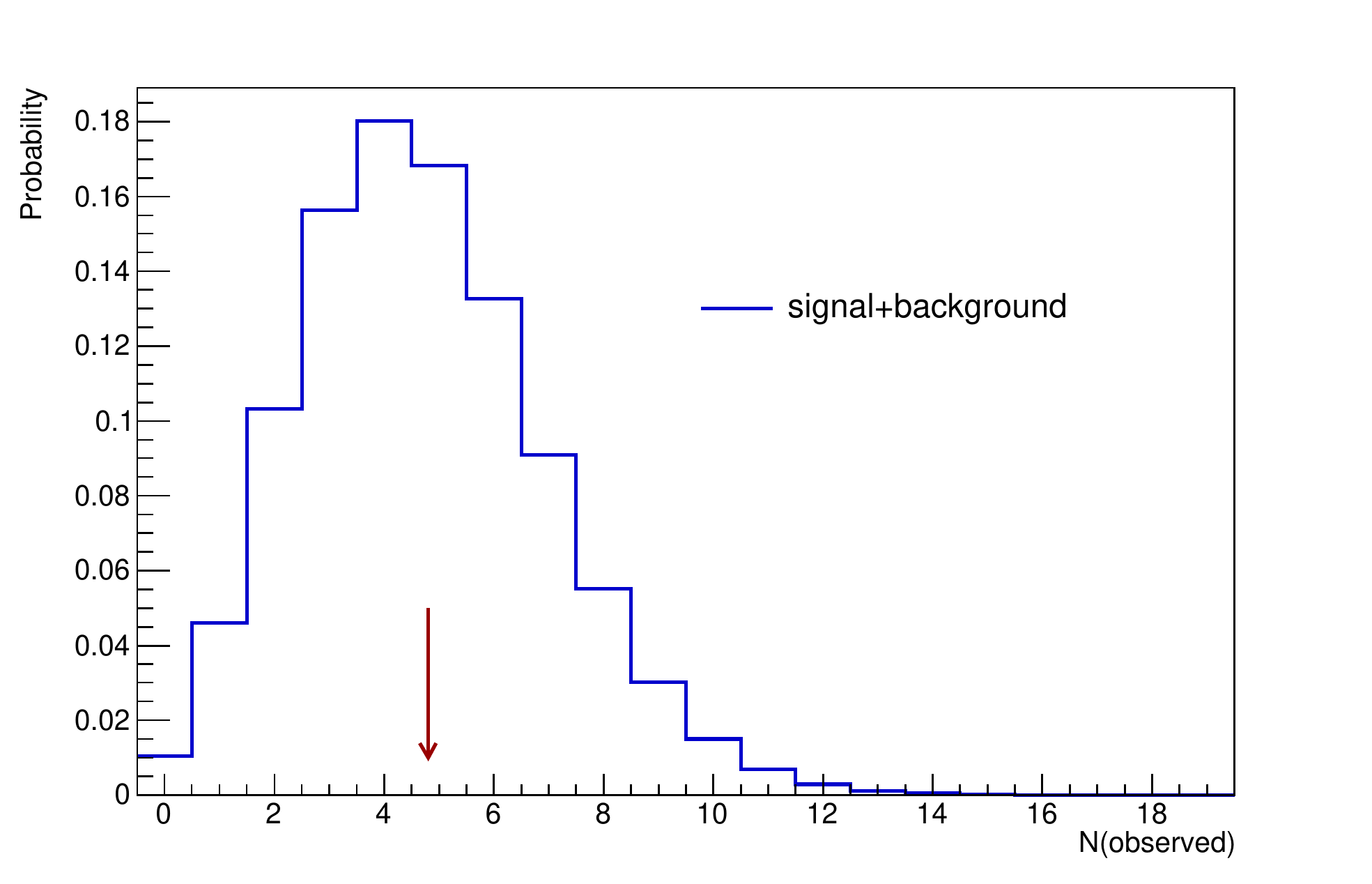}
  \caption{
    \label{fig:discovery_pdf}
    Probability for Mu2e to observe in Run I a given number of events for a \MuToEm\ signal
    corresponding to $R_{\mu e} = 1.2 \times 10^{-15}$. The red arrow represents the mean number
    of signal events corresponding to this $R_{\mu e}$ value.
%
%    \todo[color=gray!20, inline] { (Michael) Maybe it should be N(obs) centered bins? And is this the mean N(signal) before
%      or after marginalizing over the nuisance parameters?}
  }
\end{figure}

Table \ref{tab:optimization_summary} presents the Mu2e Run I discovery potential
and exclusion limit with and without the systematic uncertainties included. 
The 5$\sigma$ discovery $R_{\mu e} = 1.2 \times 10^{-15}$ , and claiming 
a \MuToEm\ signal requires an observation of 5 or more events.
Taking the systematic uncertainties into account degrades the expected
sensitivity values by about 10\%. 
As shown in Figure~\ref{fig:discovery_pdf}, for this $R_{\mu e}$ value ,
the observed number of events $ 2 \le N_{\rm obs} \le 7$ with a probability of about 75\% .
The background summary after the sensitivity optimization is given in Table \ref{tbl:background_summary_optimized}.

% \del{
%   Estimating the sensitivity for a fixed number of muon stops avoids complications related to one of the largest experimental uncertainties - uncertainty on the stopped muon rate, $N^{\mu^-}_{\rm POT}$. Under this assumption, variations of the stopped muon rate change only the data taking time and, through that, the cosmic ray background.
% }

Estimating the sensitivity for a fixed number of stopped muons makes the estimate
largely independent of one of the current largest experimental uncertainties,
the uncertainty on the stopped muon rate, $N^{\mu^-}_{\rm POT}$. 
A change in the stopped muon rate changes the data-taking time needed
to collect the required number of stopped muons, and through that, the cosmic ray background.
A stopped muon rate twice as low as the number used for the sensitivity estimate
would increase the running time by a factor of two and double the cosmic ray background.
However, the total background would increase by only about 50\%, changing the median discovery
$R_{\mu e}$ by less than 5\%. Moreover, a total background increase by a factor of three would
degrade the discovery $R_{\mu e}$ by only about 30\%.

Alternatively, for a constant data taking time, the discovery $R_{\mu e}$ would scale approximately as
$1/N^{\mu^-}_{\rm POT}$.

The current world's best limit on the \MuToEm\ conversion search, $R_{\mu e} < 7 \times 10^{-13}$ at 90\% CL,
% v1_01 Andy "on gold target" --> "on a gold target"
has been set by the SINDRUM II experiment on an Au target \cite{2006_CLFV_SINDRUM_II_GOLD}.
Compared to SINDRUM II, in Run I, Mu2e is expected to improve the search sensitivity
by a factor of more than 1,000.

\begin{specialtable}[H]
  \centering
  \small
  \caption{
    \label{tab:optimization_summary}
    Summary of the sensitivity optimization. The sensitivity values are given with and without the inclusion of
    systematic uncertainties.
  }
  \begin{tabular}{lccc}
    \toprule
    Configuration                         & Discovery $\Rmue$     & $\Rmue$ (90\% CL limit)   & N(discovery events)  \\
    \midrule
    No systematics                        & $1.1\times 10^{-15}$   & $5.7\times 10^{-16}$       & 5                    \\
    With systematics                      & $1.2\times 10^{-15}$   & $6.2\times 10^{-16}$       & 5                    \\
    \bottomrule
  \end{tabular}
\end{specialtable}

% Figure \ref{fig:discovery_pdf} shows the probability distribution to observe a given number
% of events for a \MuToEm\ signal corresponding to $R_{\mu e} = 1.2 \times 10^{-15}$.
% \todo[color=gray!20,linecolor=gray,tickmarkheight=10pt] { (Michael) This doesn't seem needed as the figure was introduced above}

\begin{specialtable}[H]
  \centering
  \small
  \caption{
    \label{tbl:background_summary_optimized}
    Background summary and SES using the optimized signal momentum and time window,
    $103.60 < p < 104.90$ \MeVc\ and $640 < T_0 < 1650$ ns.
    % \todo[color=gray!20,inline] { (Michael)
    %  The cosmics and DIO backgrounds disagree with the previous sections
    %  (likely some effect from a scale factor like deadtime?)
    % }
    }
  \begin{tabular}{l|c}
    \toprule
    Channel             &    Mu2e Run I                                                  \\
    \midrule
    SES                 &   $2.4  \times 10^{-16}$                                         \\
    \midrule
    Cosmic rays         &   $0.046 \pm 0.010~({\rm stat})   \pm 0.009 ~({\rm syst})$       \\
    DIO                 &   $0.038 \pm 0.002 ~({\rm stat})~ ^{+0.025}_{-0.015}~~ ({\rm syst})$ \\  % scale: 38.4/9.1   (9.1/9.6)
    Antiprotons         &   $0.010 \pm 0.003  ~({\rm stat})~ \pm 0.010 ~ ({\rm syst})$         \\  % scale (10.54/7.2)
    RPC in-time         &   $0.010 \pm 0.002 ~({\rm stat})~ ^{+0.001}_{-0.003}~ ({\rm syst})$     \\   % scale (108.7 / 6.6)          (6.6 / 7.05)
    RPC out-of-time ($\zeta =10^{-10}$) & $(1.2 \pm 0.1 ~({\rm stat})~ ^{+0.1}_{-0.3}~ ({\rm syst}))\times 10^{-3}$  \\
%       &                                              \\
    RMC                 &   $ < 2.4 \times 10^{-3}$                       \\
    Decays in flight    &   $ < 2   \times 10^{-3}$                         \\
    Beam electrons      &   $ < 1   \times 10^{-3}$                         \\
    \midrule
    Total               &   $0.105 \pm 0.032$                            \\
    \bottomrule
  \end{tabular}
\end{specialtable}

\begin{figure}[H]
  \centering
  \includegraphics[width=0.49\linewidth, trim = 10 0 55 0, clip]{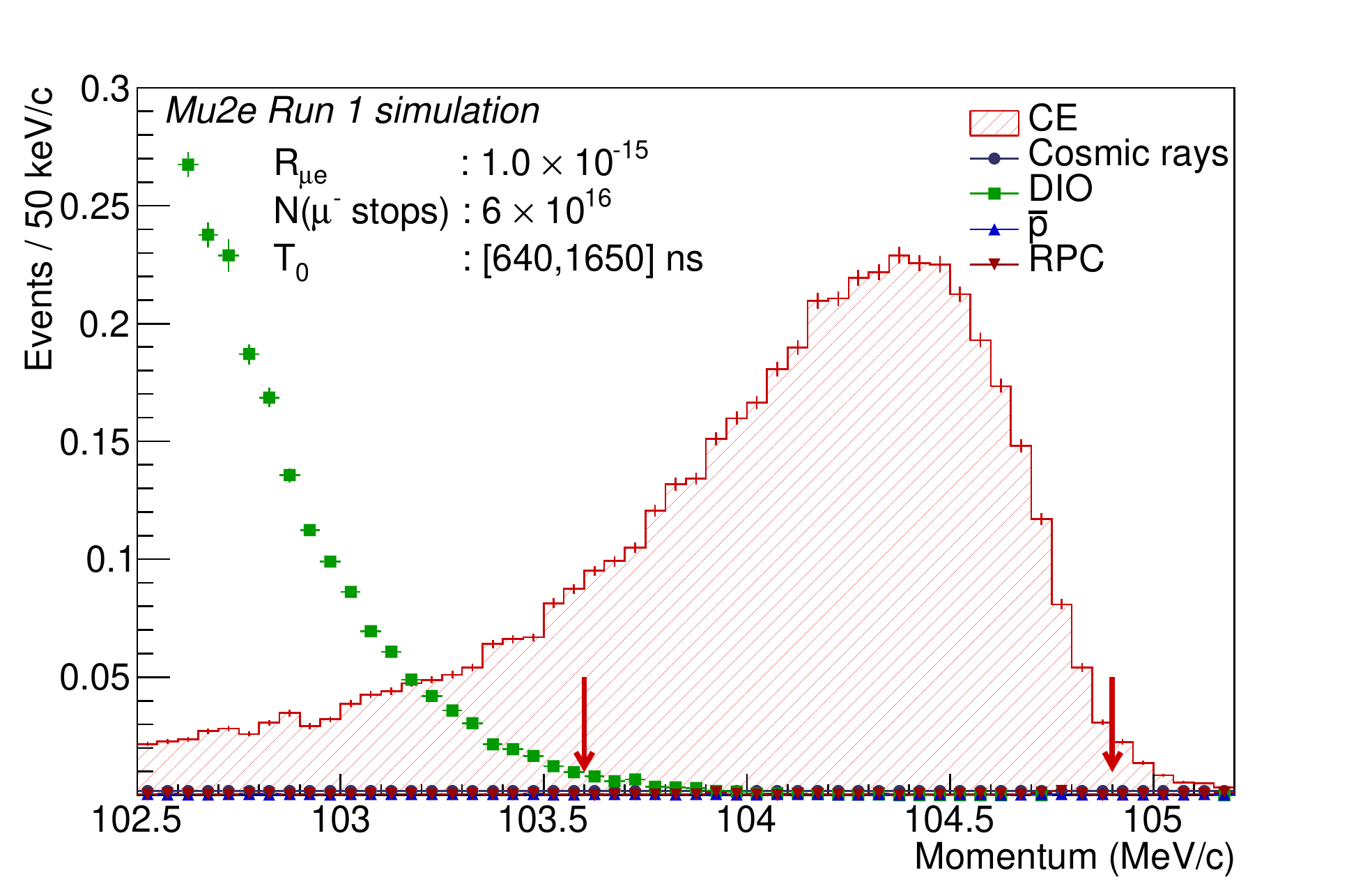}
  \includegraphics[width=0.49\linewidth, trim = 10 0 55 0, clip]{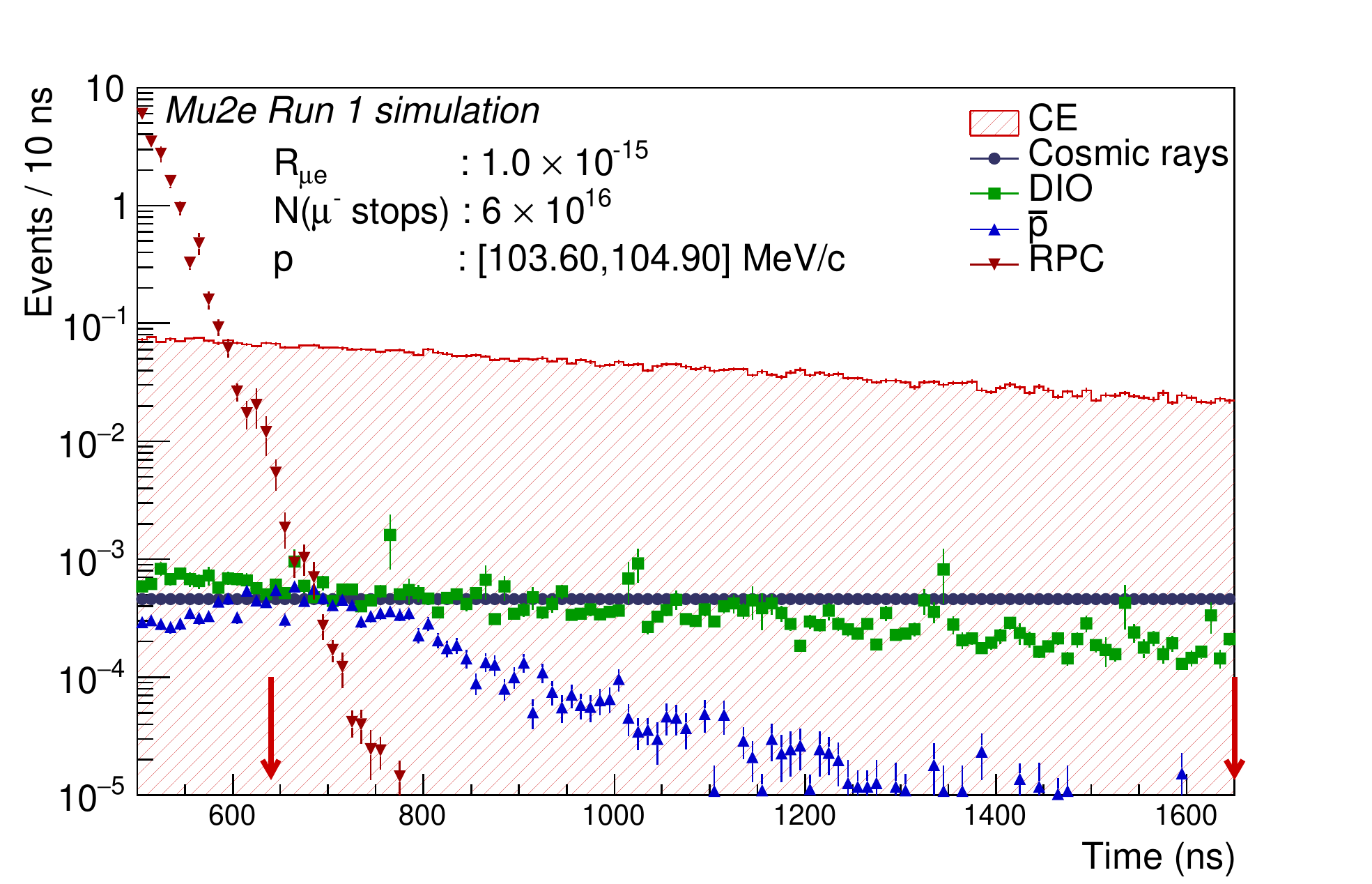}
  \caption{
  \label{fig:momentum_and_time_optimized}
  Electron momentum (left) and time (right) distributions after optimization of the signal momentum
  and time window. The CE signal distributions correspond to $R_{\mu e} = 1\times 10^{-15}$.
  The background estimate numbers are the integrals over the optimized signal window,
  $103.60 < p < 104.90$ \MeVc\ and $640 < T_0 < 1650$ ns.
  The error bars represent statistical uncertainties only.
  }
\end{figure}

\noindent
Figure~\ref{fig:momentum_and_time_optimized} shows the momentum and time distributions for the
\MuToEm\ signal and individual background processes corresponding to the optimized signal window.

%%% Local Variables:
%%% mode: latex
%%% TeX-master: t
%%% End:

\section{Summary}

We present an updated estimate of the expected Mu2e sensitivity to the
search for the neutrinoless \MuToEm\ conversion on an Al target.
Mu2e Run I, the first part of the Mu2e data-taking plan described
in Section \ref{experiment_config}, assumes an integrated flux of $6 \times 10^{16}$ stopped muons.
The discovery 
%\todo[color=gray!20,linecolor=gray,tickmarkheight=10pt] { (Michael) ``The minimum value?''}
$R_{\mu e}$ corresponding to a 50\% probability of observing the \MuToEm\ conversion signal
at a $5\sigma$ significance level is $R_{\mu e}^{5\sigma} = 1.2 \times 10^{-15}$.
Reaching the $5 \sigma$ significance level requires observing 5 \MuToEm\ candidate events
in the two-dimensional search window $103.60 < p < 104.90$ \MeVc, $640 < T_0 < 1650$ ns.
The corresponding expected background is $0.11 \pm 0.03$ events, significantly lower than one event.

In the absence of a signal, the expected 90\% CL upper limit on the \MuToEm\ conversion rate
is $R_{\mu e} ~<~ 6.2 \times 10^{-16}$, a factor of ${\sim} 10^3$ improvement
over the current experimental limit $R_{\mu e} ~<~7 \times 10^{-13}$ at 90\% CL \cite{2006_CLFV_SINDRUM_II_GOLD}.

In the second part of the data-taking plan, Run II, Mu2e is expected to improve the experimental
sensitivity of the \MuToEm\ conversion search by another order of magnitude.
\vspace{6pt}

\acknowledgments{

We are grateful for the vital contributions of the Fermilab staff and the technical staff of the participating institutions. 
This work was supported by
the US Department of Energy;
the Istituto Nazionale di Fisica Nucleare, Italy;
the Science and Technology Facilities Council, UK;
the Ministry of Education and Science, Russian Federation;
the National Science Foundation, USA;
the National Science Foundation, China;
the Helmholtz Association, Germany;
and the EU Horizon 2020 Research and Innovation Program
under the Marie Sklodowska-Curie Grant Agreement
Nos. 734303, 822185, 858199, 101003460, and 101006726.
This document was prepared by members of the Mu2e Collaboration using
the resources of the Fermi National Accelerator Laboratory (Fermilab),
a U.S. Department of Energy, Office of Science, HEP User
Facility. Fermilab is managed by Fermi Research Alliance, LLC (FRA),
acting under Contract No. DE-AC02-07CH11359
}

\conflictsofinterest{The authors declare no conflict of interest.}

%% Optional
%\sampleavailability{Samples of the compounds ... are available from the authors.}

%%%%%%%%%%%%%%%%%%%%%%%%%%%%%%%%%%%%%%%%%%
%% Only for journal Encyclopedia
%\entrylink{The Link to this entry published on the encyclopedia platform.}

%%%%%%%%%%%%%%%%%%%%%%%%%%%%%%%%%%%%%%%%%%
%% Optional
% \abbreviations{Abbreviations}{
% The following abbreviations are used in this manuscript:\\
%
% \noindent
% \begin{tabular}{@{}ll}
% MDPI & Multidisciplinary Digital Publishing Institute\\
% % DOAJ & Directory of open access journals\\
% % TLA  & Three letter acronym\\
% % LD   & Linear dichroism
% \end{tabular}}

%%%%%%%%%%%%%%%%%%%%%%%%%%%%%%%%%%%%%%%%%%%
\end{paracol}
%%%%%%%%%%%%%%%%%%%%%%%%%%%%%%%%%%%%%%%%%%
% To add notes in main text, please use \endnote{} and un-comment the codes below.
%\begin{adjustwidth}{-5.0cm}{0cm}
%\printendnotes[custom]
%\end{adjustwidth}
%%%%%%%%%%%%%%%%%%%%%%%%%%%%%%%%%%%%%%%%%%
\reftitle{References}


\externalbibliography{yes}
\bibliography{
  books,clfv_refs,dio_refs,tools,RPC,pbar,cosmics,radiative_muon_capture,
  radiative_pion_capture,statistics,trigger,pion_production,beam
}
\bibliographystyle{unsrt}
\end{document}